\documentclass[aps,pra,showpacs,amsmath,amssymb,twocolumn,10pt]{revtex4}
\usepackage{epsfig,graphicx,color}
\begin{document}
\newcommand{\Kijkl}{K_{ij,k}^{\,\,\, l}}
\newcommand{\Kikjl}{K_{ik,j}^{\,\,\, l}}
\newcommand{\Kjkil}{K_{jk,i}^{\,\,\, l}}
\newcommand{\Kiklj}{K_{ik,l}^{\,\,\, j}}
\newcommand{\Kjkli}{K_{jk,l}^{\,\,\, i}}
\newcommand{\Kklij}{K_{kl,i}^{\,\,\, j}}
\newcommand{\Kijlk}{K_{ij,l}^{\,\,\, k}}
\newcommand{\Hijkl}{H_{ij, kl}}
\newcommand{\Hikjl}{H_{ik, jl}}
\newcommand{\Hjkil}{H_{jk, il}}
\newcommand{\Hklij}{H_{kl, ij}}
\newcommand{\kijkl}{{\cal K}_{ij,k}^{\,\,\, l}}
\newcommand{\kikjl}{{\cal K}_{ik,j}^{\,\,\, l}}
\newcommand{\kjkil}{{\cal K}_{jk,i}^{\,\,\, l}}
\newcommand{\kiklj}{{\cal K}_{ik,l}^{\,\,\, j}}
\newcommand{\kjkli}{{\cal K}_{jk,l}^{\,\,\, i}}
\newcommand{\kklij}{{\cal K}_{kl,i}^{\,\,\, j}}
\newcommand{\kijlk}{{\cal K}_{ij,l}^{\,\,\, k}}
\newcommand{\hijkl}{{\cal H}_{ij, kl}}
\newcommand{\hikjl}{{\cal H}_{ik, jl}}
\newcommand{\hjkil}{{\cal H}_{jk, il}}
\newcommand{\hklij}{{\cal H}_{kl, ij}}
\newcommand{\brapsi}[1]{\left< \psi \right| }
\newcommand{\ketpsi}[1]{\left| | \psi \right> }
\newcommand{\brapsibar}[1]{\left< \overline{\psi} \right| }
\newcommand{\ketpsibar}[1]{\left| | \overline{\psi} \right> }
\newfont{\Yfont}{cmti10 scaled 2074}
\newcommand{\Y}{\hbox{{\Yfont y}\phantom.}}
\def\O{{\cal O}}
\newcommand{\bra}[1]{\left< #1 \right| }
\newcommand{\braa}[1]{\left. \left< #1 \right| \right| }
\def\Bra#1#2{{\mbox{\vphantom{$\left< #2 \right|$}}}_{#1}
\kern -2.5pt \left< #2 \right| }
\def\Braa#1#2{{\mbox{\vphantom{$\left< #2 \right|$}}}_{#1}
\kern -2.5pt \left. \left< #2 \right| \right| }
\newcommand{\ket}[1]{\left| #1 \right> }
\newcommand{\kett}[1]{\left| \left| #1 \right> \right.}
\newcommand{\scal}[2]{\left< #1 \left| \mbox{\vphantom{$\left< #1 #2 \right|$}}
\right. #2 \right> }
\def\Scal#1#2#3{{\mbox{\vphantom{$\left<#2#3\right|$}}}_{#1}
{\left< #2 \left| \mbox{\vphantom{$\left<#2#3\right|$}} \right. #3
\right> }}

\title{Binding and structure of tetramers in the scaling limit}
\author{M. R. Hadizadeh$^1$}
 \email{hadizade@ift.unesp.br}
\author{M. T. Yamashita$^1$}
 \email{yamashita@ift.unesp.br}
\author{Lauro Tomio$^{1,2}$}
 \email{tomio@ift.unesp.br}
\author{A. Delfino$^2$}
 \email{delfino@if.uff.br}
\author{T. Frederico$^3$}
 \email{tobias@ita.br}

\affiliation{$^1$Instituto de F\'{\i}sica Te\'orica,
Universidade Estadual Paulista, 01140-070, S\~ao Paulo, SP, Brazil\\
$^2$Instituto de F\'{\i}sica, Universidade Federal
Fluminense, 24210-346, Niter\'oi, RJ, Brazil\\
$^3$Instituto Tecnol\'ogico de Aeron\'autica, DCTA, 12228-900, S\~ao
Jos\'e dos Campos, SP, Brazil}

\date{\today}

\begin{abstract}
The momentum-space structure of the Faddeev-Yakubovsky (FY)
components of weakly-bound tetramers is investigated at the unitary
limit using a renormalized zero-range two-body interaction. The
results, obtained by considering a given trimer level with binding
energy $B_3$, provide further support to a universal scaling function
relating the binding energies of two successive tetramer states.
The correlated scaling between the tetramer energies comes from the
sensitivity of the four-boson system to a short-range four-body
scale. Each excited $N-$th tetramer energy $B_4^{(N)}$ moves as the
short-range four-body scale changes, while the trimer properties are
kept fixed, with the next excited tetramer $B_4^{(N+1)}$ emerging
from the atom-trimer threshold for a universal ratio
$B_4^{(N)}/B_3 = B_4^ {(N)}/B_4^{(N+1)} \simeq 4.6$, which does not
depend on $N$.
We show that both channels of the FY decomposition
[atom-trimer ($K-$type) and dimer-dimer ($H-$type)] present
high momentum tails, which reflect the short-range four-body scale.
We also found that the $H-$channel is favored over $K-$channel at
low momentum when the four-body momentum scale largely
overcomes the three-body one.
\end{abstract}
\maketitle
\section{Introduction}
\label{intro}
A limit cycle in physics~\cite{wilson71} refers to a model independent way to look
for a manifestation of a hidden scale, one in which a correlation between physical
quantities is geometrically rescaled and replicates itself. Limit cycles are shown
to exist even in simple quantum three-boson systems~\cite{amorim99,bira,mohr},
manifested for large two-body scattering lengths, as well as for short-ranged
interactions. They follow the 1970 Efimov key prediction~\cite{efimov70} of
geometrically separated weakly-bound three-boson states near the scattering
threshold, confirmed by recent experiments in cold-atom
traps~\cite{Kraemer,barontini,jochim2010}. For a more recent exposition on the
experimental status, see Ref.~\cite{phys-ferlaino}. The addition of one more
particle to the quantum three-body system has long challenged this picture~\cite{amado73,kroger-perne,adh-fonseca,tjon1}. Recent theoretical
works~\cite{meissner04,epl06,hammer07,stechernature,wang,deltuva2010,stecherjpa,Lazauskas06}
motivated by cold-atom experiments~\cite{ferlaino:140401,Zaccanti,hulet2009} have
revived this issue. The actual relevance in identifying simple universal relations
in few-body binding laws, with the corresponding experimental possibilities,
have been discussed by Modugno~\cite{modugno2009} in a recent perspective
article on this matter.

In Ref.~\cite{prl2011}, we reported a different dimension of this problem by establishing
a new universal correlation among the binding energies of two successive tetramer states
between two Efimov trimers, through precise numerical calculations within a zero-range
interaction at the unitary limit (zero two-body binding). It was shown that such tetramer
states are related to an unsuspected new limit cycle not constrained by  Efimov physics
or by three-body properties. Tetramers hitting the atom-trimer threshold, leading to
resonant recombination losses in this channel, are signatures of a new limit cycle.
Furthermore, other four-boson observables close to the unitary limit can exhibit
correlations not constrained by the low-energy two- and three-boson properties.
Recombination rates, atom-trimer or dimer-dimer scattering lengths, can move near the
Feshbach resonance (FR) independently of the two and three-body properties. The
appearance of few-body forces in the open channel, due to the coupling with the closed
channel, can drive the four-body physics~\cite{epl06}. Indeed, there is a recent
experimental evidence, reported by Nakajima et al.~\cite{nakajimaprl1}, of a three-body
force acting in the open channel near the FR in a three-component mixture of a $^6$Li
cold gas.

An appropriate way to directly probe the universal properties of a few-boson system
at low energy near the unitary limit is to present results in terms of scaling
functions correlating pair of observables, within a renormalized zero-range two-body
interaction. In this way, no effect can be claimed to be originated from a particular
form of the short-range interaction between the particles, as evidenced by the
three-body universal scaling function derived in Ref.~\cite{amorim99}. As shown in
this case, the regularization parameter in the renormalized zero-range approach can
be directly associated to a three-boson physical information, in addition to a
two-body observable, such as the scattering length.
The concept of scaling functions, which is used to evidence universal
properties of three-boson systems close to the unitary limit, is extended to
four-boson systems in Ref.~\cite{prl2011}, as well as in the present work.
Within such purpose, we study the sensitivity of tetramer properties to a
four-boson scale, by using the Faddeev-Yakubovsky (FY)
formalism~\cite{Faddeev,Yakubovsky} with zero-range two-body interactions.
Therefore, independently on the choice of the three-body scale that fixes the
trimer properties~\cite{epl06}, a four-body regularization parameter (four-body
scale) is introduced in the FY formalism.
The new short-range scale has indeed effects on four-boson physics, which appears
as universal correlations between four-boson low-energy $s-$wave observables not
constrained by the trimer properties. The four-body quantities change as long
as the new scale varies.

Previous conclusions for the nonexistence of a proper four-body
scale rely  on a strong suppression of the short-range physics,
beyond that already accounted for in the three-boson system. Indeed,
by estimating the trace of the kernel of the four-body equation in
momentum space, Amado and Greenwood~\cite{amado73} showed that there
is no infrared divergence, which led them to conclude against the
existence of Efimov effect in the case of four or more particles.
However, the momentum integrals should also implicitly have an
ultraviolet cutoff (the four-body one) to regulate them. Note that
according to Weinberg's eigenvalue criterion~\cite{weinberg} the
divergence of the trace is a necessary condition for an infinite
number of bound states. {However, in the kernel of the FY equation,
this criterion can only by applied when considering the trimer in
the ground state. More explicitly, in case of ultraviolet divergence
of the four-body kernel, by moving a four-body scale in relation to
the three-body one, an infinite number of tetramer states emerges
from the threshold, but the number is restricted by the energy
interval between trimers (when the reference trimer is an excited
one), such that it can be infinite only in case the given trimer is
the ground state.}

It is also instructive to recall other pioneering four-body calculations
that have verified the possibility of Efimov-like behavior in four-body
systems, such as the works of Kr\"oger and Perne~\cite{kroger-perne} and
Tjon and Naus~\cite{tjon1}. Conclusions drawn within the nuclear physics
context are obviously limited, in view of the strong nuclear repulsion
of the potential core, such that the possible observation of a four-body
scale, independent from the three- and two-body ones, is suppressed. It
explains why the $^4$He and the triton binding energies are strongly
correlated with a fixed slope, as verified by Tjon~\cite{tjon-line}.
In our understanding of the four-body problem, it is not
enough only two parameters, which determines trimer properties,  to
describe the four-boson system. Our results imply that the
correlation between the tetramer and trimer binding energies forms a
family of Tjon lines with slopes depending on the new four-boson
scale. To exemplify that, we have shown a four-boson universal
correlation among the energies of two successive tetramers appearing
between two consecutive Efimov states~\cite{prl2011}. The correlation
exhibits a dependence on a new scale not fixed only by the trimer
properties. In order to directly address this novel universal
behavior, we have performed a number of calculations of
tetramer properties within a zero-range model to show how the
dependence on the new short-range scale is evidenced through their
structure in momentum space.

The tetramer energies can be presented in a scaling plot where the
behavior of a four-body scale can be easily verified in comparison
with other relevant few-body scales~\cite{prl2011}.
The existence of a short-range four-body scale, which is independent
of the two- and three-body ones, is expressed in terms of scaling
functions or correlation between observables. Once the short-range
parameters are eliminated in favor of physical quantities, the renormalized
results appear as correlations between the four-body observables with fixed
dimer and trimer properties.

In the present work we provide details for the approach that lead us
to report~\cite{prl2011} a universal correlation between the energies
of two successive tetramers. As it will be thoroughly presented in
the next sections, we establish by our numerical investigations that
a new short-range parameter is necessary to fix the properties of
the four-boson system in the unitary limit. Our conclusions are
supported by precise numerical solutions of the four-boson
FY integral equations in momentum space for a
zero-range two-body interaction~\cite{epl06}, as well as by an
analysis of the four-body wave function.

The paper is organized as follows: in Sect.~\ref{FYform}, we present
the FY formalism, within a renormalized zero-range
interaction. The relevant scales are introduced in the formalism
through a subtractive renormalization approach.
In Sect.~\ref{3b-4b-scalings}, the scaling functions for the binding
energies of trimers and tetramers are discussed, following
Refs.~\cite{amorim99} and \cite{prl2011}.  Numerical results close to
the unitary limit are also reported in this section.
In Sect.~\ref{4Bstruct}, we present our main results for the structure of
universal tetramers. We include in this section an extended analysis of
the low and high momentum structures of the atom-trimer [(A+T) or $K-$type]
and dimer-dimer [(D+D) or $H-$type] components of the FY
decomposition, to pin down where the manifestation of the short-range
four-body scale is more evident. The behavior of the four-body FY components
of the wave function is presented in this section.
Finally, in Sect.~\ref{conclusions}, we present our conclusions with
perspectives in relation to possible experimental observations.
Three appendices are supplying further details on our approach:
In \ref{appendix A} and \ref{appendix B}, we have details on
the four-body formalism and total wave function in momentum space.
In \ref{appendix C}, we discuss stability and convergence of our numerical
approach, and give some details on the Lanczos-type procedure for solution
of the coupled FY equations.

\section{Faddeev-Yakubovsky formalism}
\label{FYform}
The treatment of four-body problems is quite well known in the quantum scattering theory,
following the original Faddeev formulation of the three-body systems~\cite{Faddeev}, later
on extended to $N-$particle scattering by Yakubovsky~\cite{Yakubovsky}. The actually known
Faddeev-Yakubovsky formalism has been considered by several authors in their effort to solve
different aspects of four-body systems, in nuclear and atomic physics. In addition to the
works on four-body systems already cited in the introduction, we can also mention several
other reference works concerning details of the bound-state four-body FY formalism, such as Refs.~\cite{Sasakawa76,Gibson76,Merkuriev84,Schellingerhout92,Kamada92,Zubarev95,Filikhin02,Lazauskas04,Bayegan08,Hadi2011} and references therein. Alternatively, in the solution of four-body systems with separable interactions, several authors (exemplified by Refs.~\cite{Perne77,Sofianos82,Nakaichi83,Fonseca84}) have applied the Alt-Grassberger-Sandhas formalism~\cite{AGS},
in which the t-matrix components are used instead of the wave-functions.

Here, we describe the FY formalism for bound states of four identical bosons, by
considering general separable two-body interactions. Within our approach, the form-factors
of the separable potential are conveniently replaced by point-like interactions, within
a renormalized zero-range model where the three- and four-body scales are introduced.

\subsection{Bound-state equations and notation}  \label{sec_FY-formalism}

In a four-body (4B) system (with particles $i$, $j$, $k$ and $l$) there are 18
different coordinate systems, each one associated with a specific
two-body partition (see Appendices \ref{appendix A} and
\ref{appendix B}). Basically, by considering the arrangements
of the corresponding two- and three-body sub-systems, two different
partitions are possible: $K-$type (A+T), where a single particle is bound to
a three-body subsystem; and $H-$type (D+D), where two dimers are bound.
Clearly, as we know from the Faddeev formalism, each
three-body subsystem has also three atom-dimer partitions.

We present below the FY formalism with our notation, which is used along the
next sections. The bound state of four particles which interact via pairwise forces
$V_{ij}$ is given by the Schr\"{o}dinger equation,
\begin{equation}
|\Psi\rangle=G_{0}\sum_{i<j} V_{ij}|\Psi\rangle=
\sum_{i<j}|\psi_{ij}\rangle , \label{eq.4B-Schrodinger}
\end{equation}
where $|\psi_{ij}\rangle=G_{0} V_{ij}|\Psi\rangle$,
$G_{0}=(E-H_{0})^{-1}$ is the free four-body resolvent and $H_{0}$
stands for the free Hamiltonian. The above components satisfy:
\begin{eqnarray}
|\psi_{ij}\rangle&=&G_{0}t_{ij}\sum_{kl\neq ij} |\psi_{kl}\rangle
\label{eq.6-coupled-FY}\\
&=&G_{0}t_{ij} \biggl ( |\psi_{ik}\rangle +|\psi_{il}\rangle +
|\psi_{jk}\rangle + |\psi_{jl}\rangle + |\psi_{kl}\rangle \biggr),
\nonumber
\end{eqnarray}
where the two-body transition operator is $t_{ij}$. The FY
decomposition of $|\psi_{ij}\rangle $ reads:
\begin{eqnarray}
|\Kijkl\rangle &=& G_{0}t_{ij} \biggl(|\psi_{ik}\rangle +
|\psi_{jk}\rangle \biggr), \nonumber \\*
|\Kijlk \rangle &=& G_{0}t_{ij} \biggl(|\psi_{il}\rangle
+ |\psi_{jl}\rangle \biggr), \label{eq.KH-operator-general0}
\\ |\Hijkl\rangle &=& G_{0}t_{ij}|\psi_{kl}\rangle,
\nonumber
\end{eqnarray}
where $|\Kijkl\rangle $ and $|H_{ij, kl}\rangle$ correspond,
respectively, to a $K-$type and $H-$type partitions. The Faddeev
components of the wave function
\begin{equation}
|\psi_{ij}\rangle=|\Kijkl\rangle + |\Kijlk\rangle + |\Hijkl\rangle
,\label{eq.psi_ij-definition}
\end{equation}
are reconstructed through FY components.

Every $|\psi_{ij}\rangle$ component contains two $K-$type and one
$H-$type configurations. Therefore, the total wave function
$|\Psi\rangle$ contains twelve different $K-$type chains and six
$H-$type chains, leading to eighteen independent FY components.
By considering identical bosons, the four-body wave function
$|\Psi\rangle$ has to be totally symmetric. As a consequence, all
twelve $K-$type components are identical in their functional form
with particles permuted. The same is true for the six $H-$ type
components. Thus, it is sufficient to consider
only two independent FY components corresponding to the $K-$ and
$H-$type partitions, $|K\rangle\equiv|K_{ij,k}^{\,\,\, l}\rangle$ and
$|H\rangle\equiv|H_{ij,kl}\rangle$.
The 18 coupled FY equations, for identical bosons, shrink to two
coupled homogeneous equations,
\begin{eqnarray}
|\Kijkl\rangle &=& G_{0}t_{ij}P \biggl[\Bigl(1+P_{kl} \Bigr)|\Kijkl\rangle+|\Hijkl\rangle \biggr],
\nonumber \\*
|\Hijkl\rangle &=& G_{0}t_{ij}\tilde{P}\biggl[\Bigl(1+P_{kl} \Bigr)|\Kijkl\rangle+|\Hijkl\rangle \biggr],
\label{eq.KH-operator-general}
\end{eqnarray}
where $P_{kl}$ is the permutation operator for the pair $(kl)$, with $P$ and $\tilde{P}$
defined by
\begin{eqnarray}
P = (P_{ij}+P_{ik})P_{jk}
\;\;\; {\rm and}\;\;\; \tilde{P} = P_{ik}P_{jl}.
\label{eq.permutations-operator}
\end{eqnarray}

The symmetry property of $|K\rangle$ under exchange of particles $i$
and $j$, and $|H\rangle$ under separate exchanges of particles $i,j$
and $k,l$ guarantee that the full wave function,
\begin{eqnarray}
|\Psi\rangle &=& \Bigl(1+P+P_{kl}P+\tilde{P}\Bigr)\biggl[ \Bigl(1+P_{kl} \Bigr)|\Kijkl\rangle+|\Hijkl\rangle \biggr] \nonumber \\
&=& \Bigl(1+ (1+P)P_{kl}\Bigr)\Bigl(1+P \Bigr)|\Kijkl\rangle
\nonumber\\
&+&\Bigl(1+P \Bigr)\Bigl(1+\tilde{P} \Bigr)|\Hijkl\rangle,
\label{eq.4B-wave-funtion}
\end{eqnarray}
is totally symmetric. In order to get insight on how the short-range
four-body scale is reflected in the bosonic wave function for a
zero-range force, we will analyze the momentum structure
of the $K$ and $H$ components. Following this strategy we will be
able to map where the new scale is more relevant to built
the tetramer wave function. For that purpose, we still simplify the
problem using $s$-wave one-term separable potentials. This class
contains, in particular, the contact interaction.

\subsection{One-term separable potential with $s-$wave projection}
\label{sec_FY-momentum-space}

\begin{figure}[ht]
\begin{center}
\includegraphics[width=3.2in]{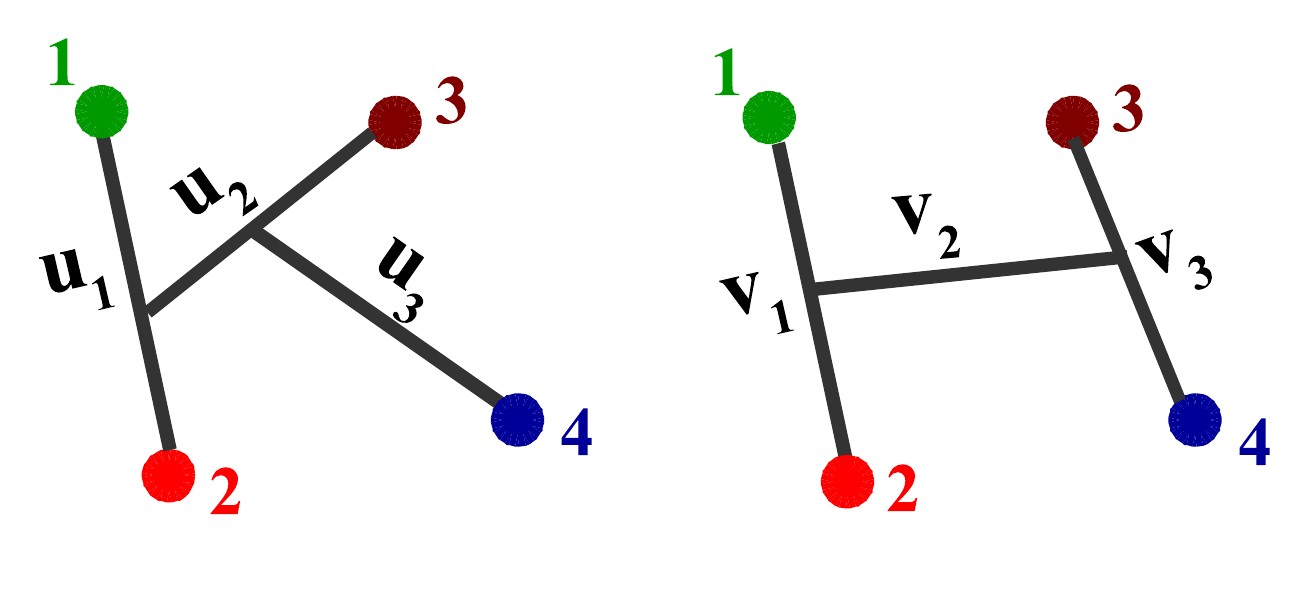}
\end{center}
\vspace{-0.5cm}
\caption{(Color on-line)
Definition of the four-body Jacobi momenta corresponding to
the $K-$ and $H-$type fragmentations.} \label{K-H}
\end{figure}

In this subsection, to simplify the notation, we use the particle-labels
$1, 2, 3$ and $4$. So, we rewrite Eq.(\ref{eq.KH-operator-general}) as:
\begin{eqnarray}
|K\rangle &=& G_{0}t_{12}P \biggl[\Bigl(1+P_{34} \Bigr)|K\rangle+|H\rangle \biggr],
\nonumber \\*
|H\rangle &=& G_{0}t_{12}\tilde{P}\biggl[\Bigl(1+P_{34} \Bigr)|K\rangle+|H\rangle \biggr].
\label{eq.KH-operator}
\end{eqnarray}
In order to solve the coupled Eqs.~(\ref{eq.KH-operator}) in
momentum space one should project these equations into standard sets
of Jacobi momenta, corresponding to both, $K-$type ($|\,\textbf{u}_1 \,
\textbf{u}_2 \, \textbf{u}_3 \,\rangle$) and $H-$type ($|\,
\textbf{v}_1 \, \textbf{v}_2 \, \textbf{v}_3 \,\rangle$) partitions,
as represented in Fig.~\ref{K-H}.

%FIG1
The standard Jacobi momenta for the 4B system can be defined in terms
of the single particle momentum variables ${\bf k}_{i=1,2,3,4}$, in
two possible configurations ($K-$ and $H-$type), as shown in
Fig.~\ref{K-H}:
\begin{eqnarray}
\left\{
\begin{array}{ll}
{\bf u}_1 = (1/2)\bigl({\bf k}_1-{\bf k}_2 \bigr),\\
{\bf u}_2 = (1/3)\left[2{\bf k}_3-\bigl({\bf k}_1+{\bf k}_2\bigr)\right],\\
{\bf u}_3 = (1/4)\left[3{\bf k}_4-\bigl({\bf k}_1+{\bf k}_2+{\bf k}_3\bigr)
\right],
\end{array}
\right.
\label{eq.u-basis}\\
\left\{
\begin{array}{ll}
{\bf v}_1 = (1/2)\bigl({\bf k}_1-{\bf k}_2\bigr),\\
{\bf v}_2 = (1/2)\left[\bigl({\bf k}_1+{\bf k}_2\bigr)-
\bigl({\bf k}_3+{\bf k}_{4}\bigr)\right],\\
{\bf v}_3 = (1/2)\bigl({\bf k}_3-{\bf k}_{4}\bigr),
\end{array}
\right.
\label{eq.v-basis}
\end{eqnarray}
where we assume identical particles, with mass $m =$1.
Since we are interested in the $s$-wave channel contribution, we
introduce the partial-wave representation of the four-body projection
operators corresponding to each Jacobi momenta set, as
\begin{equation}
|u\rangle \equiv|u_1\,u_2\,u_3\rangle,\ |v\rangle
\equiv|v_1\,v_2\,v_3\rangle. \label{eq.uv-basis-abbreviation}
\end{equation}
Within the $s$-wave projection, we consider the following
completeness relation for both basis sets:
\begin{eqnarray}
\int D^{3}U \,\,|U\rangle \langle U|=1,
\label{eq.completness}
\end{eqnarray}
where $U$ indicates each one of $u$ and $v$ sets, with
$D^{3}U \equiv U_1^2\,dU_1\;U_2^2\,dU_2\;U_3^2\,dU_3 $.
Clearly, the projection operators $|u\rangle $ and
$|v\rangle$ are adequate to expand $|K\rangle$ and
$|H\rangle$ components, respectively. Consequently,
the projection of the coupled equations (\ref{eq.KH-operator})
is given by
\begin{eqnarray}
 \langle u|K\rangle &=&
\langle u |G_{0}t P(1+P_{34})|K\rangle + \langle u |G_{0}t P|H\rangle,
  \nonumber \\*
\langle v |H\rangle &=&  \langle v |G_{0}t \tilde{P}(1+P_{34})| K\rangle
  +  \langle v |G_{0}t \tilde{P}| H\rangle.
 \label{eq.KH-in-uv}
\end{eqnarray}
In the following, the units are such that $\hbar=1$ and $m=1$.

By considering a one-term separable two-body potential operator,
$V = \lambda | \chi \rangle \langle \chi | $,
the $s-$wave two-body $t-$matrix elements can be written,
in the $K-$ and $H-$representations, by
{\small
\begin{equation}
\langle U|t|U'\rangle=4\pi\; \frac{\delta \bigl(U'_2-
U_2\bigr)}{U^2_2} \frac{\delta \bigl(U'_3- U_3 \bigr)}{U^2_3}
\chi(U_1) \chi(U'_1)\;\tau(\epsilon_U)
\label{eq.t-matrix-u},
\end{equation}
}
where the corresponding $K-$type and $H-$type Jacobi sets,
$\{u_1,u_2,u_3\}$ and $\{v_1,v_2,v_3\}$, are represented by
$\{U_1,U_2,U_3\}$.
In the above, $\tau(\varepsilon_U)$ is the reduced scattering amplitude,
given by
\begin{eqnarray}
\tau({\cal E}_U) &\equiv &\biggl[ \frac{1}{\lambda} - \int d^{3}p \,
\frac{\chi^2(p)}{{\cal E}_U-p^{2}+{\rm i}\delta}\biggr ]^{-1}
,\label{eq.t-integral}
\end{eqnarray}
where, for the $K-$type and $H-$type configurations, respectively,
we have,
\begin{eqnarray}
{\cal E}_u &\equiv& E-\frac{3u_2^{2}}{4}-\frac{2u_3^{2}}{3}\;\;\; {\rm and}
\;\;\; 
{\cal E}_v \equiv E-\frac{v_2^{2}}{2}-{v_3^{2}}.
\label{en-uv}
\end{eqnarray}
Note that, ${\cal E}_u$ and ${\cal E}_v$ have different expressions
in the $K-$ and $H-$representations, but are identical when written in
terms of the single particle momentum coordinates ${\bf k}_i$.

The final expressions for the FY components,  
derived in the Appendix \ref{appendix A} through
Eqs.~(\ref{eq.KH-in-uv-expanded}-\ref{eq.vu-final}), are:
\begin{widetext}
{\small
\begin{eqnarray}
&&K \bigl(u_1, u_2, u_3 \bigr) = 4 \pi \, G_{0} \bigl( u_1, u_2,
u_3\bigr) \, \chi(u_1)\,  \tau({\cal E}_u)  \int_0^\infty  du'_2 u'^2_2 \int_{-1}^1  dx \,
\chi \biggl(\Pi_1\bigl(u_2,u'_2,x \bigr)\biggr)\,
\Biggl[K \biggl(\Pi_1 \bigl(u'_2,u_2,x \bigr) , u'_2, u_3 \biggr)+
\frac{1}{2}\int_{-1}^1 dx'\,
\nonumber \\ &&\times
\left\{ K \biggl(\Pi_1 \bigl(u'_2,u_2,x \bigr) , \Pi_2\bigl(u'_2,u_3,x' \bigr),
\Pi_3\bigl(u'_2,u_3,x' \bigr) \biggr) +
H \biggl(\Pi_1 \bigl(u'_2,u_2,x \bigr) ,
\Pi_4\bigl(u'_2,u_3,x' \bigr), \Pi_5\bigl(u'_2,u_3,x' \bigr) \biggr)
\right\} \Biggr], 
\label{eq.KH-int} \\
&&H \bigl(v_1, v_2, v_3 \bigr) = 4 \pi \, G_{0} \bigl( v_1, v_2,
v_3\bigr) \, \chi(v_1)\,  \tau({\cal E}_v)  \int_0^\infty
dv'_3 v'^2_3 \, \chi \bigl(v'_3 \bigr)
\Biggl[\int_{-1}^1  dx\, K \biggl(v_3, \Pi_6
\bigl(v_2,v'_3,x \bigr), \Pi_7 \bigl(v_2,v'_3,x \bigr)  \biggr)
 + H \biggl(v_3, v_2, v'_3 \biggr) \Biggr],
\nonumber
 \end{eqnarray}
}
\end{widetext}
where the functions $\Pi_{J=1,2...7}$ are defined by the Eqs.~(\ref{Pi1}),
(\ref{Pi2}), (\ref{Pi3}), (\ref{Pi4}), (\ref{Pi5}), (\ref{Pi6}) and
(\ref{Pi7}).
Eqs.~(\ref{eq.KH-int}) can be rewritten within new definitions
for the FY components, such as
\begin{eqnarray}
\left\{ \begin{array}{ll}
K \bigl(u_1, u_2, u_3 \bigr) \equiv  G_{0} \bigl( u_1, u_2, u_3\bigr) \,
\chi(u_1)\, {\cal K}\bigl(u_2, u_3 \bigr), \\*
H \bigl(v_1, v_2, v_3 \bigr) \equiv G_{0} \bigl( v_1, v_2, v_3\bigr) \,
\chi(v_1)\, {\cal H}\bigl(v_2, v_3 \bigr).
\end{array}\right.
\label{eq.KH-redifinition}
 \end{eqnarray}
Furthermore, by considering a zero-range potential, with $\chi=1$, the above
coupled equations (\ref{eq.KH-int}) are reduced to the following, where the
need of regularization in the momentum integrals is explicit:
\begin{widetext}
{\small
\begin{eqnarray}
{\cal K}\bigl(u_2, u_3 \bigr)&=& 4 \pi\,\tau({\cal E}_u)\int_0^\infty  du'_2 u'^2_2 \int_{-1}^1 dx
\Biggl[ G_{0} \biggl(\Pi_1 \bigl(u'_2,u_2,x\bigr),
 u'_2,u_3\biggr) {\cal K} \biggl(u'_2, u_3 \biggr)
\nonumber \\  &+&
\frac{1}{2}  \int_{-1}^1  dx'
\left\{G_{0} \biggl( \Pi_1\bigl(u'_2,u_2,x\bigr), \Pi_2\bigl(u'_2,u_3,x' \bigr),
\Pi_3\bigl(u'_2,u_3,x' \bigr) \biggr) {\cal K} \biggl(\Pi_2\bigl(u'_2,u_3,x' \bigr), \Pi_3\bigl(u'_2,u_3,x' \bigr) \biggr)
\right. \nonumber \\  &+&  \left.
G_{0} \biggl(\Pi_1\bigl(u'_2,u_2,x\bigr) , \Pi_4\bigl(u'_2,u_3,x' \bigr),
\Pi_5\bigl(u'_2,u_3,x' \bigr) \biggr) {\cal H}
\biggl(\Pi_4\bigl(u'_2,u_3,x' \bigr), \Pi_5\bigl(u'_2,u_3,x' \bigr)
\biggr) \right\}
\Biggr]\; , \nonumber \\ \label{eq.KH-integral-revised}\\
{\cal H} \bigl(v_2, v_3 \bigr) &=& 4 \pi  \,  \tau({\cal E}_v) \int_0^\infty  dv'_3 v'^2_3
\Biggl[ \int_{-1}^1  dx \,  G_{0} \biggl(v_3, \Pi_6
\bigl(v_2,v'_3,x \bigr), \Pi_7 \bigl(v_2,v'_3,x \bigr) \biggr)
\nonumber \\  &\times&
{\cal K} \biggl( \Pi_6 \bigl(v_2,v'_3,x \bigr), \Pi_7 \bigl(v_2,v'_3,x\bigr)\biggr)
+  G_{0} \biggl(v_3, v_2, v'_3 \biggr) {\cal H}
\biggl(v_2, v'_3 \biggr) \Biggr]. \nonumber
 \end{eqnarray}
}
\end{widetext}
Following an appropriate renormalization scheme for zero-range interactions,
with the regularization parameters directly associated to observables~\cite{amorim99},
in the regularization of the above coupled equations we introduce the
short-range three- and four-body scales, as we are going to discuss in
the next subsection.
\subsection{Four-boson zero-range bound-state model}
\label{sec_zero-range-interaction}
In order to help the understanding of the zero-range model in the four-boson
formalism, we shortly review the main steps to deal with two- and three-boson
equations, when we have two-body zero-range interactions.
The momentum space representation of a zero-range interaction
$V(\textbf{r})= (2\pi)^{3} \, \lambda \, \delta(\textbf{r})$, is characterized
by the two-body coupling constant $\lambda$ and a separable potential
with constant form factor ($\langle \textbf{p}|\chi\rangle=1$).
The matrix element of the two-body transition operator, $\langle
p^\prime |t({\cal E})|p\rangle= \tau({\cal E})$, is obtained by
analytical integration of Eq.~(\ref{eq.t-integral}). The coupling
constant $\lambda$ is fixed by one physical input, e.g., the
position of the two-body pole at $E_2$:
\begin{eqnarray}
 \lambda^{-1} = \int d^{3}p \, \frac{1}{E_2-p^2} =
 - \int d^{3}p \, \frac{1}{|E_2|+p^2}.
\label{eq.lambda}
\end{eqnarray}
We consider bound or virtual two-body energies, when $E_2=-|E_2|$.
Bound in the case the pole is in the upper part of the complex
imaginary axis of the momentum plane
(${\rm i} \kappa_B$), with corresponding energy in the principal
Riemann sheet ($E_2 = - \kappa_B^2 $);
and virtual when the pole is in the lower part of the complex
imaginary axis (-${\rm i} \kappa_v$), with the energy in the
second Riemann sheet ($E_2 = - \kappa_v^2$). As we are going to
consider one case {\it or} the other, both $\kappa_B$ and $\kappa_V$
will be labelled as $\sqrt{B_2}$.
Therefore, the corresponding scattering lengths are given by
$a= \pm 1/\sqrt{B_2}$ ($+$ for bound, and $-$ for virtual), with
the renormalized two-body $t$-matrix given by
\begin{eqnarray}
\tau({\cal E}) =
\frac{1/(2\pi^{2})}{\sqrt{|E_2|}-\sqrt{-{\cal E}}} =
\frac{1/(2\pi^{2})}{ \pm{1}/{a}-\sqrt{|{\cal E}|}}.
\label{eq.t-final}
\end{eqnarray}
The above procedure is enough to render finite the scattering amplitude,
providing the cancelation of the linear divergence of the momentum integral.

\subsubsection{Subtractive regularization technique}

The subtraction technique used to define the two-body scattering amplitude
can be generalized to three-boson systems. The need for a new parameter
beyond $a$ is demanded by the Thomas collapse~\cite{ThPR35} of the three
particle ground state, when the two-body interaction range $r_0$ goes to
zero.  Such collapse is also associated to the Efimov phenomenon by
observing that in both the cases $|a|/r_0\to\infty$~\cite{th-efimov1988}.
The Thomas collapse is avoided at the expense of regularizing the kernel
of the three-body equation, which is done by introducing a subtraction at
an energy scale $-\mu_3^2$. Together with the two-body scattering length,
this procedure determines the low-energy three-boson properties.
The regularization of the momentum integration is done by the subtracted
form $G_0(E)-G_0(-\mu_3^{2})$ in substitution to the free three-body
Green's function $G_0(E)$.  The regularizing energy parameter
$\mu_3^{2}$ is also called ``three-body scale'' in view of its direct
association to a three-body physical observable (scale) in the
renormalization procedure.

By adding a fourth particle to the three-body system, within the
same zero-range two-body interaction, another regularization is required
in the corresponding formalism, due to new terms in the coupled integral
equations, not directly identified with the three-body kernel. This
regularization is followed in a similar way as done in the three-boson
case, by introducing one more regularizing parameter (``four-body scale''),
$\mu_4^2$, in the FY formalism.
This new scaling parameter appears in the integrands associated to the
presence of the fourth particle, in order to allow the complete
regularization of all the momentum integrals.
Among the 18 FY components, only the first three (i.e. $ijk+l, jki+l$
and $kij+l$) will fully describe the three-body (3B) system $(ijk)$, where the
3B scaling parameter $\mu_3^2$ enters in the subtracted form of
the free Green's function. The 4B scaling parameter $\mu_4^2$ enters in
the subtracted form of the Green's function that are present in the
remaining 15 components. Therefore, in the three components associated to
the 3B system, the regularization is done by
\begin{equation}
G_0(E) \longrightarrow G_{0}^{(3)}(E) \equiv G_0(E)-G_0(-\mu_{3}^{2}),
\label{G03}\end{equation}
with the new scale appearing in the regularization of the other fifteen
components:
\begin{equation}
G_0(E) \longrightarrow G_{0}^{(4)}(E)\equiv G_0(E)-G_0(-\mu_{4}^{2})
.\label{G04}\end{equation}

Summarizing our regularizing approach:
      The physical scales of the four-body problem are the energies of the dimer, one
      reference trimer and one reference tetramer.
      The subtractive procedure adds the regularization parameters
      $\mu_3$ and $\mu_4$ to the kernel of the FY equations,
      which are correlated to the physical three- and four-body scales.
      The dependence on the subtraction points is eliminated in favor of two
      observables, which in our case are the values of the trimer
      and tetramer reference energies. The elimination of the dependence on the subtraction
      point is possible as evidenced by the existence of correlations between four-boson
      $s-$wave observables.
      When presented in terms of dimensionless quantities, the correlation between two
      consecutive tetramer energies is indeed given by a limit cycle.
      This correlation, presented in Ref.~\cite{prl2011} and further explored in the present
      work, is in fact consistent with results obtained by other models.

The present four-body regularization strategy keeps fixed the minimal
requirement of the three-body properties, while the further freedom is used to
introduce the new scale. Therefore, in the regularization procedure of
Eqs.~(\ref{eq.KH-integral-revised}), by introducing
the required three- and four-body regularization parameters as explained above,
using (\ref{G03}) and (\ref{G04}), we obtain the corresponding zero-range
subtracted FY equations, which are given by:
{\small \begin{widetext}
\begin{eqnarray}
{\cal K} \bigl(u_2, u_3 \bigr) &=& 4 \pi \,  \tau({\cal E}_u) \int_0^\infty du'_2
u'^2_2 \int_{-1}^1  dx
\Biggl[ G_{0}^{(3)} \biggl( \Pi_1\bigl(u'_2,u_2,x\bigr), u'_2, u_3\biggr) \,
 {\cal K} \biggl(u'_2, u_3 \biggr)
\nonumber \\  &+&
 \frac{1}{2}  \int_{-1}^1  dx' G_{0}^{(4)} \biggl( \Pi_1\bigl(u'_2,u_2,x\bigr),
 \Pi_2\bigl(u'_2,u_3,x' \bigr), \Pi_3\bigl(u'_2,u_3,x' \bigr)
\biggr)
 {\cal K} \biggl(\Pi_2\bigl(u'_2,u_3,x' \bigr),
\Pi_3\bigl(u'_2,u_3,x' \bigr) \biggr)
\nonumber \\ &+&
 \frac{1}{2}  \int_{-1}^1  dx'
G_{0}^{(4)}\biggl(\Pi_1\bigl(u'_2,u_2,x \bigr) ,
\Pi_4\bigl(u'_2,u_3,x' \bigr), \Pi_5\bigl(u'_2,u_3,x' \bigr) \biggr)
 {\cal H} \biggl(\Pi_4\bigl(u'_2,u_3,x' \bigr),
\Pi_5\bigl(u'_2,u_3,x' \bigr) \biggr) \Biggr],
 \nonumber \\ \label{eq.KH-integral-final} \\
{\cal H} \bigl(v_2, v_3 \bigr) &=& 4 \pi  \,  \tau({\cal E}_v) \int_0^\infty
dv'_3 v'^2_3\, \Biggl[
G_{0}^{(4)} \biggl(v_3, v_2, v'_3 \biggr) {\cal H} \biggl(v_2, v'_3
\biggr)
\nonumber  \\ &+& \int_{-1}^1  dx \,
G_{0}^{(4)} \biggl(v_3, \Pi_6 \bigl(v_2,v'_3,x \bigr),
\Pi_7 \bigl(v_2,v'_3,x \bigr)  \biggr) {\cal K} \biggl( \Pi_6
\bigl(v_2,v'_3,x \bigr), \Pi_7 \bigl(v_2,v'_3,x \bigr)
\biggr)\Biggr] . \nonumber
\end{eqnarray}
\end{widetext}}
The physical picture behind the regularization leading to
(\ref{eq.KH-integral-final}) can be described as follows. The
three-body scale parameterizes the short-range physics in the
virtual propagation of the interacting three-boson subsystem within
the four-body system. The four-body scale parameterizes the
short-range physics beyond the three-body one, explored in the
four-boson virtual-state propagation between two different fully
interacting three-body or disjoint two-body clusters.
It is worthwhile to stress that the first term in the right hand
side of the equation for ${\cal K} \bigl(u_2, u_3 \bigr)$ is the
three-boson bound-state equation, when all the other terms are dropped
out. So, it is expected that such term should carry the 3B scale,
while one could think that the other terms do not require regularization.
However,  as shown in \cite{epl06}, still all the other terms need to be
regularized, in order to avoid the 4B ground state collapse.
Although the same regularizing parameter could also be used in all the terms,
the freedom to choose it differently is at our hands, and the constraint
of introducing the trimer properties does not require any further
assumption, but the given form of the first term in the right-hand side
of the equation for ${\cal K} \bigl(u_2,u_3 \bigr)$.
The physical conditions in cold atom traps that allow us to explore a
short-range four-boson independent scale, when approaching the
Feshbach resonance, are discussed in Ref.~\cite{prl2011}.

\section{Binding Energies of Universal Trimers and Tetramers}
\label{3b-4b-scalings}

In this section, before considering our numerical results obtained for
the FY bound-state equations (\ref{eq.KH-integral-final}),
we provide a brief discussion on scaling functions, starting with the
three-boson case (next subsection) and moving to tetramer systems,
in subsection \ref{tetramer-scaling}.
After we set up the concept of a scaling function, we will present
detailed numerical results for the tetramer binding energies, with a
comparison with other recent available calculations.
The Lanczos-type procedure for solution
of the coupled FY equations is shown in Appendix \ref{appendix C},
where we also give some details on stability and convergence
of our numerical approach.

\subsection{Trimers in the scaling limit} \label{trimers-scaling}

The three-body system is sensitive to the physics at short ranges,
which is parameterized by $\mu_3$.  After this scale was
recognized~\cite{amorim99}, it was shown that the change of the
three-body scale in respect to the two-body one (e.g., $1/a$) can be
clearly revealed by expressing the energies of the Efimov trimers in
a single curve, which  defines a scaling function ${\cal
F}_3^{(N)}$. It allows to build the energies of all the sequence of
weakly bound trimers. Thus, the energies of successive trimers are
correlated by~\cite{amorim99}
\begin{equation}
\sqrt{\frac{B_3^{(N+1)}-\overline{B}_{2}} {B_3^{(N)}}} \equiv {\cal
F}_3^{(N)}\left(\pm\sqrt{\frac{B_{2}}{B_3^{(N)}}}\right),
\label{eq1}
\end{equation}
written in terms of dimensionless quantities. For convenience, in
(\ref{eq1}), we define $\overline{B}_2\equiv B_2$ for bound two-body
systems (plus sign, or $a>0$), and $\overline{B}_2\equiv 0$ for
virtual states (minus sign, or $a<0$).

Few cycles are enough to reach a universal function independent on
$N$; i.e., in the limit $N\to\infty$ it reaches a
renormalization-group invariant limit cycle~\cite{wilson71}. At the
critical values, $B_3^{(N+1)}={B}_{2}$ (for bound two-body) and
$B_3^{(N+1)}=0$ (for virtual two-body), the scaling function
(\ref{eq1}) vanishes:
${\cal F}_3^{(N)}\left(\pm\sqrt{\frac{B_{2}}{B_3^{(N)}}}\right)=0 $ .
In this limit, the solution for a bound two-body system ($+$ sign)
is given by $B_3^ {(N)}\simeq 6.925 \;B_2$; and the solution for a
virtual two-body system ($-$ sign) is given by $B_3^ {(N)}\simeq
1141 B_{2}^v$, where $B_{2}^v$ is the virtual state
energy~\cite{amorim99,laser11}.
We observe that the scaling law (\ref{eq1}) is one among many
possible model independent correlations between three-body
observables for short-ranged interactions.
We should also point out that range effects can become relevant
as $|a|/r_0$ decreases. This is evidenced by the finite-range
numerical results from other authors, presented in the scaling
plot displayed in Ref.~\cite{amorim99}. (On range effects and
universal properties of three-body systems, see also Ref.~\cite{limit-cycle-rev}.)
Therefore, range effects will affect Eq.~(\ref{eq1}) and
the above critical values of $B_3^ {(N)}$.

Evidence of Efimov cycles, with a period of 22.7, in the values of
the two-body scattering length, $a$, at the peak of the three-body
recombination, were observed in a cold-atom experiment by the
Innsbruck group~\cite{ferlaino:140401}, as well as in other recent
experiments reported in \cite{phys-ferlaino}. These results gave a
response to a quite old discussion on the possible realization of
the Efimov effect in real physical systems. The Efimov effect was
formulated in the nuclear physics context, when considering the
solutions of the three-body Faddeev equations in the limit of
infinite large two-body scattering length. Since then, the possible
existence of Efimov states in nature have been considered in several
works, where the most promising one was the calculations done by
Cornelius and Gl\"ockle for a system with three atoms of
$^4$He~\cite{cornelius}. The revival of this matter came recently
with the advances in cold-atom laboratories, where by using Feshbach
resonance techniques it is possible to vary the two-body scattering
length from zero to infinite (positive or negative) values.

\subsection{Tetramers in the scaling limit}  \label{tetramer-scaling}

Following a complete analogy with the three-body case, we now show
evidence for the existence of a limit cycle for the binding energies
of successive tetramers. Therefore, by considering a shallow dimer
(bound or virtual), we introduce a scaling function for the energies
of two successive tetramers between two trimers, as
\begin{equation}
\sqrt{\frac{B_4^{(N+1)}-{B}_{3}}{B_4^{(N)}}} \equiv {\cal F}_4^{(N)}
\left(\sqrt{\frac{B_3}{B_4^{(N)}}};
\pm\sqrt{\frac{B_2}{B_3}}\right). \label{eq2}
\end{equation}
For the present purpose, the calculations are restricted to bound
tetramers below the ground trimer, although the results are valid
for tetramer attached to different trimers\cite{prl2011}.

The  scaling function (\ref{eq2}) has the dependence on a four-body
scale independent of the three- and two-body ones.  Due to the
flexibility of our model it is transparent how to  obtain it. The FY
equations (\ref{eq.KH-integral-final}) are solved for $\mu_3$ fixed
while $\mu_4$ is varied. In the unitary limit, the solutions of
(\ref{eq.KH-integral-final}) depends only on the ratio
$\mu_4/\mu_3$, with observables given in units of $\mu_3$. Eliminating
the dependence of the tetramer energies on the ratio $\mu_4/\mu_3$
by writing the  energy of the $(N+1)-$th state as a function of the
energy of the $N-$th tetramer, the scaling function (\ref{eq2}) is
constructed, where the dependence on $\mu_3$ is removed in favor of
$B_3$. As no other scales are present in
(\ref{eq.KH-integral-final}), a scaling function like (\ref{eq2})
seems possible. It remains to find whether Eq.~(\ref{eq2}) is
independent on $N$, which will be discussed in the next subsection.

We observe that, tetramers below a generic trimer are constrained to
be between the two successive Efimov trimers. However, below the
ground state trimer the tetramer collapses as $\mu_4/\mu_3\to\infty$
\cite{epl06}, as it is exemplified in
Table~\ref{Tabletetramers-our-results} (A),
for scale ratios above 200 in the unitary limit.

\subsection{Numerical results close to the unitary limit}

The numerical calculations with the renormalized zero-range model
requires the definition of the relevant scales. They correspond to
the two-, three- and four-body scales given by
$a^{-1}$, $\mu_{3}$ and $\mu_{4}$, and obviously for fixed
scattering length and trimer scale one can move $\mu_{4}$ to
investigate its effect on the spectrum of tetramer ground and
excited states.

\begin{table}[ht]
\caption{
Binding energies of ground and excited tetramer states, for different
four-body scales and for (A) ${B_2}=0$; (B) bound dimer, with
${B_2}={0.02\; B_3}$; and (C) virtual dimer, with
$\sqrt{B_2}= -\sqrt{0.02\; B_3}$. In (A), we verify
that a third excited tetramer emerges for $\mu_4/\mu_3 \approx$ 240.
In (B) and (C), for non-zero two-body binding, we have only
presented results for the ground and first excited state binding
energies.} \label{Tabletetramers-our-results}
\begin{center}
{\Large \textbf{(A)} }
\begin{tabular}{lccccccccccccccl}
\hline\hline
&& $\mu_{4}/\mu_{3}$ && $B_4^{(0)}/B_3$ && $B_4^{(1)}/B_3-1$ && $B_4^{(2)}/B_3-1$ \\
\hline
 &&   1 && 3.10  \\
 && 1.6 && 4.70   && 7.10$\times 10^{-4}$ \\
 &&   5 && 12.5   && 0.531 \\
 &&  10 && 24.6   && 1.44 \\
 &&  21 && 63.5   && 3.62 && 3.20$\times 10^{-4}$&&  \\
 &&  40 && 184    && 7.65 && 0.203 \\
 &&  50 && 275    && 9.80 && 0.365 \\
 &&  70 && 520   && 12.9 && 0.629 \\
 && 100 && 1.04$\times 10^{3}$   && 20.5 && 1.17 \\
 && 200 && 4.06$\times 10^{3}$   && 50.8 && 2.86 \\
 && 300 && 9.11$\times 10^{3}$   && 102 && 4.53 \\
 && 400 && 1.62$\times 10^{4}$   && 153 && 6.28 \\
\hline\hline
\end{tabular}\\
\vspace{0.5cm}
{\Large \textbf{(B)} }
\begin{tabular}{lcccccc}
\hline\hline
&& $\mu_{4}/\mu_{3}$  &&   $B_4^{(0)}/B_3$ &&   $B_4^{(1)}/B_3-1$\\
\hline
 &&    1  && 2.66  \\
 &&  1.76 && 4.24   && 9.8$\times 10^{-4}$  \\
 &&    5  && 10.0   && 0.421 \\
 &&   20  && 45.9   && 2.77 \\
 &&   40  && 139    && 6.10 \\
 &&   80  && 506    && 13.0 \\
 &&  200  && 2.86$\times 10^{3}$   && 39.5  \\
 &&  300  && 6.00$\times 10^{3}$   && 69.3 \\
 &&  400  && 9.81$\times 10^{3}$   && 104 \\
 \hline\hline
 \end{tabular}\\
\vspace{.5cm}
{\Large \textbf{(C)} }
 \begin{tabular}{lcccccc}
\hline\hline
&& $\mu_{4}/\mu_{3}$  &&   $B_4^{(0)}/B_3$ &&   $B_4^{(1)}/B_3-1$\\
\hline
 &&   1    &&  3.62  \\
 &&  1.7   &&  5.91 && 0.014 \\
 &&    5   &&  15.4 && 0.658 \\
 &&   20   &&  74.8 && 4.18 \\
 &&   40   &&  236 && 9.46 \\
 &&   80   &&  873 && 20.6\\
 &&  200   &&  5.02$\times 10^{3}$ && 64.5  \\
 &&  300   &&  1.06$\times 10^{4}$ && 115  \\
 &&  400   &&  1.73$\times 10^{4}$ && 174  \\
\hline\hline
\end{tabular}
\end{center}
\end{table}
In Table \ref{Tabletetramers-our-results}, we have listed our
numerical results for tetramer ground and excited state binding
energies. In part (A) the results at unitary limit $B_2=0$ are given
for different scale ratios from ${\mu_{4}}/{\mu_{3}}=1$ to $400$.
According to the obtained results for tetramer binding energies,
${\mu_{4}}/{\mu_{3}} \simeq 1.6$ is the threshold for the first
tetramer excited state, ${\mu_{4}}/{\mu_{3}} \simeq 21$ is the
threshold for second tetramer excited state and close to
${\mu_{4}}/{\mu_{3}} \approx 240$ the third tetramer excited state
should appear. At these critical values of scale ratios  where
$B_4^{(N+1)}=B_3$ with $B_2=0$, the scaling function (\ref{eq2})
vanishes
 \begin{equation}
{\cal F}_4^{(N)} \left(\sqrt{\frac{B_3}{B_{4-c}^{(N)}}}; 0\right)=0
\ , \label{eq2.1}
\end{equation}
with the solutions approaching ${B_{4-c}^{(N)}}\simeq 4.6\; B_3$, for
each cycle, i.e., $N=$0 and 1, as we have calculated up to three
tetramers below the trimer. The limit cycle is approached quite
fast, as verified in the case of three-boson~\cite{amorim99}.
In the critical condition (\ref{eq2.1}) we have one tetramer at the
atom-trimer scattering threshold, which allows for a resonant
atom-trimer relaxation in trapped cold-atom gases at the Feshbach
resonance.

In Table~\ref{Tabletetramers-our-results} (B) and (C), we show the
numerical results for positive and negative scattering lengths, respectively,
with $\sqrt{B_2/B_3} = \pm \sqrt{0.02}$.
From these results it is clear that, by comparing with the case of
infinite scattering length and for the same scale ratios, the tetramer
binding energies for positive and negative scattering lengths have
smaller and larger values, respectively. The critical condition for
atom-trimer resonant relaxation near the Feshbach resonance, has to
be corrected as
 \begin{equation}
{\cal F}_4^{(N)} \left(\sqrt{\frac{B_3}{B_{4-c}^{(N)}}};
\pm\sqrt{\frac{B_2}{B_3}}\right)=0 \ , \label{eq2.2}
\end{equation}
$a=\pm 1/\sqrt{B_2}$ does not vanishes. By performing a linear expansion
around $a^{-1}=0$, from the results given in
Table~\ref{Tabletetramers-our-results} (B) and (C), one has
\begin{equation}
B_{4-c}^{(N)}\approx 4.6\; B_3 \left[1 - 0.8\;(a\sqrt{B_3})^{-1}\right].
\end{equation}

%FIG2
\begin{figure}[ht]
\begin{center}
\includegraphics[width=3.5in]
{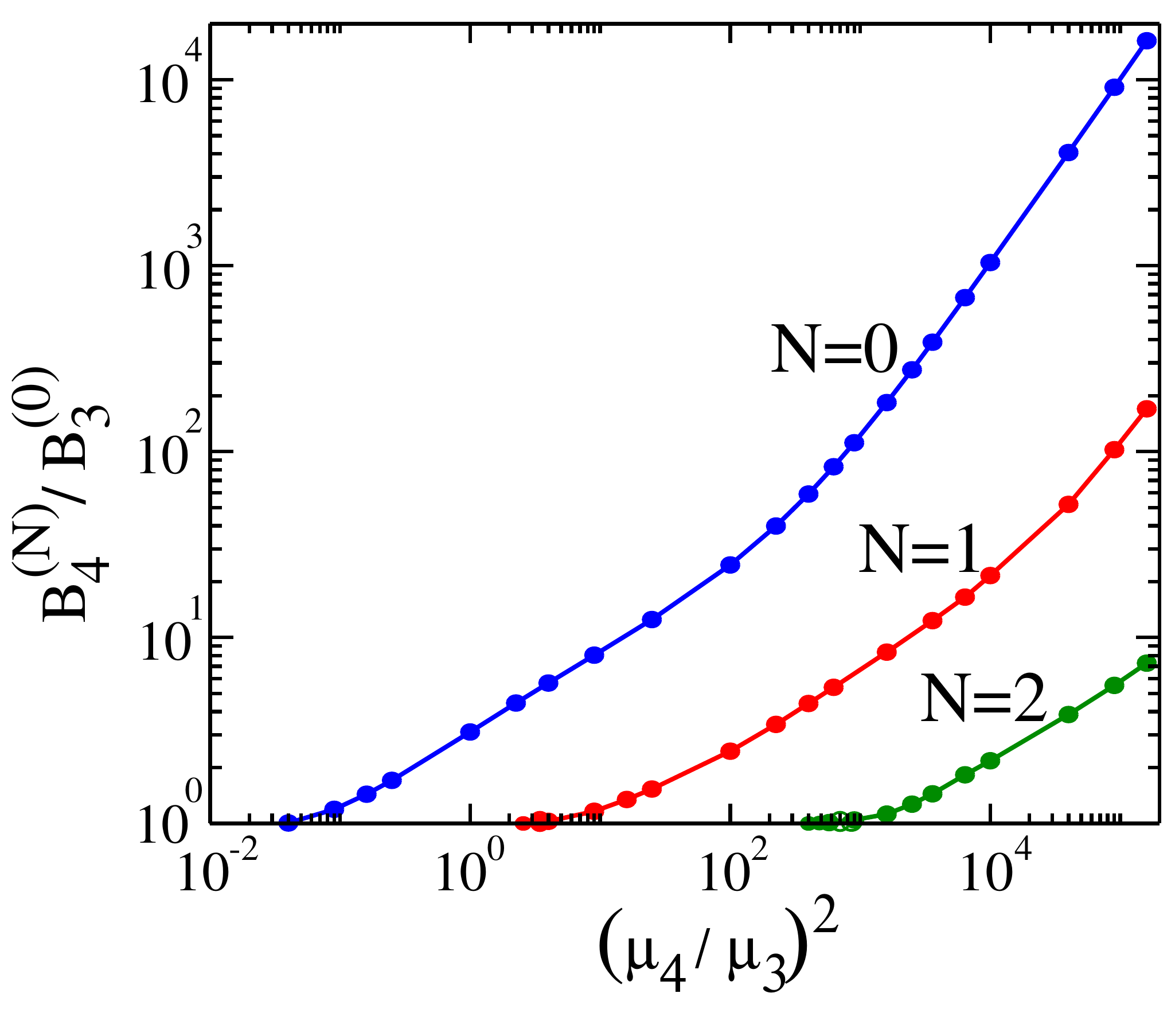}
\end{center}
\vspace{-0.5cm}
\caption{(Color on-line) The first three tetramer energy levels at unitary limit, in units of
the trimer ground state, are shown as functions of $(\mu_4/\mu_3)^2$,
where $\mu_4$ and  $\mu_3$ are, respectively, the four- and three-body
regularizing scaling parameters.} \label{Energy-Scales}
\end{figure}

The dependence of the tetramer energies with the ratio $\mu_4/\mu_3$
in the unitary limit is presented in Fig.~\ref{Energy-Scales} by
using results of Table~\ref{Tabletetramers-our-results} (A). The
collapse of the tetramer states by increasing the short-range
four-body momentum scale in respect to the three-body one is seen.
As for $1/a=0$, only two scales define the tetramer state, the
four-boson binding increases in respect to the trimer one, by either
increasing $\mu_4$ or decreasing $\mu_3$. The three tetramer states
for $N=0$, 1 and 2 show an increase of the binding energies roughly
as $\mu_4^2$. A similar scaling appears for the trimer binding
energy with $\mu^2_3$. The ratio of $B_4/B_3$ gives  the slope of
the Tjon line and it is not a constant as thought before. As a
matter of fact it can be considered as a parameter measuring the
four-body scale.

%FIG3
\begin{figure}[ht]
\centerline{\includegraphics[width=3.5in]{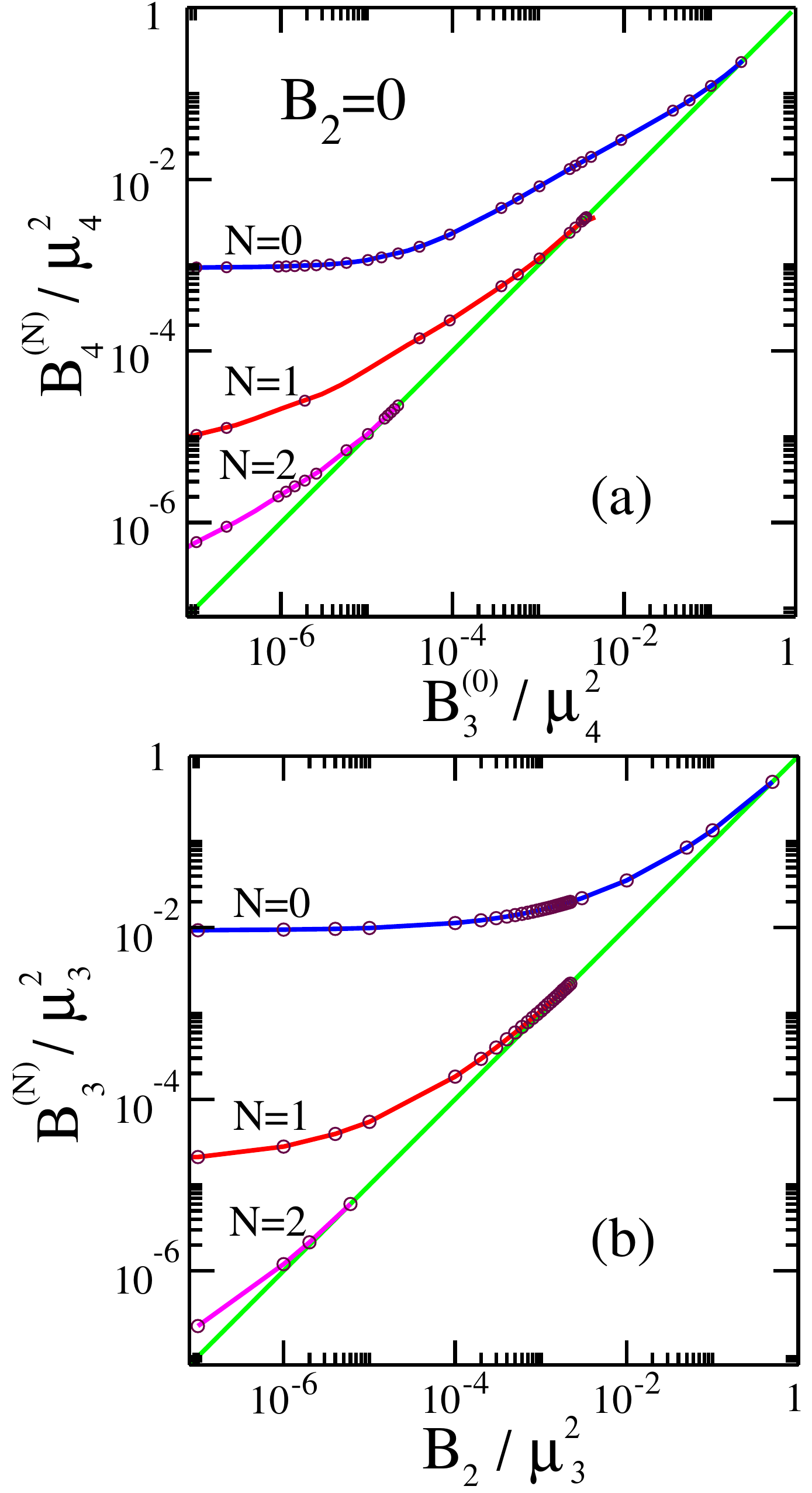}}
\vspace{-0.5cm}
\caption{(Color on-line)
In (a) we have the tetramer energy levels $N=0,1,2$, as functions of the trimer
ground-state energy for $B_2=0$. For comparison, in (b), we have the first
Efimov trimer levels as functions of $B_2$.
The energy units are $\mu_4^2$, in (a) (where $\mu_3^2 = 1$);
and $\mu_3^2$, in (b).
In both cases, the threshold is given by the diagonal line.}
\label{E4xE3-E3xE2}
\end{figure}

In order to compare directly the numerical results for the three-
and four-body limit cycle, we present the Fig.~\ref{E4xE3-E3xE2}
with two panels, (a) and (b). The three-body Efimov states are shown in 
(b), with the energies given in units of the trimer scaling parameter $\mu_3^2$. 
The corresponding four-body Efimov-like states
are given in (a), when considering a ground-state trimer in the unitary limit. 
The dependence of $B_4^{(N)}$ with $B_3$ are
given with the energies in units of the tetramer energy parameter
$\mu_4^2$, for $N=$0, 1 and 2 (with fixed trimer parameter $\mu_3 = 1$).
We illustrate the main qualitative phenomenon of tetramers emerging
from the atom plus trimer threshold: by decreasing $B_3$, an increasing
number of tetramers become bound. A similar phenomenon is shown in the
right panel for trimers in terms of a dimer energy, which is the Efimov
effect.
In our calculations, we first consider tetramers below
a trimer in the ground state, such that the tetramer spectrum is not
limited from below and can collapse as we increase the four-body
scale in relation to the three-body one. In this case, the number of
excited tetramer levels can increase with no limit. However, for a
given general excited trimer, the applicability of our results are
restricted to a region where the trimer energy varies by a factor of
about 515 (the Efimov ratio in the unitary limit). Therefore,
between two successive trimer states at most three tetramers can be
found.

%FIG4
\begin{figure}[ht]
\centerline{\includegraphics[width=8cm]{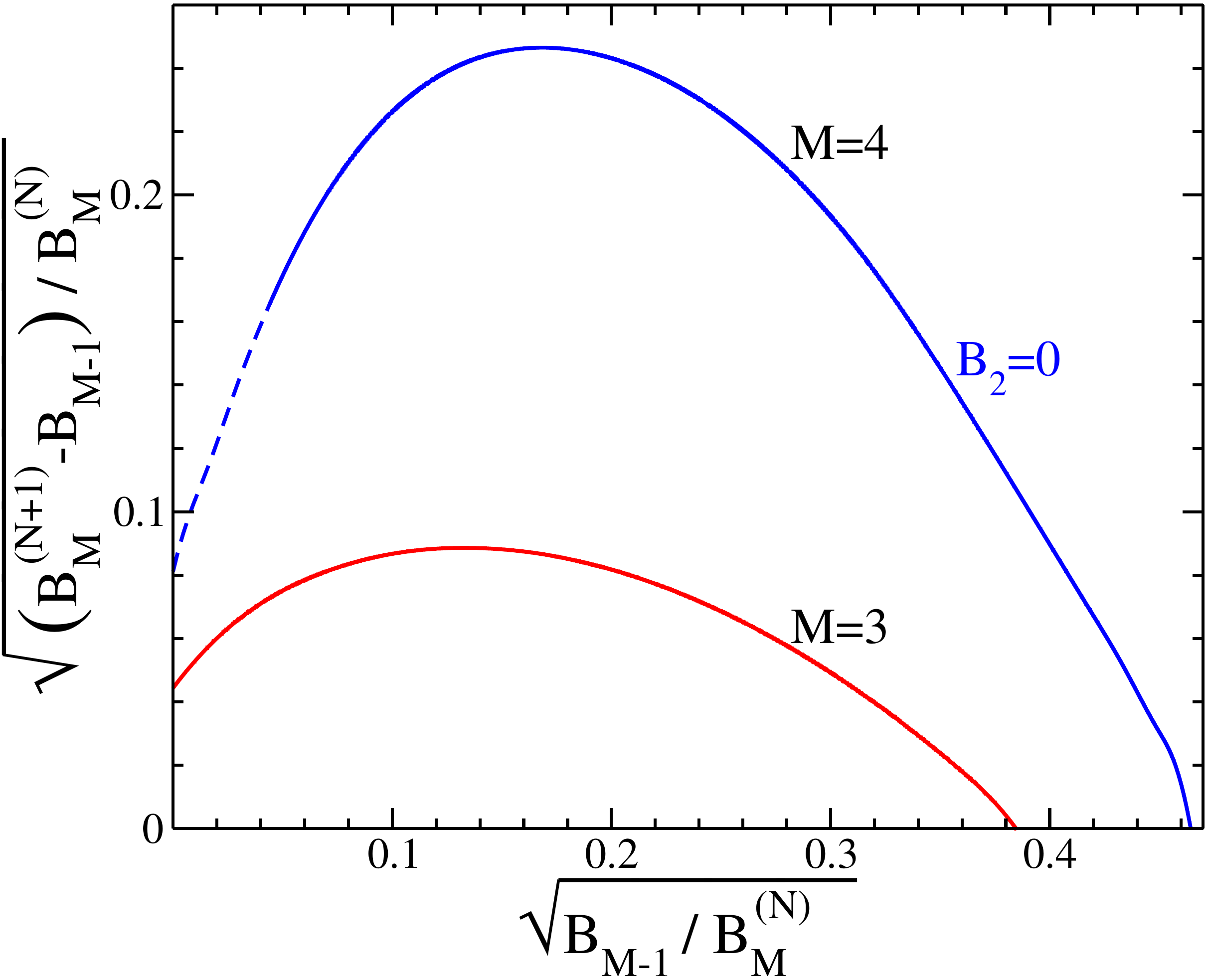}}
\vspace{-.2cm}
\caption{(Color on-line)
The four- and three-body scaling functions are represented, respectively, by
$M=4$ and $M=3$. The energies are rescaled in terms of $B_M^{N}$, where $N$
represents the corresponding energy level in both the cases ($N=0$ is the
ground state).
The behavior of excited trimer ($M=3$) and tetramer ($M=4$) energies
are shown as the corresponding subsystem energy ( $B_{2}$ or $B_{3}$ )
is varying.
In the case of $M=4$, $B_2$ is fixed to zero and the dashed part is only
reachable by tetramers below the ground-state trimer.}
\label{scalings43}
\vspace{-0.3cm}
\end{figure}
The results for the four-boson binding energies, plotted in
Fig.~\ref{E4xE3-E3xE2}(a), exhibit a limit cycle, which expresses
the universal behavior of the energies with a moving four-body scale
parameter. The curves shown in this figure reduces to a single one,
when they are plotted as the correlation between successive tetramer
energies, as presented in Fig.~\ref{scalings43} for $M=4$, where we
consider the scaling plot in the unitary limit. The scaling function
${\cal F}_4^{(N)}
\left(\sqrt{B_3/B_4^{(N)}}; 0\right)$ was built considering up to
the third excited state. As in the case of trimer Efimov cycles, the
numerical results for tetramers also present a very rapid
convergence towards a {\it four-boson limit cycle}. For comparison,
we also present ${\cal F}_3^{(N)}\left(\sqrt{B_2/B_3^{(N)}}\right)$,
which has the value of 1/22.7 when $B_2$ vanishes. The three-body
scaling function, identified inside the plot by $M=3$, was derived
in Ref.~\cite{amorim99}, being compared directly with the
corresponding four-body scaling function, identified by $M=4$.
{The dashed part of the tetramer scaling curve presented in
Fig.~\ref{scalings43}, where $\sqrt{B_3/B^{(N)}_4}< 1/22.7$, is not
accessible for the excited tetramer energies belonging to the
tetramer spectrum that is obtained for a given {\it excited}
trimer. This restriction implies that a meaningful infinite number
of tetramer levels is only possible when considering the trimer
in the ground-state level.}

Let us consider the unitary limit (when we have the geometrically
spaced trimer levels), to resume the global picture suggested by our model.
By increasing the tetramer scale in respect to the trimer one, as a new excited
tetramer emerges from the threshold, the less excited ones will slide down,
moving in direction of the next deeply bound trimer. Actually these tetramers,
between two trimers, are resonances decaying to an atom and a trimer. They
will dive into the complex plane through the atom-trimer scattering cut with
their width increasing as the four-body scale is incremented. This qualitative
discussion deserves further investigations from the point of view of the
complex analytical structure of the four-body scattering equations. Our picture
suggests that, in the unitary limit, at most three tetramer resonances lie
between two successive trimers.

\begin{widetext}
\hspace{-5cm} .\vspace{-0.2cm}
%FIG5
\begin{figure}[ht]
\centerline{
\includegraphics[width=6.in]{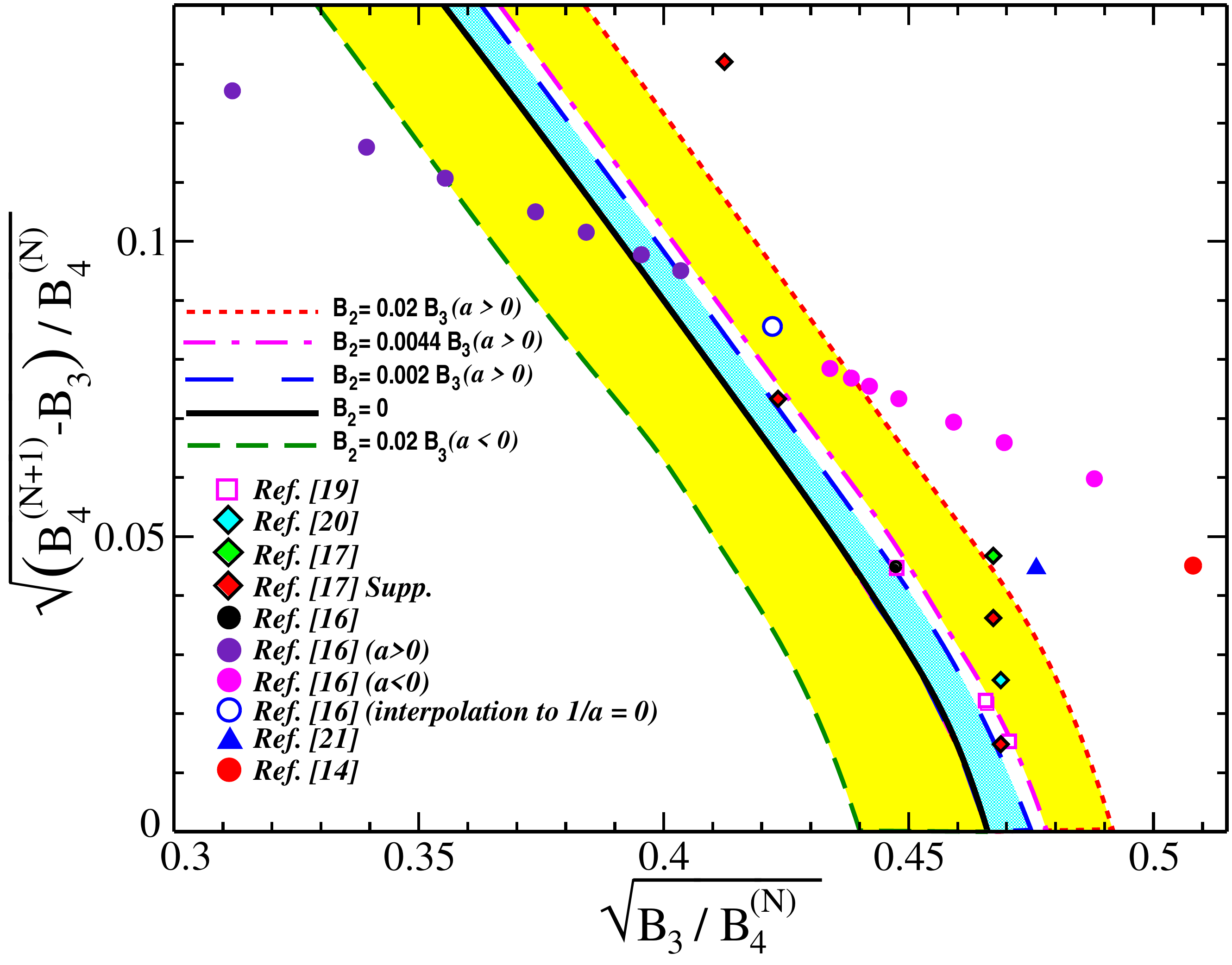}}
\caption{(Color on-line)
Our results for the four-boson scaling function, reported in \cite{prl2011}, are
displayed in the limited region where more recent results are available. The plots
are shown for $B_2=0$ (solid-black curve) and in four cases with $B_2\ne 0$
[three cases with bound dimers, $B_2/B_3 =$ 0.02, 0.0044, 0.002, and one
case with virtual dimer, $\sqrt{B_2/B_3}= -\sqrt{0.02}$], as indicated inside
the frame.
In the case of $B_2=0$, this figure refers to the right-hand-side corner of Fig.~\ref{scalings43} for $M=4$.
The results were explicitly verified numerically for $N=0$ and $N=1$, considering the
energies $B_4^{N=0,1,2}$, in case $B_2=0$; and for $N=0$ in the other cases where
$B_2\ne 0$. The symbols refer to other model calculations (as indicated) near
the unitary limit.} \label{scaling-section}
\end{figure}
.\end{widetext}
In Fig.~\ref{scaling-section}, to verify the universality of
the four-boson scaling function (\ref{eq2}) and for comparison, we include results
obtained in other recent calculations. In Table~\ref{Table.tetramers-all-groups}
we have listed the available results of other authors, for binding energy ratios
of ground and excited tetramers close to the unitary limit. Within those results
we should also point out that the ones given in Refs.~\cite{stechernature,deltuva2010}
are for more than one trimer levels.
In our results with zero-range interaction, we include not only the exact limit
$B_2=0$, but also non-zero two-body energies (bound or virtual), as shown
in Tables~\ref{Tabletetramers-our-results} (A), (B) and (C). The
sensibility of the scaling function with variation of the two-body
conditions is shown when non-zero two-body energies are used. As it
is shown, the exact unitary limit scaling plot is shifted to the
right-hand side (left-hand side) when considering non-zero two-body
bound (virtual) state energies.

The results of available calculations for tetramer
energies plotted in the way we are suggesting  put in evidence the
effect of the four-body scale. Consistent with our findings, they
slide along the universal correlation shown in Fig.~\ref{scaling-section}.
The sensitivity of our defined scaling plot, in respect to variations of
a two-body observable, as the scattering length or range effects,
is also consistent with the results obtained by other authors.

In order to verify the sensitivity of the results (obtained for $B_2=0$)
with small changes in the dimer energies, we calculate the corresponding
scaling functions for a few cases with $B_2\ne 0$. In the given results,
we have three cases for bound two-body systems, where
$B_2/B_3 = $ 0.02, 0.0044, 0.002, and one case for a virtual two-body state,
with $\sqrt{B_2/B_3} = -\sqrt{0.02}$.
In Fig.~\ref{scaling-section}, we have also included the Hammer and Platter
results for both virtual and real dimer close to unitary limit (See
Fig. 2 of Ref.~\cite{hammer07}). Despite of the observed
difference in the position of both results in our plot, it is
remarkable that they both fit nicely in our scaling curve, meaning
that the four-boson model of Ref.~\cite{hammer07} is sensitive
to a four-body scaling parameter, carried out implicitly by the
momentum cut-offs used in their calculations.

Although we calculate tetramer energies below the trimer ground
state, such that the dashed part of Fig.~\ref{scalings43} is also
verified, the results for tetramers attached to any other Efimov state,
does not change the present conclusions on the existence of a proper
four-body scale or its universal manifestation through the scaling
function (\ref{eq2}), calculated with the zero-range model
regularized within our scheme. This fact is evidenced in
Fig.~\ref{scaling-section}. Distinct short-ranged interaction
models~\cite{hammer07,stechernature,deltuva2010,stecherjpa,Lazauskas06}
show that the energies of successive tetramers linked to the tower
of Efimov states scale according to the plot, and thus verifying the
universality of the four-boson limit cycle.

\begin{table}[ht]
\caption[dummy0]{Binding energy ratio of tetramer ground and excited energies
to a trimer level,
$\left[{B_4^{(0)}}/{B_3},{B_4^{(1)}}/{B_3}-1\right]$, where
the results of Refs.\cite{stechernature,deltuva2010} refer to different trimers.\label{Table.tetramers-all-groups}}
\begin{tabular}{cccccccccc}
\hline\hline
Ref.\cite{blume00}&&Ref.\cite{meissner04}&&
Ref.\cite{Lazauskas06}&&Ref.\cite{hammer07}&&Ref.\cite{stecherjpa}\cr
\hline\hline
[4.46,0.06]&&[4.075,0.003]&&[4.41,0.01]&&[5.0,0.01]&&[4.55,0.003]\cr
\hline\hline \\
\end{tabular}
\begin{tabular}{cccc}
\hline\hline
Ref.\cite{stechernature}&&Ref.\cite{deltuva2010} \cr
\hline\hline
[4.58,\;\;0.01]&&[4.5175,\;\;0.00106]\cr
[5.88,\;\;0.10]&&[4.6041,\;\;0.00217]\cr
[4.58,\;\;0.006]&&[4.6104,\;\;0.00227]\cr
[5.58,\;\;0.03] &&[4.6108,\;\;0.00228]\cr
[4.55,\;\;0.001]&&[4.9929,\;\;0.00997]\cr
     -      &&[4.6114,0.00228]\cr
\hline\hline
\end{tabular}
\end{table}

\section{Structure of Universal Tetramers}
\label{4Bstruct}

The effect of the four-body scale on the structure of tetramers at
the unitary limit is detailed in this section. We show results for
the FY components in momentum space for different scale ratios as
well as for momentum probability densities. A close inspection to
the reduced FY components, $\cal K$ and $\cal H$, will show how
$\mu_4$ manifests through their dependence on the different Jacobi
momenta. An analogous study will be presented for the momentum
probability densities.

\subsection{Reduced Faddeev-Yakubovsky Components}
\label{sec_structure-FY}
\begin{figure}[ht]
\begin{center}
\includegraphics[width=3.5in]{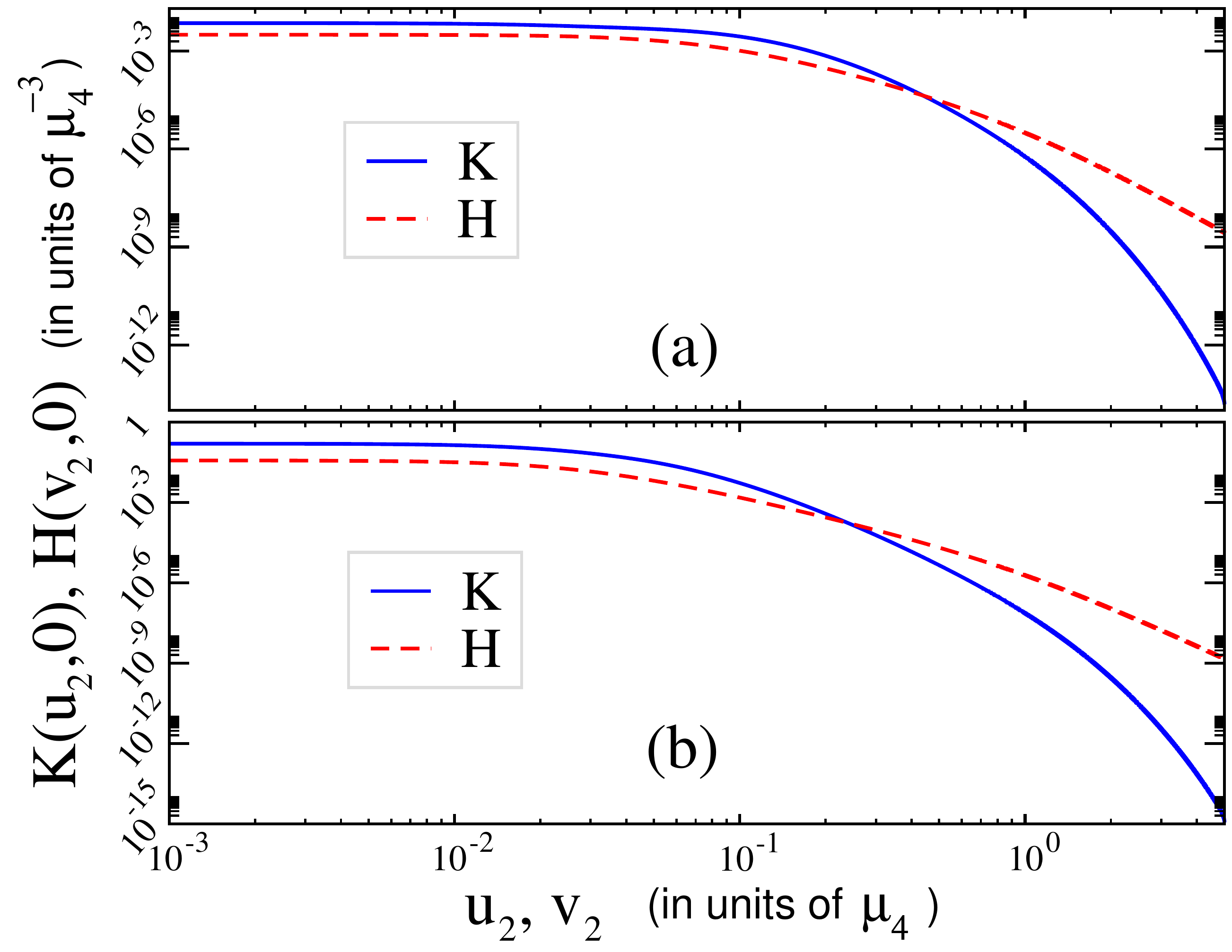}
\hspace{.2cm}
\includegraphics[width=3.5in]{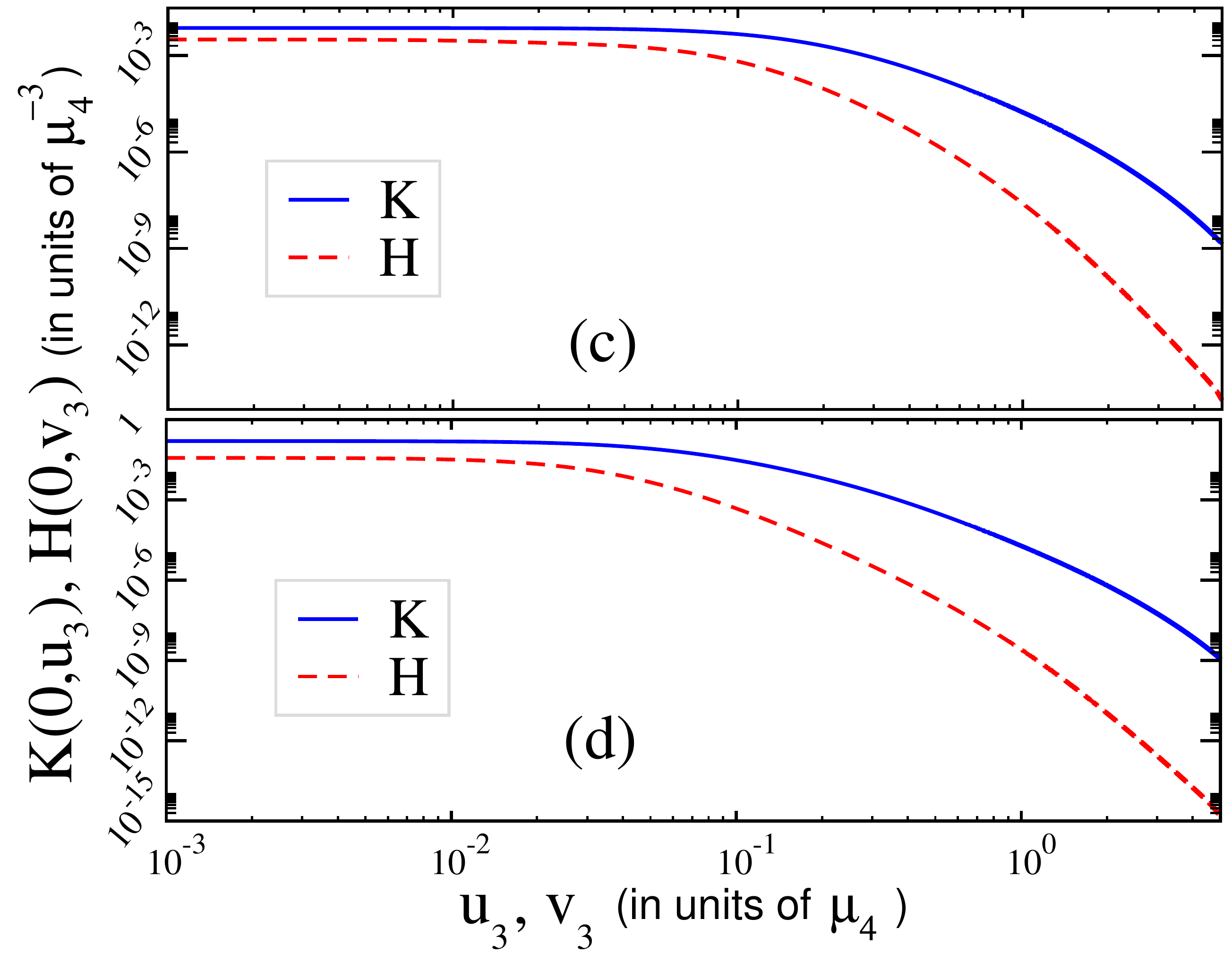} 
\end{center}
\vspace{-0.5cm}
\caption{(Color on-line) The Yakubovsky components ${\cal K}$ and ${\cal H}$, as
functions of the Jacobi momenta for scale ratio
${\mu_4}/{\mu_3}=5$, when only one four-body excited state
exists. In the frames (a) and (b) the components are shown as a function
of $u_2, v_2$ where $u_3, v_3=0$ and in the frames (c) and (d) are shown
as function of $u_3, v_3$ where $u_2, v_2=0$. In (a) and (c) we have
results for the ground-state level; and in (b) and (d) for the first
excited-state level. The normalization is such that
$\int_{0}^\infty  du_2 \, u_2^2 \int_{0}^\infty
du_3 \, u_3^2 \,\, {\cal K}^2(u_2,u_3)+ \int_{0}^\infty  dv_2 \, v_2^2 \int_{0}^\infty  dv_3
\, v_3^2 \,\, {\cal H}^2(v_2,v_3)=1$.
}\label{fig:KH-5}
\end{figure}

\begin{figure}[ht]
\begin{center}
\includegraphics[width=3.5in]{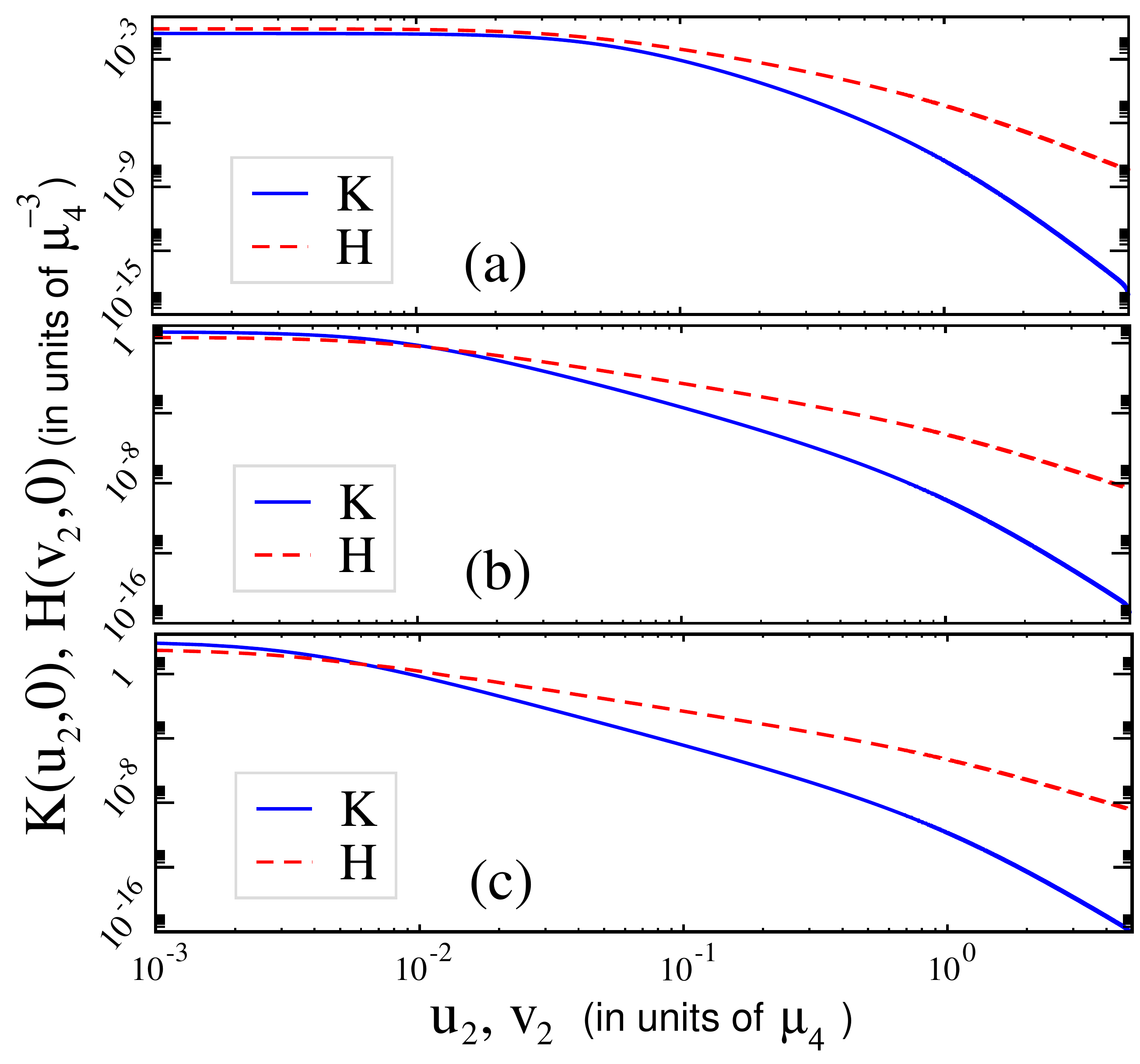}
\includegraphics[width=3.5in]{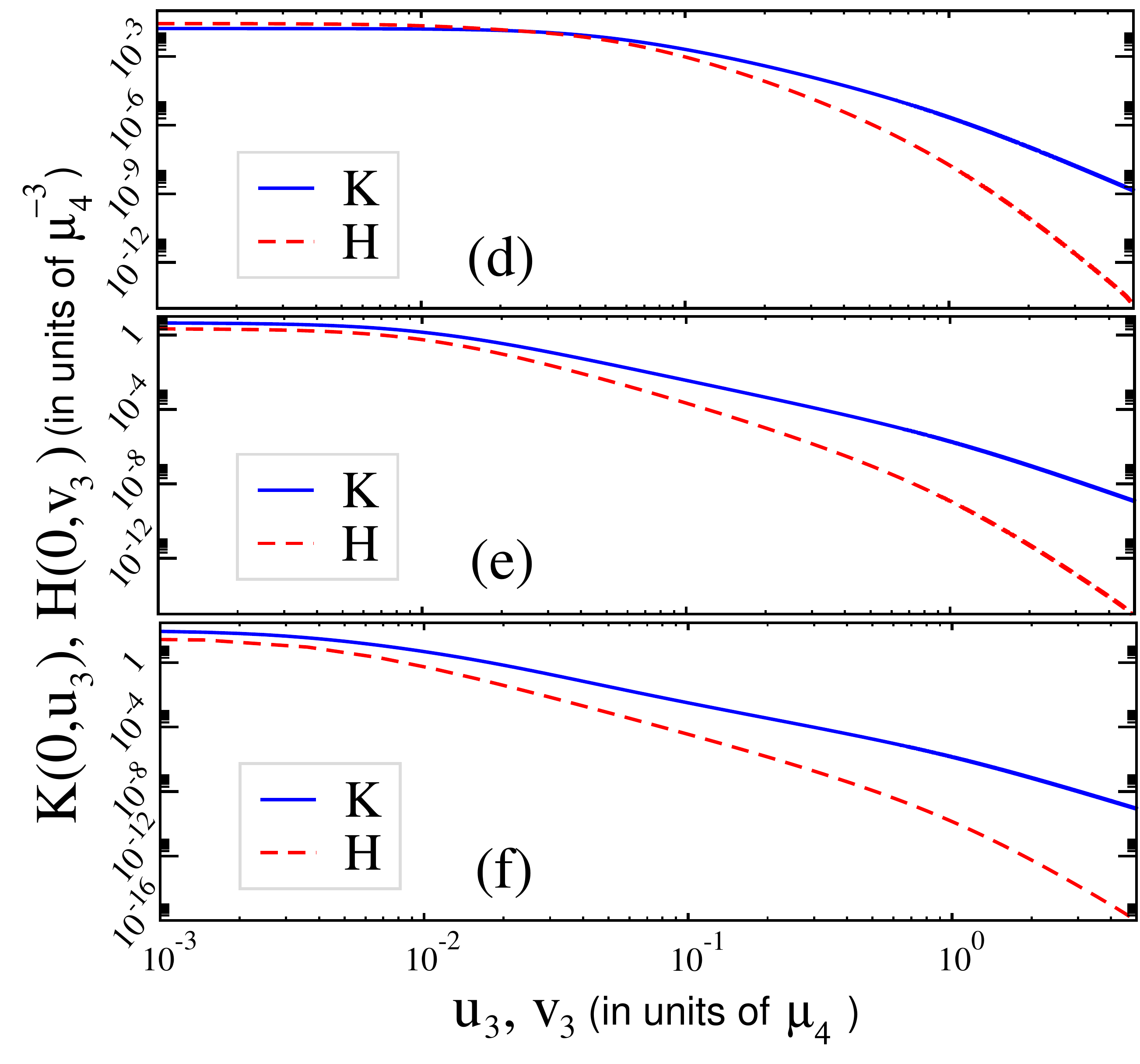}  
\end{center}
\vspace{-.5cm}
\caption{(Color on-line) Following the caption of  Fig.~\ref{fig:KH-5}, here we
have the same quantities for ${\mu_4}/{\mu_3}=50$. In the frames (a) and (d)
we have the ground state; in (b) and (e), the first excited state;
and in (c) and (f), the second excited state.}
\label{fig:KH-50}
\vspace{-0.5cm}
\end{figure}

The reduced FY components ${\cal K}(u_2,u_3)$ and ${\cal H}(v_2,v_3)$,
are shown to spread out up to momentum of the order 1, due to the relevance
of the four-body momentum scale $\mu_4$ to regularize the kernel of the set
of FY equations (\ref{eq.KH-integral-final}) at short distances.
Both components have narrow peaks
appearing at small momenta,
which are even more pronounced for higher scale ratios $\mu_4/\mu_3$
and more excited states.
They follow a typical tetramer momentum scale
($\sqrt{B^{(N)}_4}/\mu_4$).
These features are evident in Figs.~\ref{fig:KH-5} and \ref{fig:KH-50}.
For a given ratio $\mu_4/\mu_3$, the ground state wave function is
more expanded in momentum space in comparison to excited states. The
$K-$ and $H$-channels, and consequently the total wave function for
smaller tetramer bindings, should extend to larger distances as
compared to the corresponding ones with larger binding.

%FIG6

%FIG7

The peaks at small momentum are due to the nearest trimer pole in
the three-boson interacting resolvent, appearing in the $K-$channel
equation (\ref{eq.KH-integral-final}). This leads to a dominance of
$\cal K$ over $\cal H$ for small momentum when the tetramer energy
approaches $B_3$.

In Figs.~\ref{fig:KH-5} and \ref{fig:KH-50} 
we detail the sections of
${\cal K}(u_2,u_3)$ and ${\cal H}(v_2,v_3)$ with one of the Jacobi
momenta being zero for scale ratios of 5, 50 and 200. The
interpretation of the Jacobi momentum in respect to the relevant
scales is necessary to proceed in the analysis of these plots. The
variable $u_2$ is the relative momentum of particle 3 in respect to
the center of mass of the pair (1,2), which belongs to a trimer
configuration (see Fig.~\ref{K-H}). Note that in our derivation of
the FY equation for the zero range potential, we have chosen to
factorize out the dependence of the relative momentum of the pair
(1,2) in the definition of the reduced amplitudes, as given by
Eq.~(\ref{eq.KH-redifinition}).
Naively, by invoking kinematical reasons, it is reasonable
to expect that, within the tetramer, the average distance between two
bosons is larger than the average distance between the center-of-mass
(c.m.) of the virtual pair to the third particle, which is larger than the
average distance of the virtual trimer to the fourth particle.
The ordering effect comes because the clusters being heavier should have
their c.m. closer to the c.m. of the tetramer.
This size ordering has a correspondence in respect to
the momentum variables.  In ${\cal K}(u_2,u_3)$, the momentum of the
fourth particle in respect to the virtual trimer, $u_3$,
should explore larger momentum regions than $u_2$, which is the
relative momentum of the third particle in respect to a pair.
An analogous reasoning suggests that the momentum dependence of
${\cal H}(v_2,v_3)$ is such that $v_2$ explores larger momentum regions
than  $v_3$ (the relative momentum of two bosons). These qualitative
properties are verified in Figs.~\ref{fig:KH-5} and \ref{fig:KH-50}.

The plots in Fig.~\ref{fig:KH-5} show the reduced FY
components for the ground, frames (a) and (c), and excited state,
frame (b) and (d), for $\mu_4/\mu_3=5$. The high momentum tails of
${\cal H}(v_2,0)$ and ${\cal K}(0,u_3)$ in respect to ${\cal
H}(0,v_3)$ and ${\cal K}(u_2,0)$ are visible by comparing the frames
(a) and (c) in the case of the ground state, and (b) and (d) for
the excited state. By inspecting the pair of frames (a) against (b);
and (c) against (d), we also note that all the FY components for the
excited state are more concentrated at lower momentum than the
corresponding ones for the ground state.
Although the binding energy ratio $B_4^{(1)}/B_3$ for the excited state
is close to 1, the high momentum tail bringing the four-body scale is
present, giving to this state the possibility to move as $\mu_4$ is
changed in respect to $\mu_3$. 
These qualitative features are supported by the results we have obtained
for larger scale ratio, such as $\mu_4/\mu_3=$50 and 200, where three 
tetramer states are possible. 
This behavior is clearly shown in Fig.~\ref{fig:KH-50} 
for $\mu_4/\mu_3=$50.

In all the cases we analyzed, the dependence of $u_3$ in ${\cal
K}(u_2,u_3)$ and $v_2$ in ${\cal H}(v_2,v_3)$ show that these two
variables explore larger momentum regions than $u_2$ and $v_3$,
respectively. As we have shown ${\cal K}(u_2,0)$ and ${\cal
H}(0,v_3)$ have a less prominent tail at large momentum than ${\cal
K}(0,u_3)$ and ${\cal H}(v_2,0)$. If, for any reason, the high
momentum dependence that appears in the $K-$ and $H-$ channel are
dropped out, the dependence on the four-body scale will be minimized
(may be even completely removed!).
Therefore, unreasonable selection of cut-off values in the mapping
of momentum variables, which can happen in particular when a fixed
cut-off is being used for increasing values of the scale ratios,
can lead to convergence in the four-body binding energies rather
than collapse.
For the discretization of momentum variables one should not only
consider large enough cut-off values, consistent with the used
four-body scale, but also consider a reasonable number of mesh
points in the interval near zero-momentum. Since the iteration of
the coupled Eqs.~(\ref{eq.KH-integral-final}) requires a very large
number of multi-dimensional interpolations on the Yakubovsky
components, we have used Cubic-Hermite Splines to reach high
computational accuracy.

From the above analysis, represented in Figs.~\ref{fig:KH-5} and 
\ref{fig:KH-50}, we conclude that, against the general belief stating that in
four-body atomic and nuclear calculations, with model potentials
in the FY scheme, the $K-$channel is always dominant, we show
some cases where the $H-$channel is dominant even at low momentum.
This is shown in particular for high scale ratios $\mu_4/\mu_3$.

\subsection{Momentum probability densities}
\label{sec_momentum-distribution-functions}

The tetramer wave function  $\Psi \bigl ( u_1, u_2, u_3 \bigr)$
fully symmetrized is built from Eq. (\ref{eq.4B-wave-funtion}) using
the definitions (\ref{eq.KH-redifinition}) of the $K$ and $H$
components in terms of ${\cal K}(u_2,u_3)$ and ${\cal H}(v_2,v_3)$.
In order to simplify our analysis of the wave function, and to
obtain an insight on how the momentum is shared among the Jacobi
coordinates, we introduce the momentum probability densities
$n(u_i)$ as:
\begin{eqnarray}
n(u_i) &=& u_i^2 \int_{0}^\infty  du_j u_j^2 \int_{0}^\infty
du_k u_k^2 \,\, \Psi^2 \bigl (
u_1, u_2, u_3 \bigr);\label{eq.nui}\\
&&\int_{0}^\infty  du_i \, n(u_i) =1,
\nonumber \end{eqnarray}
with $(i,j,k)$ = (1,2,3), (2,3,1) and (3,1,2). For reference we
use the variables $u_i$ of the $K$ configuration, where $u_1$ is the
relative momentum of a pair, $u_2$ the relative momentum of the
third boson to the C.M. of the pair, and $u_3$ the relative momentum
of the fourth boson to the C.M. of the other three (see Fig.
\ref{K-H}). As we have discussed in sect. \ref{sec_structure-FY},
the average values of the momentum variables are ordered as $\langle
u_1\rangle < \langle u_2\rangle < \langle u_3\rangle$, which is
reflected in momentum densities, independent of the degree of
excitation of the tetramer.

The momentum probability densities $n(u_1)$, $n(u_2)$ and $n(u_3)$
for scale ratio ${\mu_4}/{\mu_3}=50$, for ground, first and
second excited tetramer states are presented in Fig.~\ref{fig:nui_50}.
In the left panel we show our results for the probability densities
in units of $\mu_4$. Clearly for all three
momentum densities, the second excited tetramer leads to a very
sharp peak close to zero momentum, whereas for the first excited and
ground tetramers the peak is more wide, which is evident as the
size is inversely correlated to the tetramer binding energy. For the
ground tetramer state the FY components and consequently the 4B
total wave function are expanded to higher momentum region and for
excited tetramers they are more restricted to zero momentum region.
In the right panel of Fig.~\ref{fig:nui_50} we present the
probability densities in momentum units $\beta_N\equiv \sqrt{B_4^{(N)}}$,
for $\mu_4/\mu_3=50$. By considering this scaling factor corresponding
to the energy of each tetramer state, which from
Table~\ref{Tabletetramers-our-results} varies two order of magnitude, the
densities are amazingly close. The tetramer energy can be considered
as the physical scale, which correlates other observables.

One feature of the universal properties of the four-body wave
function is presented in Fig.~\ref{fig:nui_all} , where we have
compared the momentum distribution functions $n(u_i)$ of the two
shallowest tetramers, for two very different values of $\mu_{4}/\mu_{3}$ equal
to 5 and 50. The functions $n(u_i)$ were rescaled to the
same maximum value to make transparent their universal form. The
comparison between them clearly confirms a
universal dependence of the momentum distributions on
the tetramer binding energy. Note that, these quantities
are plotted in terms of dimensionless quantities in
natural units of the length associated with the tetramer energy.

In Fig.~\ref{fig:nui_all} we have shown the renormalized form of the
rescaled distribution functions, considering two different scale
ratios $\frac{\mu_4}{\mu_3}=5$ and $50$, whereas the distributions
are renormalized to the same maximum value. As we have shown the
distribution functions have universal shape, independent of the
value of scale ratio. We should also add a remark that, in order to
simplify the presentation of Fig.~\ref{fig:nui_all}, we have
included only the momentum distribution functions for the two more
excited states in case of $\frac{\mu_4}{\mu_3}=50$, whereas the same
universal shape is also obtained for the momentum distribution
functions of the ground state (see Fig.~\ref{fig:nui_50}). The
ground state energy ratio is $B_4^{(0)}/B_3$ = 275 (see
Table~\ref{Tabletetramers-our-results}), which means that
$B_4^{(0)}$ is about 2.7$\mu_3^2$, while for the two excited states,
$B_4^{(1)}$ and $B_4^{(2)}$ is much smaller than $\mu_3^2$, thus the
three-body regularization scale is far more important to the ground
state than for the excited states, which approach an universal form
for the momentum distribution functions.

\subsection{Tetramer wave function}

The calculation of the FY amplitudes $K$ and $H$, given by
Eq.~(\ref{eq.KH-redifinition}) for
form factor unity, with the reduced amplitudes $\cal K$ and $\cal H$
from the numerical solution of the
zero-range FY integral equations  (\ref{eq.KH-integral-final}), allows
to build the four-boson wave function fully symmetric by boson exchange.
The necessary permutations to reconstruct the 4B wave function from the
computed $K(u_1,u_2,u_3)$ and $H(v_1,v_2,v_3)$ are given  by introducing
permutation operators $\cal O$ with the corresponding FY amplitudes denoted
by $K_{\cal O}$ and $H_{\cal O}$ (see Appendix \ref{appendix B}).
Computing $K_{\cal O}$ and $H_{\cal O}$ for different ${\cal O}$,  the 4B
total wave function can be obtained by using Eq. (\ref{eq.4B-wave-function}) or (\ref{eq.K-H-in-momentum}).

For our purposes of presenting graphically the total wave function we will show plots
for the angle averaged wave function, where the dependence on the relative angles between
Jacobi momentum $\vec u_i$ ($i=(1,2,3)$) is integrated. Our notation is such that we still
use $\Psi\bigl ( u_1, u_2, u_3 \bigr)$ for the angle averaged wave function.
Instead of the usual normalization, we consider
{\small \begin{eqnarray}
\langle \Psi | \Psi \rangle &\equiv&
\int_{0}^\infty  du_1 u_1^2 \int_{0}^\infty  du_2 u_2^2 \int_{0}^\infty
du_3 u_3^2 \,\, \Psi^2 \bigl ( u_1, u_2, u_3 \bigr)
\nonumber\\ &=& 1.
\label{eq.4B-WF-normalization}
\end{eqnarray}}

In Figs. \ref{Fig:WF_u1u2_50}-\ref{Fig:WF_u2u3_50} we present our
numerical results for the angle averaged total wave function with
scale ratio ${\mu_4}/{\mu_3}=50$, for the ground, first and second
excited states. The wave functions are shown as functions of two
Jacobi momenta, when the third one is chosen to be zero. As we did
before, the magnitudes of Jacobi momenta are re-scaled by the momentum
factor ${\beta_N}\equiv \sqrt{B_4^{(N)}}$ ($N=0,1,2$). In view of the
normalization  condition (\ref{eq.4B-WF-normalization}), this
momentum scaling implies in a rescaling of the magnitude of the
corresponding total wave function by the factor of $\beta_N^{9/2}$.

The universal form of the 4B wave function is evidenced in Figs.
\ref{Fig:WF_u1u2_50}-\ref{Fig:WF_u2u3_50} by a proper rescaling with
the four-body energy. Correspondingly, the contour plots are also
presented as functions of two non-vanishing Jacobi momenta, given in
units of $\beta_N$. Also the momentum density distributions, for two
excited states, present a universal form, i.e., independent on the
binding energies, when the momentum is measured in units of
$\beta_N$, as shown in Fig.~\ref{fig:nui_50}. Note that each of the
choices of the vanishing momentum implies in a very long wavelength
limit for the fourth particle, indicating the dominance of a
three-body cluster configuration within the four-body system. This
configuration carries the four-body scale information, as clearly
shown by the plots for the wave-function dependence on the two
non-vanishing momentum in units of $\beta_N$.

\section{Conclusions and Perspectives}\label{conclusions}
The momentum-space structure of the FY
components of weakly-bound tetramers was thoroughly analyzed at the
unitary limit using a renormalized zero-range two-body interaction,
with three and four-body scales. Our regularization scheme allows to
introduce independently these two scales providing general framework
to investigate the universal properties of tetramers, which has
extended tails in the classically forbidden region. The step to
renormalized results for the observables is found by introducing
scaling functions, which are written only in terms of dimensionless
physical quantities. These universal scaling functions achieve
numerically a limit cycle.

The universal scaling function correlating two successive tetramer
energies attached to one trimer comes from the sensitivity of the
four-boson system to a short-range four-body scale. Each excited
$N-$th tetramer energy $B_4^{(N)}$ moves as the short-range
four-body scale changes while the trimer properties are kept fixed.
We suggest that the ratio of $B_4/B_3$, which gives  the slope of
the Tjon line, and it is not a constant as thought before, as a
parameter measuring the four-body scale. At the unitary limit, the
successive $(N+1)-$th  tetramer state emerges from the atom plus trimer
threshold
for a universal ratio $B_4^{(N)}/B_3 = B_4^ {(N)}/B_4^{(N+1)} \simeq
4.6$, which does not depend on $N$. The atom-trimer relaxation
resonates when the tetramer hits the scattering threshold.

We also find out that other model results obtained at the unitary
limit or close to
it~\cite{hammer07,stechernature,deltuva2010,stecherjpa,Lazauskas06},
are quite consistent with our four-boson scaling plot, giving
confidence on the universality of our proposed scaling function. As
evidenced by our results, the independent behavior of the four-body
scale is verified in particular when a universal excited four-boson
state pumps out from the atom-trimer threshold as the four-boson
parameter is driven to short distances or to the ultraviolet
momentum region. In the case that both scales are similar, we
confirm that our scaling approach is consistent with the results of
other model calculations.  We note that the results obtained by other
groups appear near the threshold region due to the model assumptions
that have been considered, which are much too restrictive when
compared to the flexibility of our model and need to be relaxed to
allow a wider variation between the four- and three-body properties.
The suggested scaling plot offers a model independent way to view
the relation between successive tetramer states, which can be
verified by experiments exploring two-body scattering lengths very
close to the Feshbach resonance, where multi-boson forces are
expected to be active~\cite{nakajimaprl1}.
{A four-boson short-range interaction can drive the four-body
scale independently from the three-body one. Short-range three-body
forces accompanied by different off-shell behaviors of the two-body
interaction, which kept the low-energy two- and three-body
properties unaltered, could provide another possibility to move the
four-body scale in practice.}

The attent reader should realize the power of the four-body
scaling plot in pinning down the independent behavior of a four-body
scale (for a fixed three-body one), in a similar way as the
three-body scaling plot, given in Ref.~\cite{amorim99}, resumes all
the Efimov states obtained by changing the three-body scale. Next,
we should emphasize that the scaling curve relating two successive
tetramer energies is valid not only for the stable tetramers
attached to the ground-state trimer, but also for the
unstable tetramers between two successive trimers. For such
conclusion we have also performed resonance calculations of
tetramers attached to excited trimers, which are under detailed
numerical analysis to be included in a future work. 
We avoid to present such results in the present work  
mainly because we have considered other independent and precise 
calculations supporting our claim of model independence. 
In this regard, the recent calculations performed by 
Deltuva~\cite{deltuva2010} are consistent with our results and give 
confidence on the model independence of the scaling plot relating 
the energies of two successive tetramers between two successive 
Efimov trimers in the unitary limit.
As we have shown, the effect of scattering length variations
is noticeable and shows a universal pattern in agreement with 
calculations shown in \cite{hammer07}, performed in the context of 
effective field theory. However, range corrections,
which were not yet explored within a general study considering the 
trimer and tetramer interwoven spectra, will certainly have their importance 
in explaining how results near the unitary limit deviate from the
calculated scaling function for a zero-range force. In this respect,
range corrections, linear in the two-body effective range, have been already
addressed within an effective theory treatment of the three-boson
problem with subtracted momentum-space integral equations
\cite{PlatterPRA09}. 

The scaling plot representation exhibiting the dependence on the
four-body scale, which is also being confirmed by other models,
shows that the model independence is not fortuitous. The reason for
that comes from the fact that the wave function of tetramers are
largely dominated by configurations where the bosons are outside the
potential range in the classically forbidden region and depend just
on two scales at the unitary limit. The sensitivity to the
short-range scales appears in both the $K-$ and $H-$channel of the
FY decomposition, which present high momentum tails for any degree
of the tetramer excitation. We also found that the $H-$channel is
favored over $K-$channel at low momentum when the four-body momentum
scale largely overcomes the three-body one.

The universal form of the wave function is putted forward at the
unitary limit by a scaling plot where the probability momentum
densities for different Jacobi momenta are shown in units such that
the tetramer binding energy is one.  We found that the shapes are
independent on the scale ratio and excitation. It only depends on
the chosen Jacobi momenta. A simple scaling rule giving the
ordering $\langle u_1\rangle < \langle u_2\rangle < \langle
u_3\rangle$, where $u_1$ is the relative momentum of a pair, $u_2$
is the relative momentum of the third particle to the pair, and
$u_3$ is the relative momentum of the fourth particle in respect to
the three-body subsystem.

Our calculations of tetramer binding energies, with corresponding
structure of momentum probability density and wave functions,
are supplying strong numerical evidences that universal tetramers
can arise from the trimer threshold as the four-body scale is moved.
These  results suggest the possibility of resonant atom-trimer recombination
process near the unitary limit, i.e., when $a_{AT}\to \pm\infty$. It
is worth noting that the problem of a new scale in the tetramer
properties has a long history, which was never clarified before in
the physics community to the extend we are doing now. The four-body
scale gives rise to new scalings of tetramer observables which are
not determined only by the two and three-body properties. We present
detailed and accurate numerical results for the binding and the
structure of tetramer states to support our claim, given conveniently
within the framework of scaling functions. As a matter of fact, even results
obtained near the unitary limit, by authors that are claiming that
no four-body scale is necessary, are consistent with the universal
scaling plot for the tetramer binding energies.

In our numerical results of four-boson bound states, we first
consider the exact unitary limit ($1/a=0$), and we next consider
deviations of this limit to bound ($a>0$) and virtual ($a<0$)
two-body branches. As shown in Figs.~\ref{E4xE3-E3xE2}-\ref{scaling-section},
universal tetramers can hit the atom-trimer threshold leading to a resonant
relaxation in this channel; a crucial phenomenon not considered in
other cited recent numerical investigations.
In fact, in those analysis, the authors could also have verified the occurrence
of tetramer states merging to an atom-trimer threshold, particularly if
they had considered the possibility of a four-body independent
scaling behavior near a Feshbach resonance. This could be done by
including, for example, tunable three or four-body potentials, which
would allow an independent change of the three and four-body spectra.
The position of the atom-trimer resonance is  not only a function of the
atom-atom scattering and the three-body scale, but it also depends
on the new four-body scale.
Our results do not exclude the resonant
dimer-dimer recombination~\cite{incaoprl2009} but also add the
possibility of a resonant atom-trimer recombination.

Finally, we remark that the four-boson scale can be driven near the
Feshbach resonance by induced four-body forces (coming from the
one-channel reduction of the atomic interaction)
\cite{epl06,nakajimaprl1}. Therefore, in this case the Efimov ratio
percolating the tetramer observables is not assured anymore. Other
universal scaling functions can be derived correlating properties of
tetramers, in particular the one that correlates the binding
energies of tetramers attached to different trimers, and as well
as by extending our framework to the scattering region. The possibility
that tetramers could be formed and driven in cold-atom laboratories,
as have been achieved by radio frequency association techniques
applied successfully to measure trimer energies \cite{jochim2010},
will allow to verify experimentally the universal scaling relations
between tetramer properties exhibiting a new scale beyond the trimer
one.

\section*{Acknowledgments}
We acknowledge partial financial support from the Brazilian agencies
Funda\c c\~ao de Amparo \`a Pesquisa do Estado de S\~ao Paulo and
Conselho Nacional de Desenvolvimento Cient\'\i fico e Tecnol\'ogico.

\appendix
\section{Momentum space representation of FY equations}
\label{appendix A}
In the following we present some details on the FY components in the
momentum space representation. By considering the completeness relations,
Eq.(\ref{eq.completness}), between the permutation operators, we have
\begin{widetext}
\begin{eqnarray}
\langle u|K\rangle &=&
\int D^{3}u'\,\langle u |G_{0}t P|u'\, \rangle\langle u' |K\rangle
+ \int D^{3}u'\, D^{3}v\,\langle u |G_{0}t P| u'\rangle  \langle u'|v\rangle\langle v|H\rangle
 \nonumber \\*  &+& \int D^{3}u'\, D^{3}u''\, \langle u |G_{0}t P |u' \rangle \langle u'| P_{34}|u'' \rangle
\langle u'' |K\rangle , \nonumber \\*
\label{eq.KH-in-uv-expanded} \\*
\langle v | H\rangle &=&
\int D^{3}v'\,\langle v |G_{0}t \tilde{P}|v'\rangle\langle v'|H\rangle
+ \int D^{3}v'\,D^{3}u\,\langle v |G_{0}t \tilde{P} | v' \rangle  \langle v'|u\rangle \langle u|K\rangle
\nonumber \\*
&+&\int D^{3}v'\,D^{3}u\,D^{3}u'\,  \langle v |G_{0}t
\tilde{P} | v'\rangle  \langle v'|u\rangle \langle u|P_{34}|u'\rangle
\langle u'|K\rangle.
  \nonumber
\end{eqnarray}
\end{widetext}
Therefore, to evaluate the above coupled equations
(\ref{eq.KH-in-uv-expanded})
we need to obtain the following matrix elements:
\begin{equation}
\langle u |G_{0}tP| u' \rangle, \;\;\;\;
\langle v |G_{0}t\tilde{P}|  v' \rangle, \label{eq.G0tpbar-operator}
\end{equation}
\begin{equation}
\langle u|P_{34}| u'\rangle,\;\;\;\;
\langle u'| v\rangle,\;\;\;\; \quad \langle v'|u\rangle.
\label{eq.uv-sandwich-operator}
\end{equation}
By considering that, in the expressions (\ref{eq.G0tpbar-operator})
we have to insert the two-body matrix elements (\ref{eq.t-matrix-u}), and that
\begin{eqnarray}
\langle u| P| u'\rangle&=& \int_{-1}^1  dx \,
\frac{\delta\biggl(u_1-\Pi_1(u_2,u'_2,x)\biggr)}{u^2_1}
\label{eq.p-matrix} \\* &\times&
\frac{\delta\biggl(u'_1-
\Pi_1(u'_2,u_2,x)\biggr)}{u'^2_1} \, \frac{\delta\biggl(u_3-
u'_3\biggr)}{u^2_3},
\nonumber
\end{eqnarray}
\begin{eqnarray}
\Pi_1(u_2,u'_2,x) &\equiv& \biggl|\frac{1}{2} \textbf{u}_2 +
\textbf{u}'_2 \biggr|
\label{Pi1}\\&=&
\sqrt{\frac{1}{4}u^2_2+u'^2_2+u_2 u'^2_2 x},
\nonumber
\end{eqnarray}
\begin{eqnarray}
\langle   v | \tilde{P}| v'\rangle
&=& \frac{\delta \bigl(v'_1- v_3\bigr)}{v'^2_1} \, \frac{\delta
\bigl(v'_3- v_1\bigr)}{v'^2_3} \, \frac{\delta \bigl(v'_2-
v_2\bigr)}{v'^2_2}, \label{eq.pbar-matrix}
\end{eqnarray}
the matrix elements in (\ref{eq.G0tpbar-operator})
are given by the following:
{\small
\begin{eqnarray}
&&\langle u|G_{0}t P| u'\rangle = G_{0} \bigl( u_1, u_2, u_3\bigr)
\int D^{3}u''\langle u|t| u''\rangle  \langle u''| P|u'\rangle
\nonumber\\*
&&= 4 \pi \, G_{0} \bigl( u_1, u_2, u_3\bigr) \chi(u_1) \tau({\cal E}_u)
\frac{\delta\biggl(u'_3- u_3\biggr)}{u'^2_3}
\label{eq.G0tp-final}\\*
&&\times
\int_{-1}^1 dx \chi \biggl(\Pi_1(u_2,u'_2,x) \biggr)\,
\frac{\delta\biggl(u'_1- \Pi_1(u'_2,u_2,x)\biggr)}{u'^2_1}\; ;
\nonumber
\end{eqnarray}
\begin{eqnarray}
&&\langle  v |G_{0}t \tilde{P} |v'\rangle =
G_{0} \bigl( v_1, v_2, v_3\bigr)
 \int D^{3}v'' \, \langle  v |t |  v'' \rangle
\, \langle  v'' | \tilde{P}|  v'\rangle,
\nonumber\\*
&&= 4 \pi \, G_{0} \bigl(
v_1, v_2, v_3\bigr) \, \chi(v_1)\, \chi(v'_3)\, \tau({\cal E}_v)
\label{eq.G0tpbar-final}\\ &&\times
\frac{\delta \bigl(v_3- v'_1\bigr)}{v^2_3}
\frac{\delta \bigl(v_2- v'_2\bigr)}{v^2_2} \; ;
\nonumber
\end{eqnarray}
\begin{eqnarray}
\langle u|P_{34}| u'\rangle &=&
\frac{\delta \bigl(u_1-u'_1\bigr)}{u'^2_1}\frac{1}{2}
\int_{-1}^1 dx \label{eq.p34-final}\\* &\times&
\frac{\delta\biggl(u'_2- \Pi_2 \bigl(u_2,u_3,x \bigr) \biggr)}{u'^2_2}
\frac{\delta \biggl(u'_3- \Pi_3 \bigl(u_2,u_3,x\bigr)\biggr)}{u'^2_3} ,
\nonumber \end{eqnarray}
where
\begin{eqnarray}
\Pi_2 \bigl(u_2,u_3,x \bigr)
&\equiv& \biggl| \frac{1}{3}\textbf{u}_2+\frac{8}{9}\textbf{u}_3
\biggr| \label{Pi2} \\*
&=& \sqrt{\frac{1}{9}u^2_2 + \frac{64}{81}u^2_3 +
\frac{16}{27}u_2 u_3 x}, \nonumber \\*
\Pi_3\bigl(u_2,u_3,x \bigr) &\equiv&
\biggl| \textbf{u}_2- \frac{1}{3}\textbf{u}_3 \biggr|
\label{Pi3}\\*
&=& \sqrt{ u^2_2 + \frac{1}{9}u^2_3 -
\frac{2}{3}u_2 u_3 x}
 \nonumber\; ;
\end{eqnarray}
\begin{eqnarray}
\langle u'| v\rangle &=& \frac{\delta \bigl(v_1 - u'_1
\bigr)}{v^2_1} \, \frac{1}{2}\int_{-1}^1  dx
\label{eq.uv-final} \\
&\times&
\frac{\delta \biggl(v_3-\Pi_4 \bigl(u'_2,u'_3,x\bigr)
\biggr)}{v^2_3} \frac{\delta
\biggl(v_2- \Pi_5 \bigl(u'_2,u'_3,x\bigr) \biggr)}{v^2_2},
\nonumber
\end{eqnarray}
where
\begin{eqnarray}
\Pi_4 \bigl(u_2,u_3,x\bigr)  &\equiv& \biggl|
\frac{1}{2}\textbf{u}_2+\frac{2}{3}\textbf{u}_3   \biggr|
\label{Pi4}\\*&=&
\sqrt{\frac{1}{4}u^2_2 + \frac{4}{9}u^2_3 + \frac{2}{3}u_2
u_3 x}\; , \nonumber \\*
\Pi_5 \bigl(u_2,u_3,x\bigr) &\equiv&
\biggl| \textbf{u}_2- \frac{2}{3} \textbf{u}_3   \biggr|
\label{Pi5}\\* &=&
\sqrt{u^2_2 + \frac{4}{9}u^2_3 - \frac{4}{3}u_2 u_3 x} \; ;
\nonumber
\end{eqnarray}
and
\begin{eqnarray}
\langle v'| u\rangle &=& \frac{\delta \bigl(u_1 - v'_1
\bigr)}{u^2_1} \, \frac{1}{2} \int_{-1}^1  dx
\label{eq.vu-final}\\*
&\times& \frac{\delta \biggl(u_2-
\Pi_6 \bigl(v'_2,v'_3,x \bigr) \biggr)}{u^2_2} \frac{\delta
\biggl(u_3- \Pi_7 \bigl(v'_2,v'_3,x \bigr) \biggr)}{u^2_3} ,
\nonumber
\end{eqnarray}
where
\begin{eqnarray}
\Pi_6 \bigl(v_2,v_3,x \bigr)  &\equiv& \frac{2}{3}
\biggl| \textbf{v}_2+ \textbf{v}_3\biggr|\label{Pi6}\\*
&=&\frac{2}{3} \sqrt{v^2_2 + v^2_3 + 2 v_2 v_3 x},
\nonumber \\*
\Pi_7\bigl(v_2,v_3,x \bigr) &\equiv& \biggl| \textbf{v}_3- \frac{1}{2}
\textbf{v}_2   \biggr|\label{Pi7}\\
&=&\sqrt{ v^2_3 + \frac{1}{4} v^2_2 - v_2 v_3 x}.
 \nonumber
\end{eqnarray}
}

\section{Four-Body Total Wave Functions in Momentum Space}
\label{appendix B}

The total four-body wave function, which can be written as
\begin{eqnarray}
|\Psi\rangle &=&\Bigl(1+ P + P_{34} + PP_{34}
+ P_{34} P + P P_{34} P \Bigr)|K \rangle
\nonumber\\* &+&
\Bigl(1+P  + \tilde{P} + P \tilde{P} \Bigr)|H\rangle,
\label{eq.4B-wave-function}
\end{eqnarray}
is composed of eighteen FY components, according to the possible
arrangements of the four particles ($ijkl$).
We have 12 of $K-$type and 6 of $H-$type, as follows:
\begin{eqnarray}
    \begin{array}{ll}
  ijk+l \rightarrow
   \left\{
    \begin{array}{ll}
     ij+k+l \quad \equiv K_{ij,k}^{\,\,\, l}  \\
     jk+i+l \quad \equiv K_{jk,i}^{\,\,\, l} =P_{ij} \, P_{jk} \, K_{ij,k}^{\,\,\, l} \\
     ki+j+l \quad \equiv K_{ki,j}^{\,\,\, l} =P_{ik} \, P_{jk} \, K_{ij,k}^{\,\,\, l}
    \end{array}
    \right.
       \\
  ijl+k \rightarrow
   \left\{
      \begin{array}{ll}
    ij+l+k \quad \equiv K_{ij,l}^{\,\,\, k} =P_{kl} \, K_{ij,k}^{\,\,\, l} \\
    jl+i+k \quad \equiv K_{jl,i}^{\,\,\, k} =P_{ij} \, P_{jl} \, K_{ij,l}^{\,\,\, k} \\
    li+j+k \quad \equiv K_{li,j}^{\,\,\, k} =P_{il} \, P_{jl} \, K_{ij,l}^{\,\,\, k}
    \end{array}
    \right.
    \\
  ikl+j \rightarrow
   \left\{
    \begin{array}{ll}
    ik+l+j \quad \equiv K_{ik,l}^{\,\,\, j} =P_{jk} \, P_{kl} \, K_{ij,k}^{\,\,\, l} \\
    li+k+j \quad \equiv K_{li,k}^{\,\,\, j} =P_{il} \, P_{lk} \, K_{ik,l}^{\,\,\, j}  \\
    kl+i+j \quad \equiv K_{kl,i}^{\,\,\, j} =P_{ik} \, P_{lk} \, K_{ik,l}^{\,\,\, j}
    \end{array}
    \right.
    \\
  jkl+i \rightarrow
   \left\{
    \begin{array}{ll}
    jk+l+i \quad \equiv K_{jk,l}^{\,\,\, i} =P_{ij} \, P_{jk} \, P_{kl} \, K_{ij,k}^{\,\,\, l}\\
    lj+k+i \quad \equiv K_{lj,k}^{\,\,\, i} =P_{lk} \, P_{kj} \, K_{jk,l}^{\,\,\, i}  \\
    kl+j+i \quad \equiv K_{kl,j}^{\,\,\, i} =P_{lj} \, P_{kj} \, K_{jk,l}^{\,\,\, i}
    \end{array}
    \right.
       \\
    \end{array},
\nonumber\end{eqnarray}

\begin{eqnarray}
\begin{array}{ll}
ij+kl \rightarrow
   \left\{
    \begin{array}{ll}
     ij+k+l \quad \equiv H_{ij, kl} \\
     kl+i+j \quad \equiv H_{kl, ij} = P_{ik} P_{jl} \, H_{ij, kl}
    \end{array}
    \right.
       \\
ik+jl \rightarrow
   \left\{
    \begin{array}{ll}
     ik+j+l \quad \equiv H_{ik, jl} = P_{jk} \, H_{ij, kl} \\
     jl+i+k \quad \equiv H_{jl, ik} = P_{jk} P_{ik} P_{jl} \, H_{ij, kl}                                  \end{array}
    \right.
       \\
il+jk \rightarrow
   \left\{
    \begin{array}{ll}
     il+j+k \quad  \equiv H_{il, jk} =  P_{ij} P_{jk} P_{ik} P_{jl} \, H_{ij, kl}  \\
     jk+i+l \quad  \equiv H_{jk, il}  = P_{ij} P_{jk} \, H_{ij, kl}
    \end{array}
    \right.
\end{array}
\nonumber
\end{eqnarray}
In momentum space, we have
\begin{eqnarray}
\langle u |\Psi\rangle &=&
\langle u | 1 | K \rangle +  \langle u | P | K \rangle +
\langle u | P_{34} | K \rangle \nonumber\\*
&+&  \langle u | PP_{34} | K \rangle
+ \langle u | P_{34} P | K \rangle
+ \langle u | P P_{34} P | K \rangle
\nonumber \\* &+&
\langle u | 1 | H \rangle +
\langle u | P | H \rangle +
\langle u | \tilde{P} | H \rangle +
\langle u | P \tilde{P} | H \rangle
\nonumber \\*
&\equiv& \sum \left[ K_{\cal O} + H_{\cal O}\right];\;\;
{\cal O}=1, P, P_{34}, PP_{34}, ...
\label{eq.K-H-in-momentum}
\end{eqnarray}
where
\begin{eqnarray}
K_{\cal O} &\equiv &  \langle u | {\cal O} | K \rangle =
\int D^3u' \langle u | {\cal O} | u' \rangle \langle u' | K \rangle,
\label{eq.K-operator} \\*
&=&  \int D^3u' \langle u | {\cal O} | u' \rangle \,
G_{0} \bigl( u'_1, u'_2, u'_3\bigr) \,
\chi(u'_1)\, {\cal K}\bigl(u'_2, u'_3 \bigr),
\nonumber
\end{eqnarray}
and
\begin{eqnarray}
H_{\cal O} &\equiv & \langle u | {\cal O} | H \rangle =
\int D^3v' \langle u | {\cal O} | v' \rangle
\langle v' | H \rangle
\label{eq.H-operator} \\*
&=&  \int D^3v' \langle u | {\cal O} | v' \rangle
\, G_{0} \bigl( v'_1, v'_2, v'_3\bigr) \,
\chi(v'_1)\, {\cal H}\bigl(v'_2, v'_3 \bigr).
\nonumber
\end{eqnarray}
In the following we present the matrix elements of
$K_{\cal O}$ and $H_{\cal O}$ explicitly:
\begin{eqnarray}
K_1 &\equiv &  \langle u | 1 | K \rangle
\label{eq.K1}\\&=&
K  \bigl( u_1, u_2, u_3 \bigr) = G_{0} \bigl( u_1, u_2, u_3\bigr) \,
 {\cal K}\bigl(u_2, u_3 \bigr)\; ;
\nonumber
\end{eqnarray}

\begin{eqnarray}
K_P &\equiv &  \langle u | P | K \rangle =
\int dx_{12} \, K  \biggl( \Pi_P^{1} , \Pi_P^{2}, u_3 \biggr)
\label{eq.KP}\\ &=&  \int dx_{12} \,
G_{0} \biggl( \Pi_P^{1} , \Pi_P^{2}, u_3 \biggr) \,
 {\cal K} \biggl( \Pi_P^{2}, u_3 \biggr) \; ,
\nonumber
\end{eqnarray}
where
\begin{eqnarray}
\Pi_P^{1} &\equiv& \Pi_P^{1} \bigl(u_1, u_2,x_{12} \bigr) =
\biggl| -\frac{1}{2}\textbf{u}_1 -\frac{3}{4}\textbf{u}_2  \biggr|
\label{eq.Pi1P}\\&=&
\sqrt{\frac{1}{4}u_1^{2} + \frac{9}{16}u_2^{2}+ \frac{3}{4} u_1 u_2 x_{12}} \; ,
\nonumber \\*
\Pi_P^{2} &\equiv& \Pi_P^{2} \bigl(u_1, u_2,x_{12} \bigr) =  \biggl| \textbf{u}_1 -\frac{1}{2}\textbf{u}_2  \biggr|\label{eq.Pi2P}\\
&=&\sqrt{u_1^{2} + \frac{1}{4}u_2^{2}-  u_1 u_2 x_{12} }\; ;
\nonumber
\end{eqnarray}

\begin{eqnarray}
K_{P_{34}} &\equiv &  \langle u | P_{34} | K \rangle = \frac{1}{2} \int dx_{23} \,
K  \biggl( u_1, \Pi_{P_{34}}^{2}, \Pi_{P_{34}}^{3}  \biggr) \label{eq.KP34} \\*
&=&  \frac{1}{2} \int dx_{23} \,
G_{0}  \biggl( u_1, \Pi_{P_{34}}^{2}, \Pi_{P_{34}}^{3}  \biggr)
 {\cal K} \biggl(\Pi_{P_{34}}^{2}, \Pi_{P_{34}}^{3}  \biggr)\; ,
\nonumber
\end{eqnarray}
where
\begin{eqnarray}
\Pi_{P_{34}}^{2} &\equiv& \Pi_{P_{34}}^{2}\bigl(u_2, u_3,x_{23} \bigr)
=  \biggl| \frac{1}{3}\textbf{u}_2 +\frac{8}{9}\textbf{u}_3  \biggr|
\label{eq.Pi2P34}\\&=&
\sqrt{\frac{1}{9}u_2^{2} + \frac{64}{81}u_3^{2}+ \frac{16}{27} u_2 u_3 x_{23} },
\nonumber \\*
\Pi_{P_{34}}^{3} &\equiv& \Pi_{P_{34}}^{3} \bigl(u_2, u_3,x_{23} \bigr) =  \biggl| \textbf{u}_2 -\frac{1}{3}\textbf{u}_3  \biggr|
\label{eq.Pi3P34}\\&=&
\sqrt{u_2^{2} + \frac{1}{9}u_3^{2}- \frac{2}{3} u_2 u_3 x_{23} }\; ;
\nonumber
\end{eqnarray}

\begin{eqnarray}
K_{P P_{34}} &\equiv &  \langle u | P P_{34} | K \rangle
\label{eq.KPP34} \\
&=& \frac{1}{2} \int dx_{12}  \int dx_{12,3}\,
 K  \Biggl( \Pi_{PP_{34}}^{1}, \Pi_{PP_{34}}^{2}, \Pi_{PP_{34}}^{3}  \Biggr)
\nonumber \\*
&=&  \frac{1}{2} \int dx_{12}  \int dx_{12,3}\,
 G_{0} \Biggl( \Pi_{PP_{34}}^{1}, \Pi_{PP_{34}}^{2}, \Pi_{PP_{34}}^{3}
\Biggr)\nonumber\\&\times&
{\cal K} \Biggl( \Pi_{PP_{34}}^{2}, \Pi_{PP_{34}}^{3}  \Biggr)\; ,
\nonumber
\end{eqnarray}
where
\begin{eqnarray}
\Pi_{PP_{34}}^{1} &= &  \Pi_{P}^{1} \bigl(u_1, u_2,x_{12} \bigr),
\label{eq.PiPP34-1} \\*
\Pi_{PP_{34}}^{2} &= & \Pi_{P_{34}}^{2} \biggl(\Pi_{P}^{2} \bigl(u_1, u_2,x_{12}\bigr), u_3,x_{12,3} \biggr),
\label{eq.PiPP34-2} \\*
\Pi_{PP_{34}}^{3} &= & \Pi_{P_{34}}^{3} \biggl(\Pi_{P}^{2} \bigl(u_1, u_2,x_{12}\bigr), u_3,x_{12,3} \biggr)\; ;
\label{eq.PiPP34-3}
\end{eqnarray}

\begin{widetext}
\begin{eqnarray}
K_{P_{34}P} &\equiv &  \langle u | P_{34}P | K \rangle
= \frac{1}{2} \int dx_{23}  \int dx_{23,1}\,
 K  \Biggl( \Pi_{P_{34}P}^{1}, \Pi_{P_{34}P}^{2}, \Pi_{P_{34}P}^{3}\Biggr)
\nonumber \\*
&=&  \frac{1}{2} \int dx_{23}  \int dx_{23,1}\,
G_{0} \Biggl( \Pi_{P_{34}P}^{1}, \Pi_{P_{34}P}^{2}, \Pi_{P_{34}P}^{3}\Biggr)
{\cal K} \Biggl( \Pi_{P_{34}P}^{2}, \Pi_{P_{34}P}^{3}\Biggr),
\label{eq.KP34P}
\end{eqnarray}
where
\begin{eqnarray}
\Pi_{P_{34}P}^{1} &= &  \Pi_{P}^{1} \biggl(u_1, \Pi_{P_{34}}^{2} \bigl(u_2, u_3,x_{23}\bigr) ,x_{23,1} \biggr),
\label{eq.PiP34P-1} \\*
\Pi_{P_{34}P}^{2} &=&
\Pi_{P}^{2} \biggl(u_1, \Pi_{P_{34}}^{2} \bigl(u_2, u_3,x_{23}\bigr) ,x_{23,1}
\biggr), \;\;\;
\Pi_{P_{34}P}^{3} = \Pi_{P_{34}}^{3} \bigl(u_2, u_3,x_{23}\bigr) \; ;
\nonumber
\end{eqnarray}

\begin{eqnarray}
K_{P P_{34} P} &\equiv &  \langle u | P P_{34} P | K \rangle =
\frac{1}{2} \int dx_{12}  \int dx_{12,3}\, \int dx_{12;12,3}\,
 K  \biggl( \Pi_{P P_{34} P}^{1}, \Pi_{P P_{34} P}^{2}, \Pi_{P P_{34} P}^{3}  \biggr)
\label{eq.KPP34P} \\* &=&
 \frac{1}{2} \int dx_{12}  \int dx_{12,3}\, \int dx_{12;12,3}\,
G_{0}  \biggl( \Pi_{P P_{34} P}^{1}, \Pi_{P P_{34} P}^{2}, \Pi_{P P_{34} P}^{3}  \biggr)
 {\cal K} \biggl( \Pi_{P P_{34} P}^{2}, \Pi_{P P_{34} P}^{3}  \biggr)\; , \nonumber
\end{eqnarray}
where
\begin{eqnarray}
\Pi_{P P_{34} P}^{1} &= &  \Pi_{P}^{1} \Biggl(\Pi_{P}^{1} \bigl (u_1,u_2,x_{12}\bigr), \Pi_{P_{34}}^{2}
\biggl(\Pi_{P}^{2}\bigl (u_1,u_2,x_{12}\bigr), u_3,x_{12,3}\biggr) ,x_{12;12,3} \Biggr),
\label{eq.PiPPP1} \\*
\Pi_{P P_{34} P}^{2} &= & \Pi_{P}^{2} \Biggl(\Pi_{P}^{1} \bigl (u_1,u_2,x_{12}\bigr), \Pi_{P_{34}}^{2}
\biggl(\Pi_{P}^{2}\bigl (u_1,u_2,x_{12}\bigr), u_3,x_{12,3}\biggr) ,x_{12;12,3} \Biggr),
\label{eq.PiPPP2} \\*
\Pi_{P P_{34} P}^{3} &= & \Pi_{P_{34}}^{3} \biggl(\Pi_{P}^{2}\bigl (u_1,u_2,x_{12}\bigr), u_3,x_{12,3}\biggr)\; .
\label{eq.PiPPP3}
\end{eqnarray}
\end{widetext}
For the matrix elements of $H_{\cal O}$, we need also to
evaluate the transformation between the two different
representations, i.e., the $K-$ and $H-$type.
\begin{eqnarray}
H_1 &\equiv&  \langle u | 1 | H \rangle =   \frac{1}{2} \int dx_{23}\,
H \biggl(u_1, \Sigma_1^{2}, \Sigma_1^{3}  \biggr)
\label{eq.H1} \\* &=&   \frac{1}{2} \int dx_{23,1}\,
G_{0} \biggl(u_1, \Sigma_1^{2}, \Sigma_1^{3}  \biggr)
 {\cal H} \biggl(\Sigma_1^{2}, \Sigma_1^{3}  \biggr),
\nonumber
\end{eqnarray}
where
\begin{eqnarray}
\Sigma_1^{2} &\equiv& \Sigma_1^{2} \bigl(u_2, u_3,x_{23} \bigr) =
\biggl| \textbf{u}_2 +\frac{2}{3}\textbf{u}_3  \biggr|
\nonumber\\ &=&
\sqrt{u_2^{2} + \frac{4}{9}u_3^{2}+ \frac{4}{3} u_2 u_3 x_{23} },
\nonumber \\*
\Sigma_1^{3} &\equiv& \Sigma_1^{3} \bigl(u_2, u_3,x_{23} \bigr) =  \biggl| \frac{1}{2}\textbf{u}_2 -\frac{2}{3}\textbf{u}_3  \biggr|
\nonumber\\ &=&
\sqrt{ \frac{1}{4} u_2^{2} + \frac{4}{9}u_3^{2}- \frac{2}{3} u_2 u_3 x_{23} }\; ;
\label{eq.H1-arguments}
\end{eqnarray}

\begin{eqnarray}
H_P &\equiv &  \langle u | P | H \rangle \label{eq.HP}\\
&=&   \frac{1}{2} \int dx_{12} \int dx_{12,3} \,
H \biggl(\Sigma_P^{1}, \Sigma_P^{2}, \Sigma_P^{3}  \biggr)
\nonumber \\* &=&
\frac{1}{2} \int dx_{12} \int dx_{12,3} \,
G_{0} \biggl(\Sigma_P^{1}, \Sigma_P^{2}, \Sigma_P^{3}  \biggr)
{\cal H} \biggl( \Sigma_P^{2}, \Sigma_P^{3}  \biggr),
\nonumber
\end{eqnarray}
where
\begin{eqnarray}
\Sigma_P^{1} &=& \Pi_P^{1} \bigl(u_1, u_2,x_{12} \bigr),
\label{SigmaP1} \\*
\Sigma_P^{2} &=& \Sigma_1^{2} \biggl(\Pi_P^{2} \bigl(u_1, u_2,x_{12} \bigr), u_3,x_{12,3} \biggr),
\label{SigmaP2} \\*
\Sigma_P^{3} &=& \Sigma_1^{3} \biggl(\Pi_P^{2} \bigl(u_1, u_2,x_{12} \bigr), u_3,x_{12,3} \biggr) \; ;
\label{SigmaP3}
\end{eqnarray}

\begin{eqnarray}
H_{\tilde{P}} &\equiv &  \langle u | {\tilde{P}} | H \rangle =   \frac{1}{2} \int dx_{23}  \,
H \biggl(\Sigma_{\tilde{P}}^{1}, \Sigma_{\tilde{P}}^{2}, u_1  \biggr)
\label{eq.HPbar} \\* &=&
  \frac{1}{2} \int dx_{23}\,
G_{0} \biggl(\Sigma_{\tilde{P}}^{1}, \Sigma_{\tilde{P}}^{2}, u_1  \biggr)
 {\cal H} \biggl( \Sigma_{\tilde{P}}^{2}, u_1  \biggr),
\nonumber
\end{eqnarray}
where
\begin{eqnarray}
\Sigma_{\tilde{P}}^{1} &=& \Sigma_1^{3} \bigl(u_2, u_3,x_{23} \bigr),
\label{Sigma1Ptil} \\*
\Sigma_{\tilde{P}}^{2} &=& \Sigma_1^{2} \bigl(u_2, u_3,x_{23} \bigr)\; ;
\label{Sigma2Ptil}
\end{eqnarray}

\begin{eqnarray}
H_{P \tilde{P}} &\equiv &  \langle u | {P \tilde{P}} | H \rangle
\label{eq.HPPbar}\\&=&\frac{1}{2} \int dx_{12}  \int dx_{12,3}\,
H \biggl(\Sigma_{P \tilde{P}}^{1}, \Sigma_{P \tilde{P}}^{2}, \Sigma_{P \tilde{P}}^{3}  \biggr)
 \nonumber \\* &=&
  \frac{1}{2} \int dx_{12}  \int dx_{12,3}\,
G_{0} \biggl(\Sigma_{P \tilde{P}}^{1}, \Sigma_{P \tilde{P}}^{2}, \Sigma_{P \tilde{P}}^{3}  \biggr)
\nonumber\\&\times&
{\cal H} \biggl( \Sigma_{P \tilde{P}}^{2}, \Sigma_{P \tilde{P}}^{3}  \biggr),
\nonumber
\end{eqnarray}
where
\begin{eqnarray}
\Sigma_{P \tilde{P}}^{1} &=& \Sigma_1^{3} \biggl(\Pi_P^{2} \bigl(u_1, u_2,x_{12} \bigr), u_3,x_{12,3} \biggr),
\label{SigmaPP1} \\*
\Sigma_{P \tilde{P}}^{2} &=& \Sigma_1^{2} \biggl(\Pi_P^{2} \bigl(u_1, u_2,x_{12} \bigr), u_3,x_{12,3} \biggr),
\label{SigmaPP2} \\*
\Sigma_{P \tilde{P}}^{3} &=& \Pi_P^{1} \bigl(u_1, u_2,x_{12} \bigr) \; .
\label{SigmaPP3}
\end{eqnarray}

\section{On the numerical approach}
\label{appendix C}
The final set of homogeneous coupled integral equations in momentum
space, after discretization, defines a huge matrix eigenvalue
equation, which is solved by iteration within a Lanczos-like method.
This method is quite efficient in solving few-body systems. For
continuous momentum and angle variables discretization we have used
Gaussian-quadrature grid points with hyperbolic and linear mappings,
respectively.
After considering a convenient mesh distribution to have optimal
numerical stability, the number of grid points found necessary in
mapping the momentum variable was set up to 140.
That was achieved by studying the concentration of the points in the
different relevant regions of momentum integration. For more details
on the general numerical techniques that we are considering, see
also Ref.~\cite{hadi2007}.

Next, we also address problems concerning to the accuracy and precision
of the results in the unitary limit.

\subsection{Numerical convergence}
\label{appendix C1} The numerical convergence of our results is
exemplified with one case, where stability is more difficult to
achieve. As this happens in the excited states of large scale
ratios, we choose the case of $\mu_4/\mu_3=$ 300, in which we can
verify the existence of up to four tetramers. The number of points
and, particularly, their distribution, is quite critical for the
accuracy of the excited-state energies and corresponding FY reduced
amplitudes. In our systematic study to construct the limit cycle
shown in Fig.~\ref{scaling-section}, we are satisfied to achieve
$\sim $1\% of deviation from the converged results. To be clear, we
choose to present in Fig.~\ref{convergence} the convergence in
respect to the mesh number of points, for the first and second
excited states.

As the corresponding FY components become quite concentrated at the
momentum origin, in this case the momentum space discretization
$u_i$ is derived from the quadrature Gaussian mesh distribution
$x_i\equiv\left\{-1,+1\right\}$ by the following hyperbolic mapping:
\begin{eqnarray}
\displaystyle{u_i = \frac{1+x_i}{c_1(1-x_i)+c_2 x_i}},
\label{eq.mapping}
\end{eqnarray}
with $c_1\equiv \mu_4/\mu_3$ and $c_2=0.4$.
In general, we observe that for all the cases with smaller scaling
ratios, or for excited states that are not too close to the
continuum threshold, i.e., about to be unbound, 140 points are
enough to achieve less than 1\% of inaccuracy, within an appropriate
mesh distribution.

\subsection{Iterative Lanczos-type diagonalization
algorithm for solution of the coupled FY equations}
\label{sec_Lanczos}

The coupled FY integral equations (\ref{eq.KH-integral-final})
can be schematically represented as an eigenvalue problem:
\begin{eqnarray}
\Bbbk(E).\psi &=& \lambda(E).\psi
\label{eq.eigenvalue}
\end{eqnarray}
where the kernel of integral equations $\Bbbk(E)$ is energy
dependent,  $\lambda(E)$ and $\psi$ are its eigenvalue and
eigenvector, correspondingly. The vector $\psi$ is composed of FY
components as $\psi=\left(^{{\cal K}}_{{\cal H}}\right)$. The binding
energy of four-body bound state can be obtained when one of the eigenvalues
obtained from the solution of Eq.
(\ref{eq.eigenvalue}) for an input energy becomes one ($\lambda(E)=1$). To this aim one
should solve the eigenvalue equation (\ref{eq.eigenvalue}) for a set of input energies.

After discretization of continuous momentum and angle variables, the kernel of the
eigenvalue equation turns into a huge matrix with dimensionality of
$120\times120\times120\times40\times40\sim 10^{9}$, where we have used 120 mesh points for Jacobi momentum variables (with a hyperbolic mapping) and 40 mesh points for angle variables (with a linear mapping).
Considering that, exact solutions of huge matrices by diagonalization are not
so efficient, we avoid such procedure by using a Lanczos-type technique, which
is based on iteration. In this way, one can obtain the eigenvalues and eigenvectors
of a huge matrix, from input energies~\cite{StadlerPRC44}. In the following,
we describe some details of this technique.

The iteration procedure is performed with a properly starting vector
$\psi_{0}=\left(^{{\cal K}_{0}}_{{\cal H}_{0}}\right)$, chosen as Gaussian functions
for both ${\cal K}$ and ${\cal H}$ components. After $N$ iterations,
one obtains the set of vectors $\{\psi_{i}\} =
{\psi_{1}, \psi_{2}, ..., \psi_{N}}$
where
\begin{eqnarray}
\psi_{i} = \Bbbk(E).\psi_{i-1} \quad; i=1,2,...,N
\label{eq.eigenvalue3}
\end{eqnarray}
In the iterative diagonalization approach, an orthonormal  basis
$\{\overline{\psi}_{i}\}$ can be built up by following a
recursive procedure and by using the original vector set $\{\psi_{i}\}$,
where
\begin{eqnarray}
\overline{\psi}_{i} = c_i \biggl \{ \psi_{i} - \sum_{j=1}^{i-1}
\Bigl (  \overline{\psi}_{j} . \psi_i \Bigr ) \overline{\psi}_{j}
\biggr \}  \;\; (i=1,2,...,N) .
\label{eq.orthogonalization}
\end{eqnarray}
In the above, $c_i$ are normalization factors which can be obtained by orthogonalization
of orthonormal basis states $\overline{\psi}_{i}. \overline{\psi}_{j} = \delta_{ij}$.
By introducing a matrix $D$, obtained from the FY components as
\begin{eqnarray}
D_{ij} = \psi_i.\psi_j &=& \Bigl( {\cal K}_i  \,\, {\cal H}_i \Bigr).  \Bigl(^{{\cal K}_j}_{{\cal H}_j}\Bigr) \nonumber \\&=&
\label{eq.D-matrix}
{\cal K}_i.{\cal K}_j + {\cal H}_i.{\cal H}_j = D_{ij}^K + D_{ij}^H,
\end{eqnarray}
where
\begin{eqnarray}
D_{ij}^K &=&  {\cal K}_i.{\cal K}_j  = \int du_2 \, u_2^2 \int du_3 \, u_3^2 \,\, {\cal K}_i(u_2,u_3) \, {\cal K}_j(u_2,u_3),
\nonumber \\*
D_{ij}^H &=& {\cal H}_i.{\cal H}_j =  \int dv_2 \, v_2^2 \int dv_3 \, v_3^2 \,\, {\cal H}_i(v_2,v_3) \, {\cal H}_j(v_2,v_3), \nonumber \\*
\label{eq.multiplication}
\end{eqnarray}
the normalization factors $c_i$ can be obtained as:
\begin{eqnarray}
c_{i} = \frac{1}{\sqrt{D_{ii}-\sum_{j=1}^{i-1} N_{ij}^2}}.
\label{eq.c}
\end{eqnarray}
The matrix elements $N_{ij}$ are projection of vectors $\psi_i$ on  $\overline{\psi}_{j}$. i.e. $N_{ij}=\psi_i.\overline{\psi}_{j}$,
\begin{eqnarray}
N_{ij} = c_j \biggl \{   D_{ij} - \sum_{k=1} ^{j-1} N_{ik}.N_{jk} \biggr \}.
\label{eq.N}
\end{eqnarray}

By expanding the original vector $\psi$, in Eq. (\ref{eq.eigenvalue}),
in terms of constructed orthonormal vectors:
\begin{eqnarray}
\psi=\sum_{i=1}^{N-1} g_i \, \overline{\psi}_{i}
\label{eq.expanding}
\end{eqnarray}
and projecting the eigenvalue equation onto $\overline{\psi}_{j}$, the huge dimension
of the eigenvalue problem will be reduced to a problem of much smaller dimension $N-1$:
\begin{eqnarray}
\sum_{i=1}^{N-1}  M_{ij} \,g_i = \lambda  \, g_j
\label{eq.eigenvaue-revised}
\end{eqnarray}
where
\begin{eqnarray}
M_{ij}  \equiv   \overline{\psi}_{i}. \Bbbk .  \overline{\psi}_{j}
\label{eq.M}
\end{eqnarray}
By defining $a_{ij}$ and $b_{ij}$ as coefficients of linear expansion of original vectors in terms of orthonormal vectors and reversely,
\begin{eqnarray}
\psi_i &=& \sum_{j=1}^{i} a_{ij}  \, \overline{\psi}_{j} ,
\;\;\;\; \overline{\psi}_{i} = \sum_{j=1}^{i} b_{ij}  \, \psi_j
,\label{eq.a-b}
\end{eqnarray}
the matrix elements of $M$  can be obtained as:
\begin{eqnarray}
M_{ij}  = \sum_{k=1}^{i}  b_{ik} \, a_{k+1j}   \quad 1 \leq i,j<N
\label{eq.M2}
\end{eqnarray}
where
\begin{eqnarray}
a_{ij}  &=& \left\{  \begin{array}{lll}
    N_{ij} & &  j<i\\
    \dfrac{1}{c_i}   & &  j=i
  \end{array}
  \right. \\
\label{eq.a}
\end{eqnarray}
\begin{eqnarray}
b_{ij} &=& \left\{ \begin{array}{lll}
    c_j \bigl(  -\sum_{k=j+1}^{i}  b_{ik} \, a_{kj} \bigr) & &  j<i\\
    \dfrac{1}{c_i}   & &  j=i
  \end{array}
    \right.
\label{eq.b}
\end{eqnarray}

The reduced eigenvalue problem, Eq.~(\ref{eq.M2}), can be solved easily by
diagonalization of the matrix $M$.
For a given input energy, one can obtain a set of eigenvalues
and their corresponding eigenvectors. In order to obtain the physical binding
energy, one should look for an eigenvalue $\lambda=1$ in the obtained eigenvalue
spectrum. Clearly, the eigenvector corresponding to this eigenvalue gives
the coefficients $g_i$. Therefore, by using the Eq.~(\ref{eq.expanding}), one
can obtain the physical eigenvector or the FY components.

\begin{widetext}

%FIG8
\begin{figure}[ht]
\includegraphics[width=3.2in]{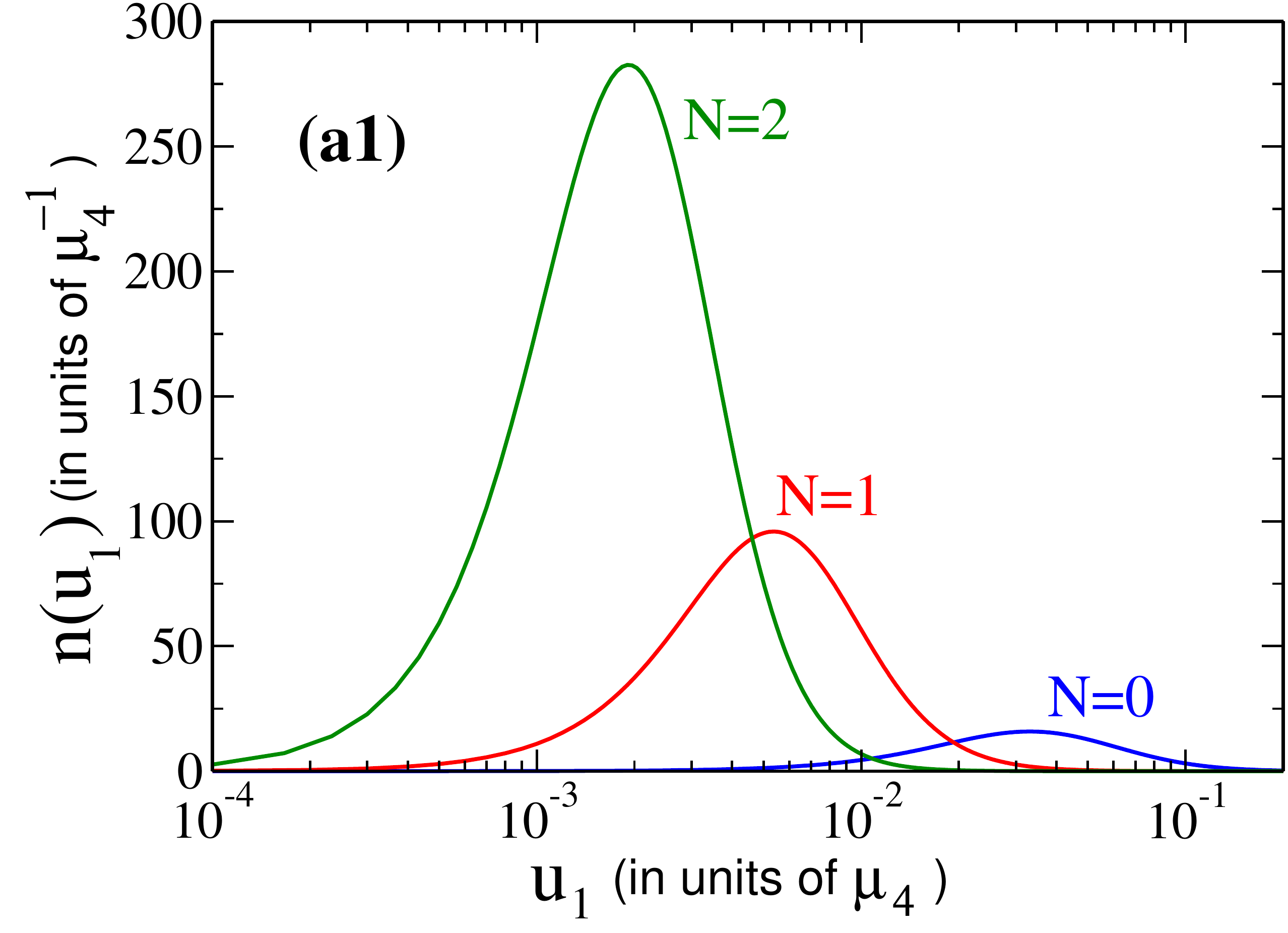}
\includegraphics[width=3.2in]{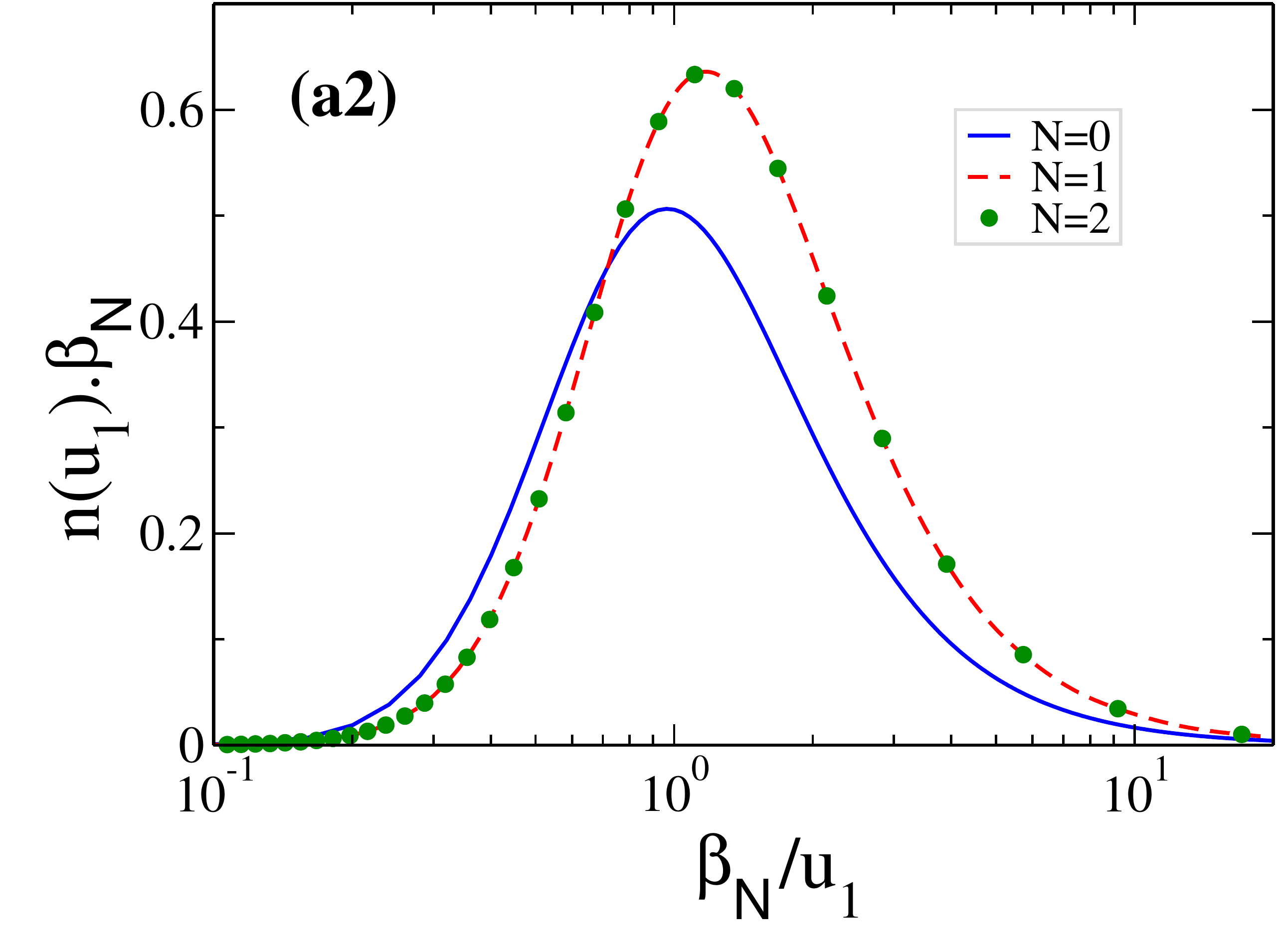} \\
\vspace{.3cm}
\includegraphics[width=3.2in]{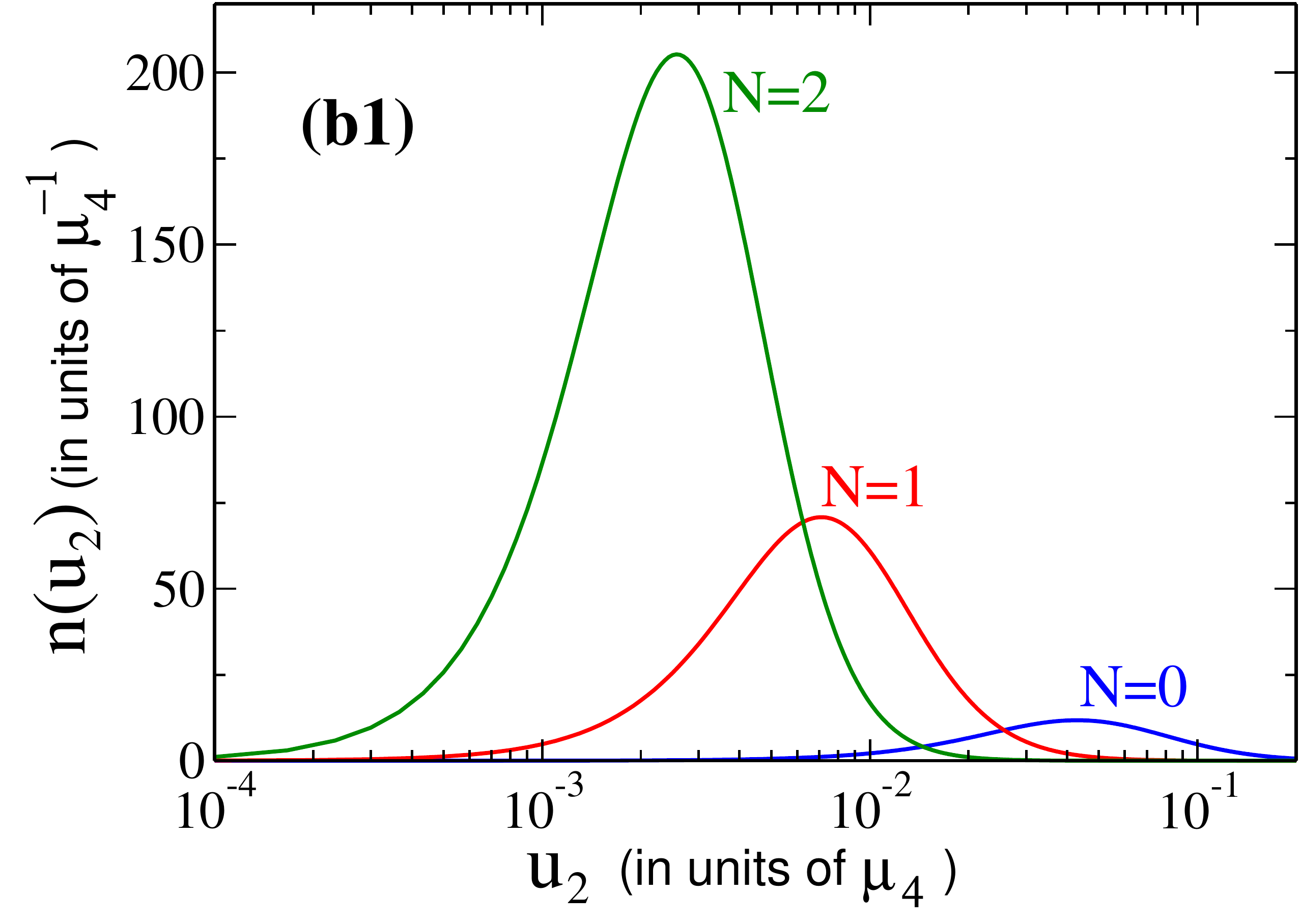}
\includegraphics[width=3.2in]{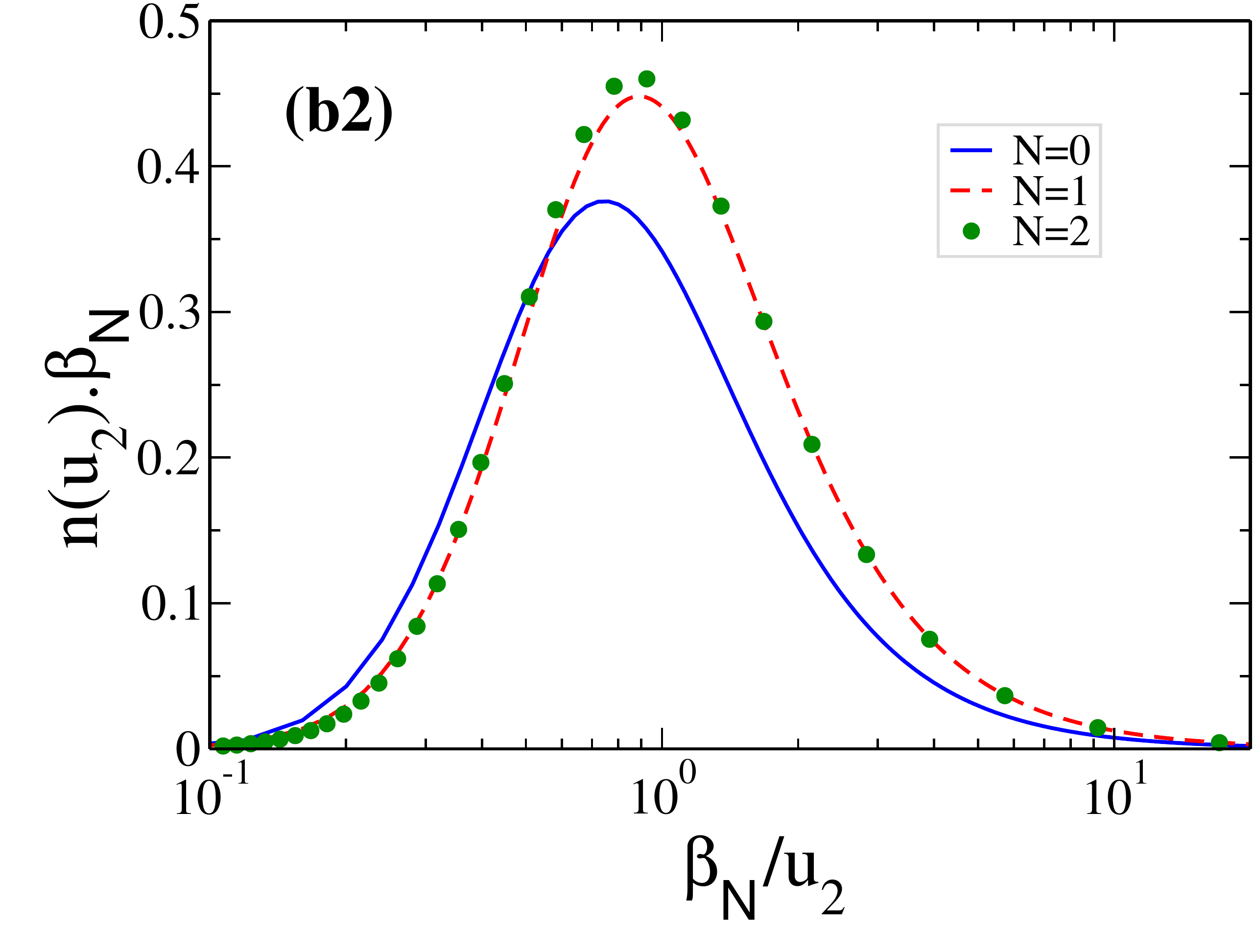}  \\
\vspace{.3cm}
\includegraphics[width=3.2in]{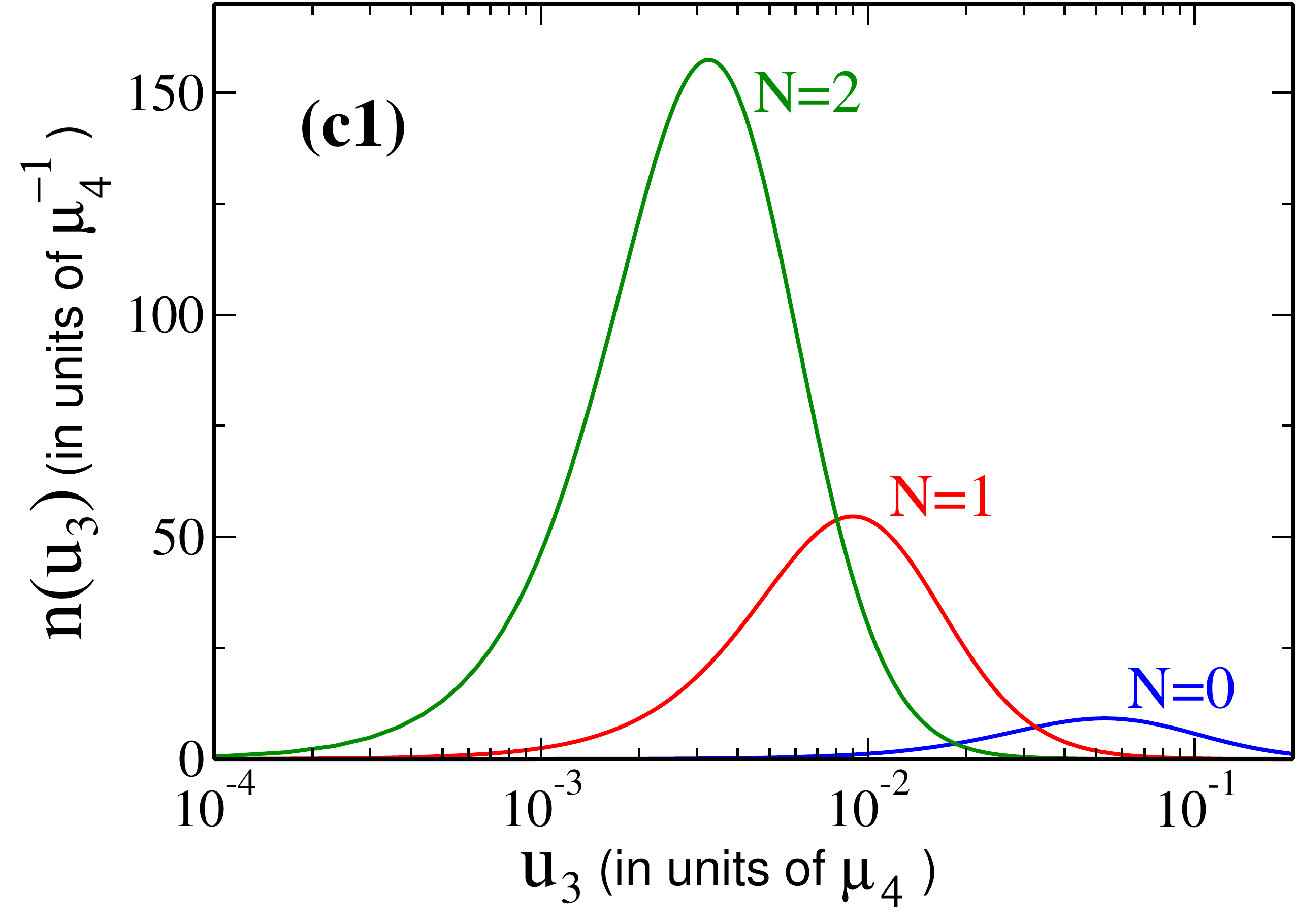}
\includegraphics[width=3.2in]{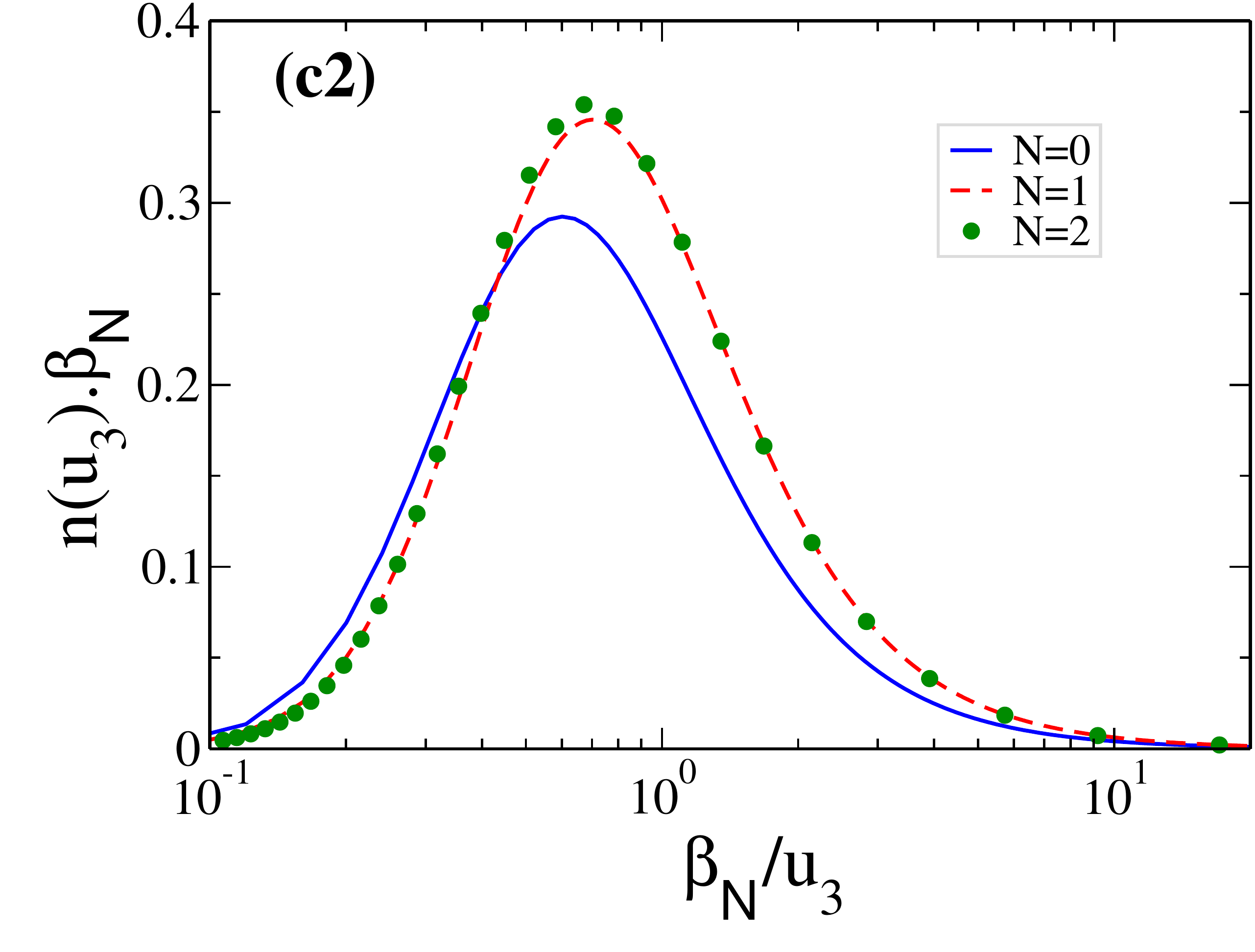}
\vspace{-.3cm}
\caption{(Color on-line) Momentum distribution functions $n(u_1)$, $n(u_2)$ and $n(u_3)$,
with scale ratio ${\mu_4}/{\mu_3}=50$, for the ground ($N=0$) and the
first two ($N=1, 2$) excited tetramer levels, normalized as shown in
Eq.~\ref{eq.nui}.
In the left panels, they are given as functions of the corresponding momentum,
where all momenta are in units of the momentum scale $\mu_4$.
In the right panels, the momentum distributions are rescaled by the momentum factor
$\beta_N\equiv\sqrt{B_4^{(N)}}$, where $N$ is the tetramer level.
} \label{fig:nui_50}
\vspace{-0.3cm}
\end{figure}

%FIG9
\begin{figure}[ht]
\includegraphics[width=7in]{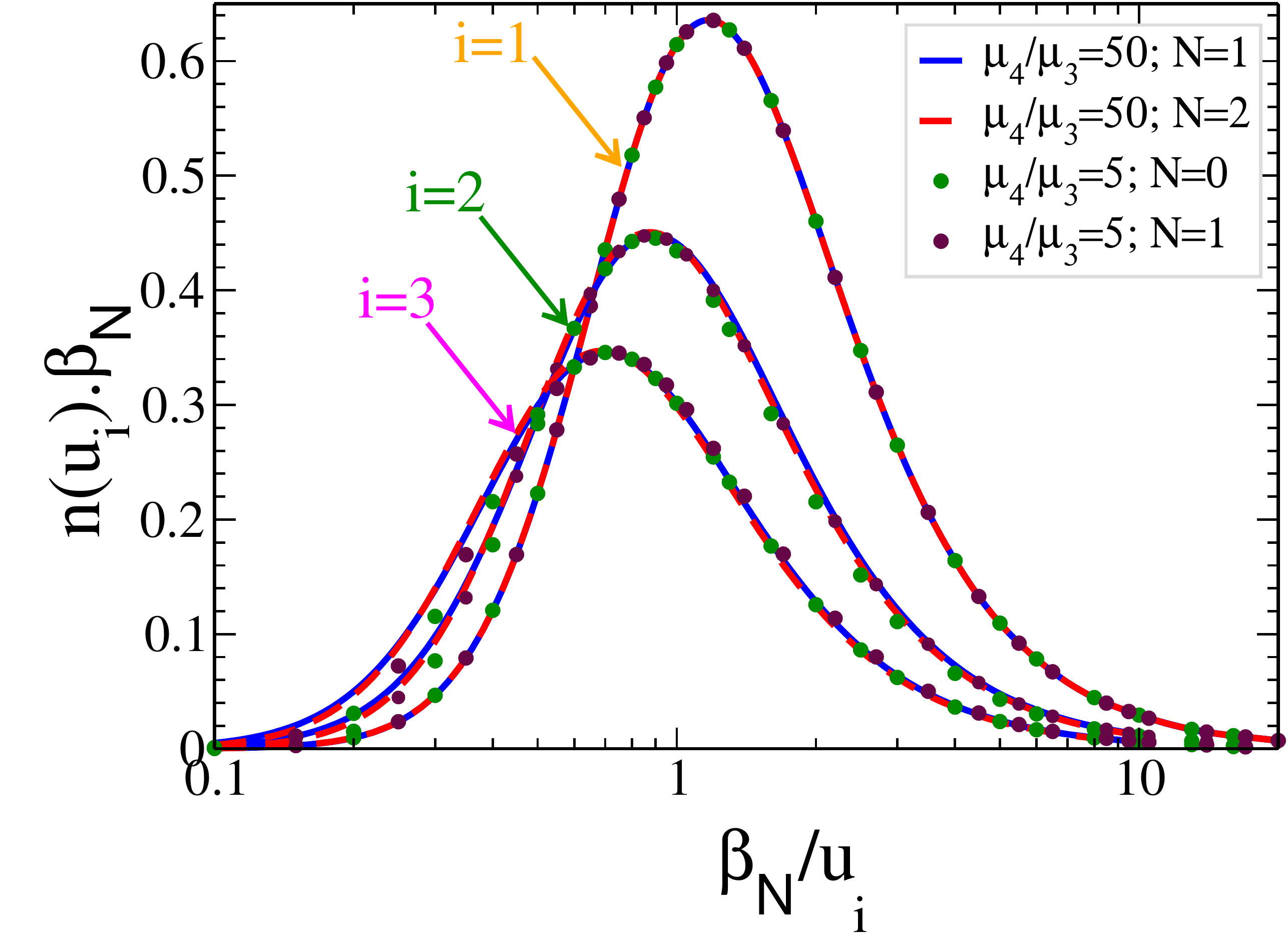}
\caption{(Color on-line) Renormalized momentum distributions as functions of the Jacobi
momenta $u_{i=1,2,3}$, considering the scaling ratios ${\mu_4}/{\mu_3}=5$, when there is
only one excited state, and ${\mu_4}/{\mu_3}=50$, when we have two excited states.
The units are given by the momentum factor $\beta_N\equiv\sqrt{B_4^{(N)}}$.
For each momentum component $u_i$, the corresponding distributions of both scale ratios, when
renormalized to the same maximum value, are shown to reduce almost to the same
final form (the labels inside the frame are just to indicate the performed
calculations).}
\label{fig:nui_all}
\end{figure}

%FIG10
\begin{figure}[ht]
\begin{center}
\includegraphics[width=3.2in]{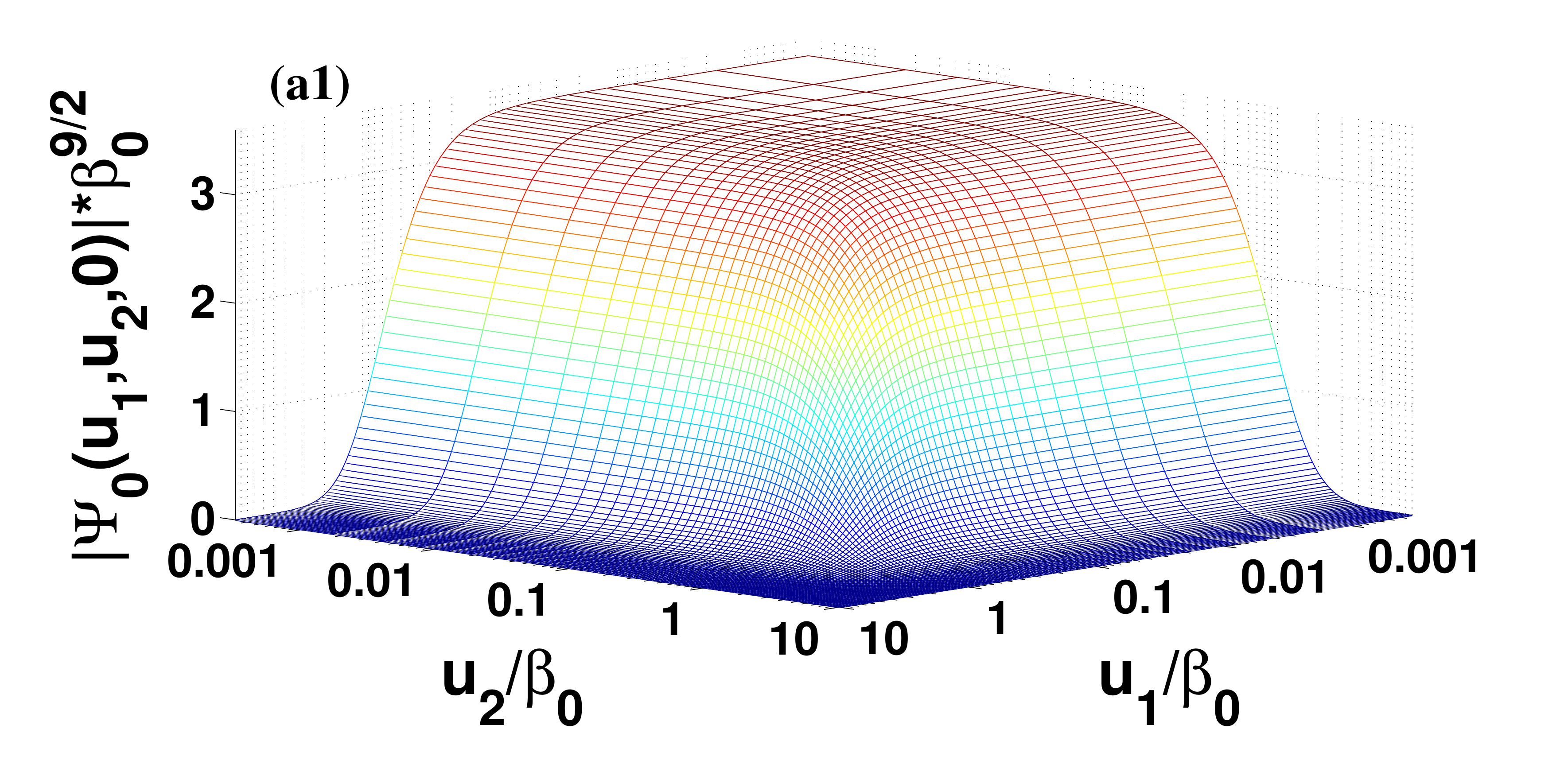}
\includegraphics[width=3.2in]{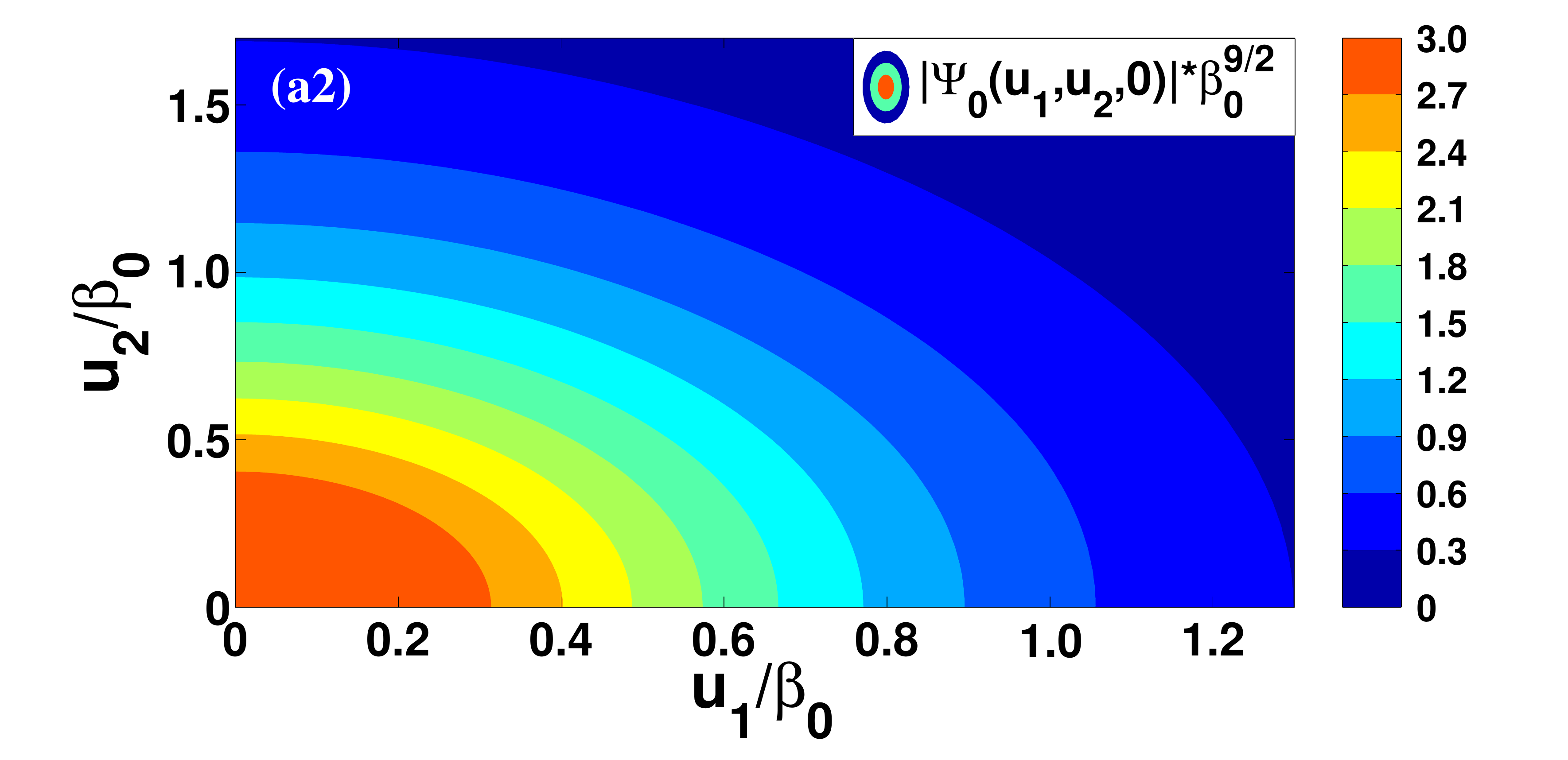}\\
\includegraphics[width=3.2in]{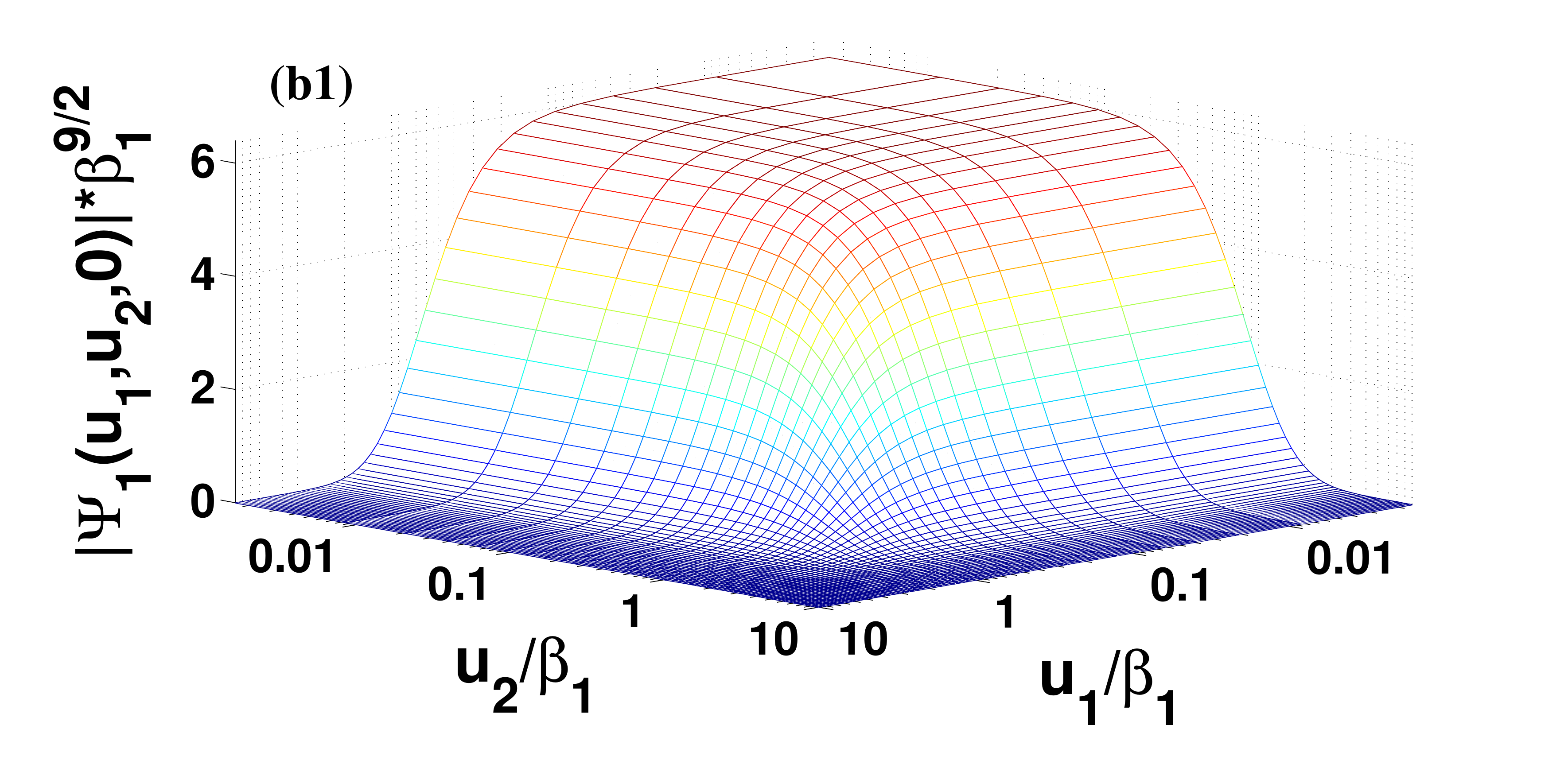}
\includegraphics[width=3.2in]{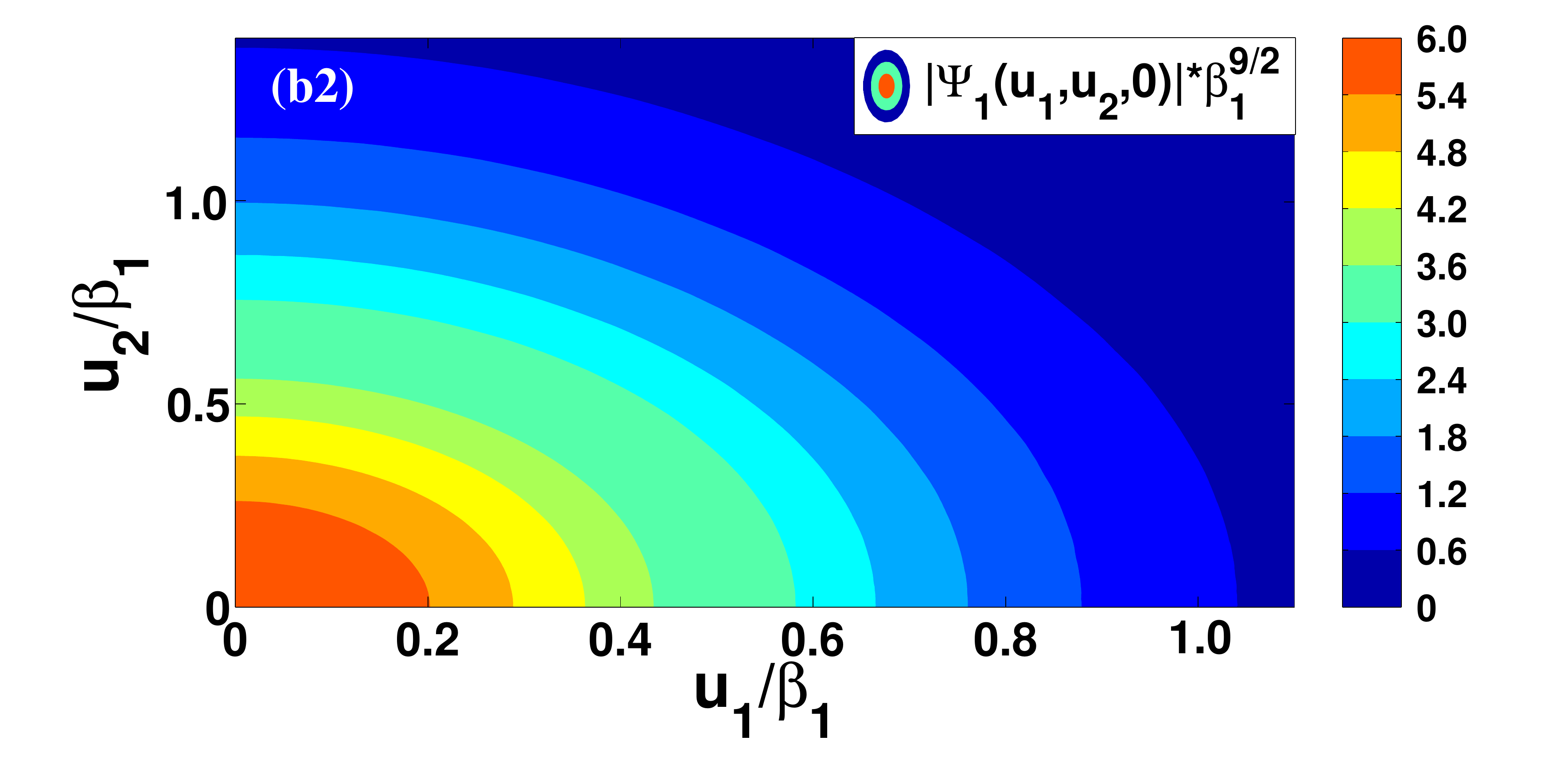}\\
\includegraphics[width=3.2in]{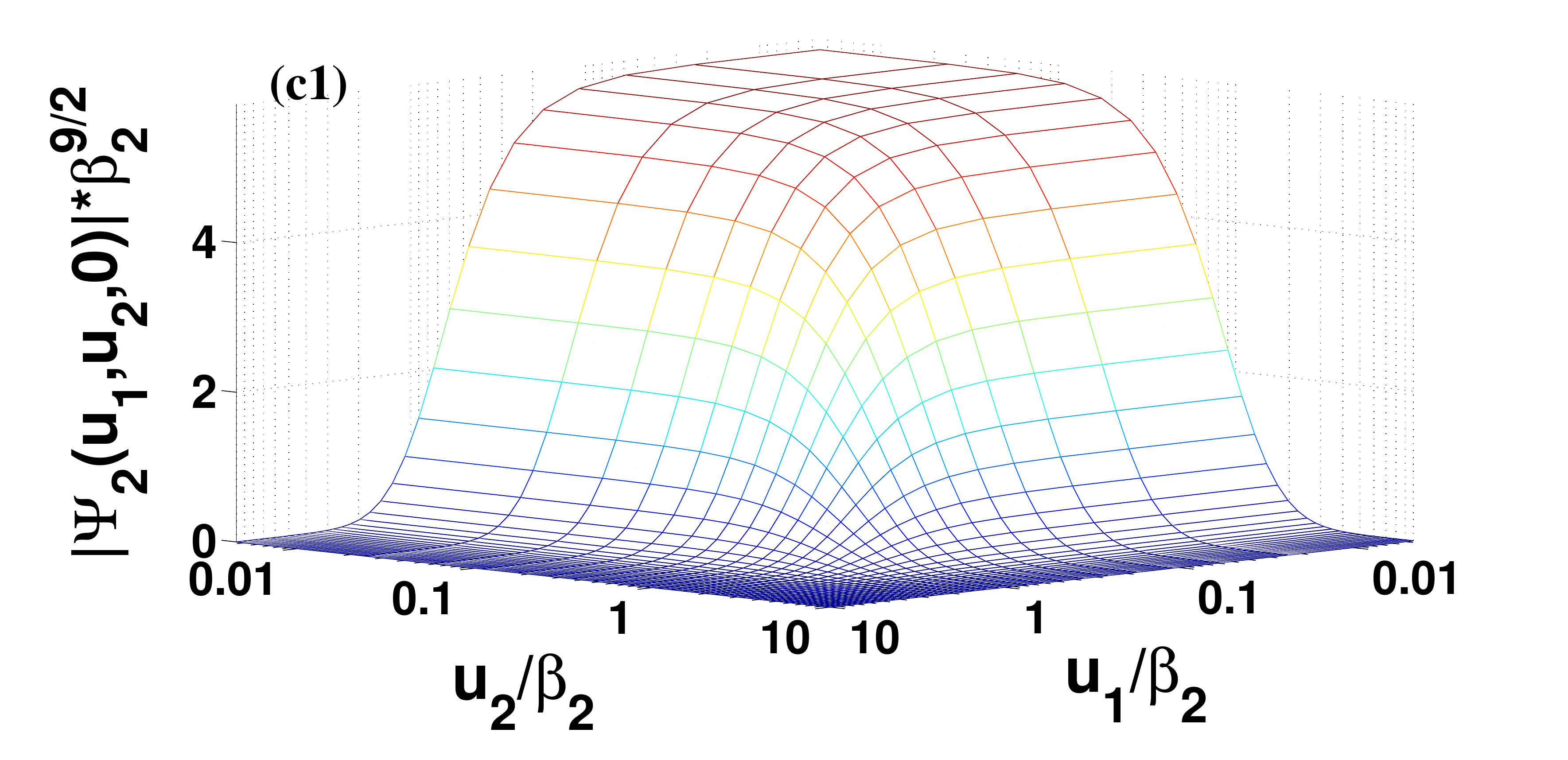}
\includegraphics[width=3.2in]{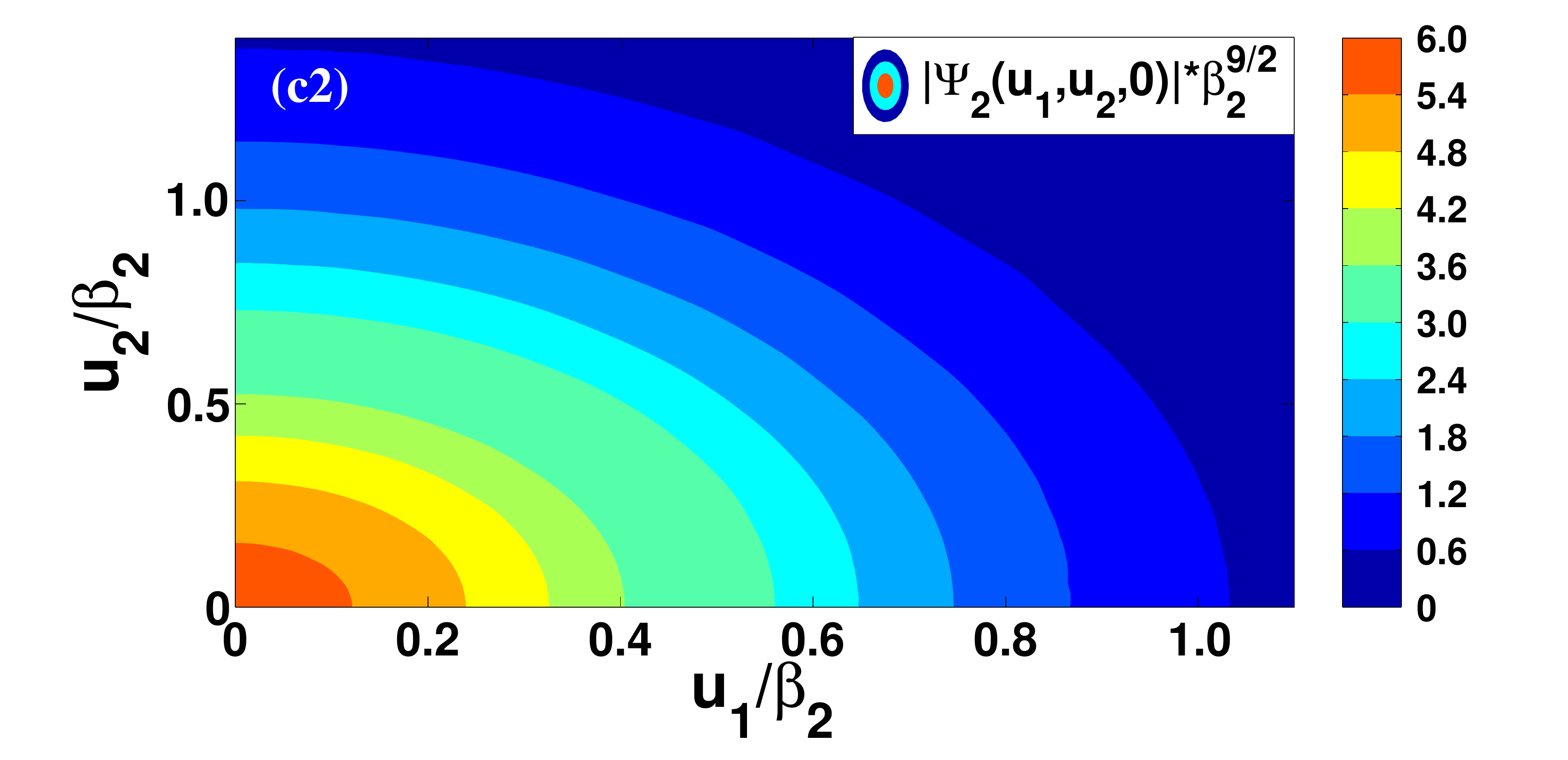}
\end{center}
\vspace{-0.5cm}
\caption{(Color on-line) The magnitude of the 4B total wave function $\Psi(u_1, u_2,
u_3)$ as a function of $u_1$ and $u_2$, with $u_3=0$, for the scale ratio
${\mu_4}/{\mu_3}=50$. The corresponding contour plots are shown for
low momentum region, near the peak of the wave function.}
 \label{Fig:WF_u1u2_50}
 \vspace{-0.5cm}
\end{figure}

%FIG11
\begin{figure}[ht]
\begin{center}
\includegraphics[width=3.2in]{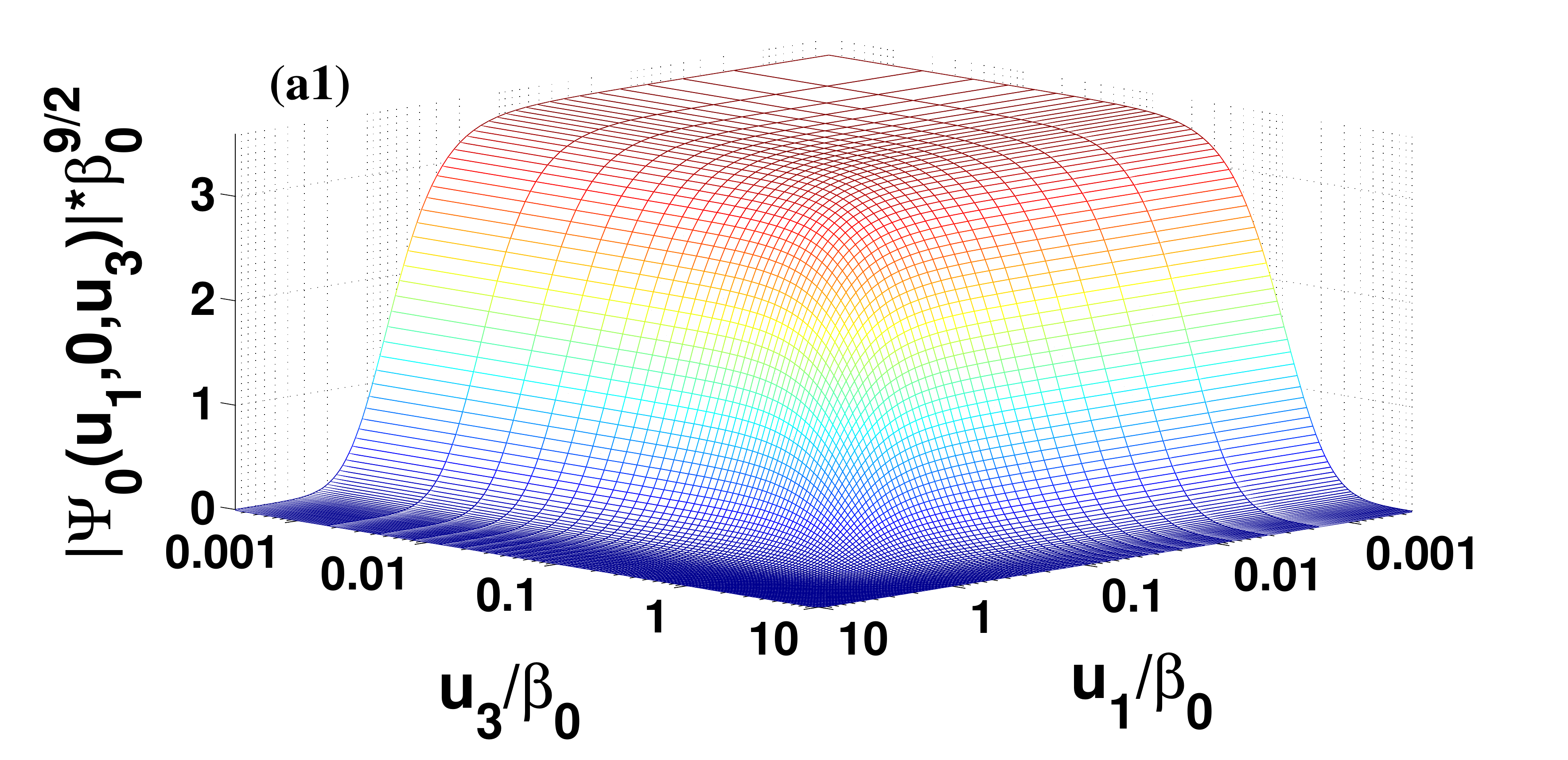}
\includegraphics[width=3.2in]{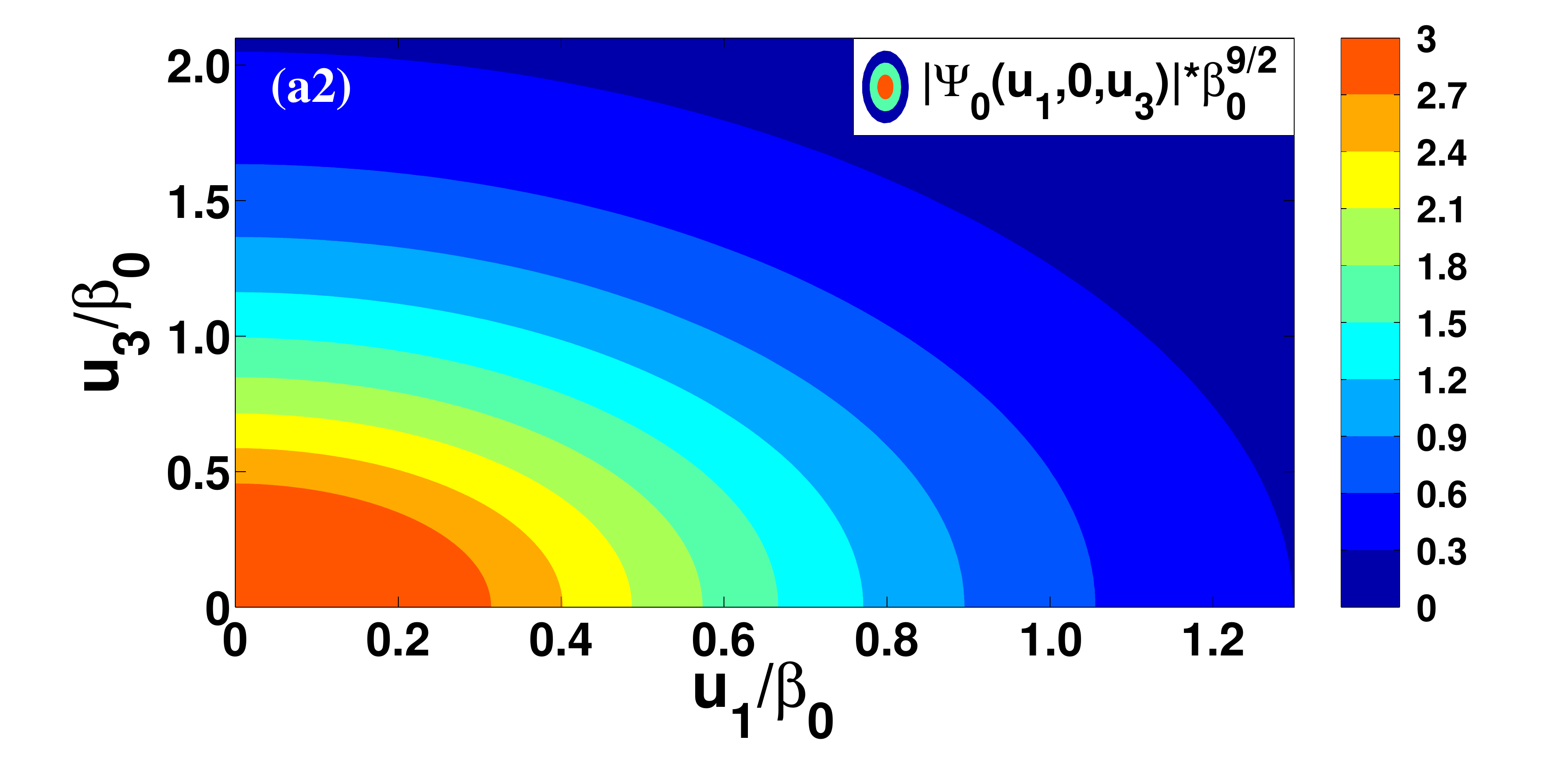} \\
\includegraphics[width=3.2in]{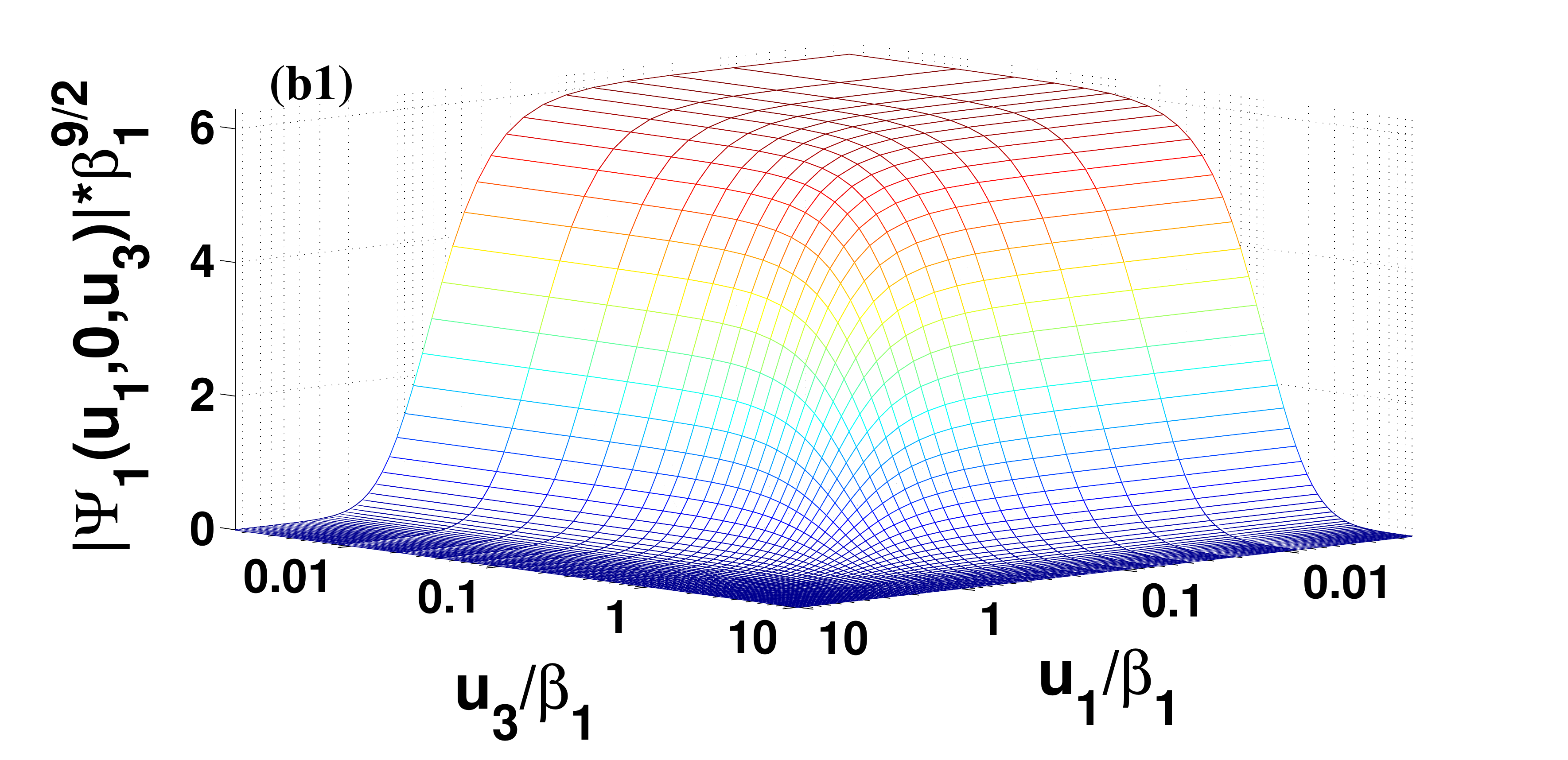}
\includegraphics[width=3.2in]{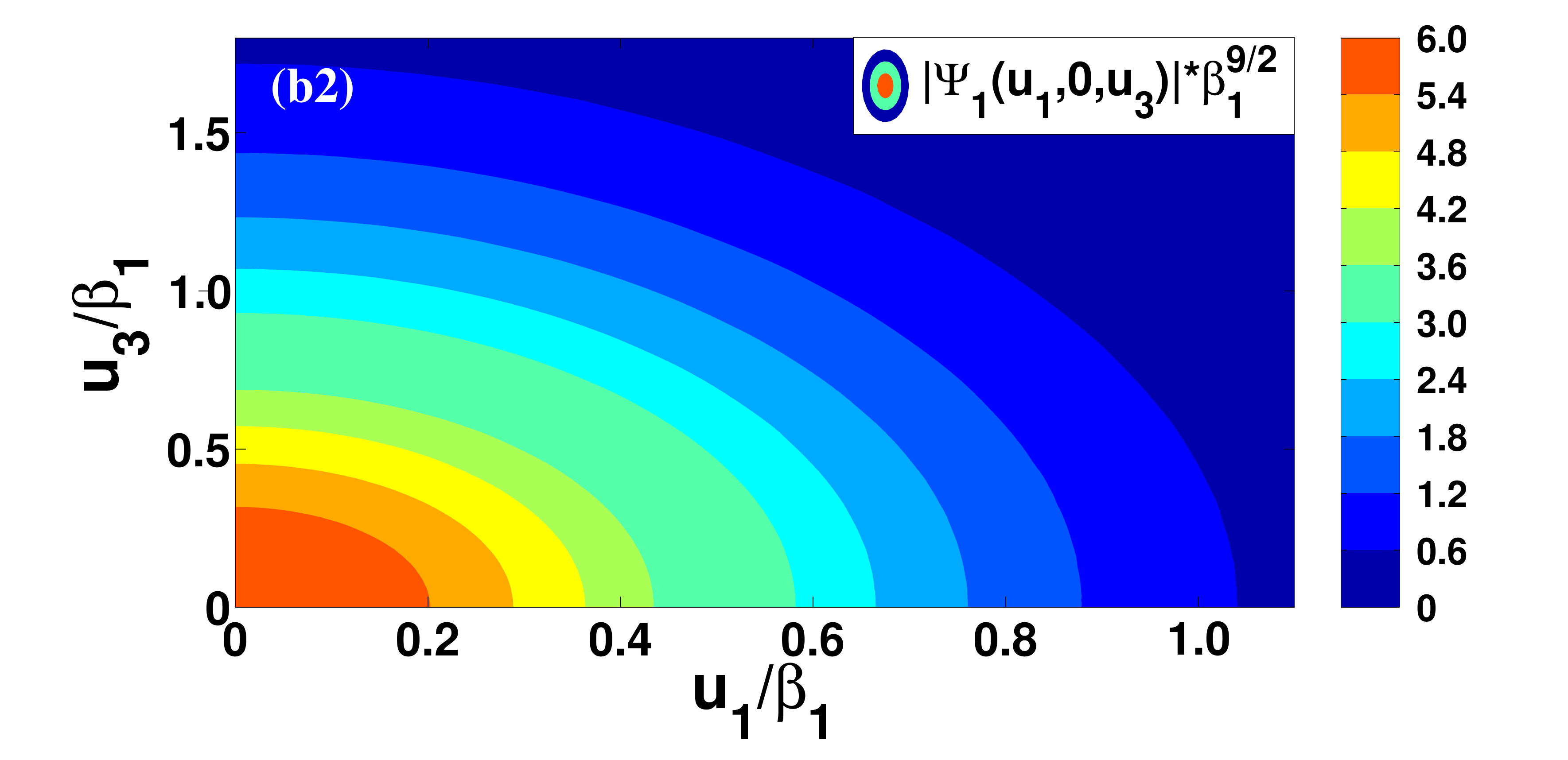} \\
\includegraphics[width=3.2in]{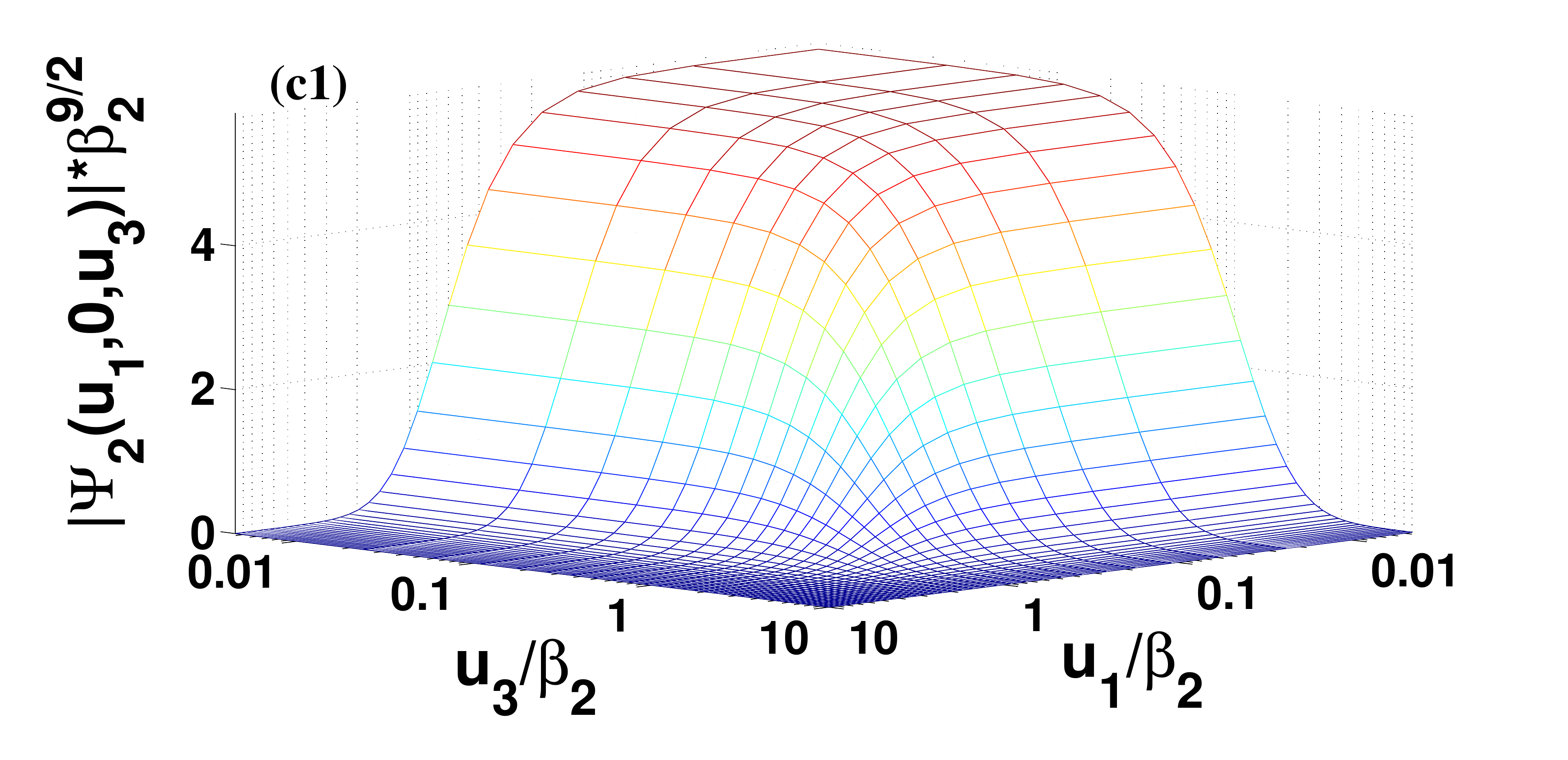}
\includegraphics[width=3.2in]{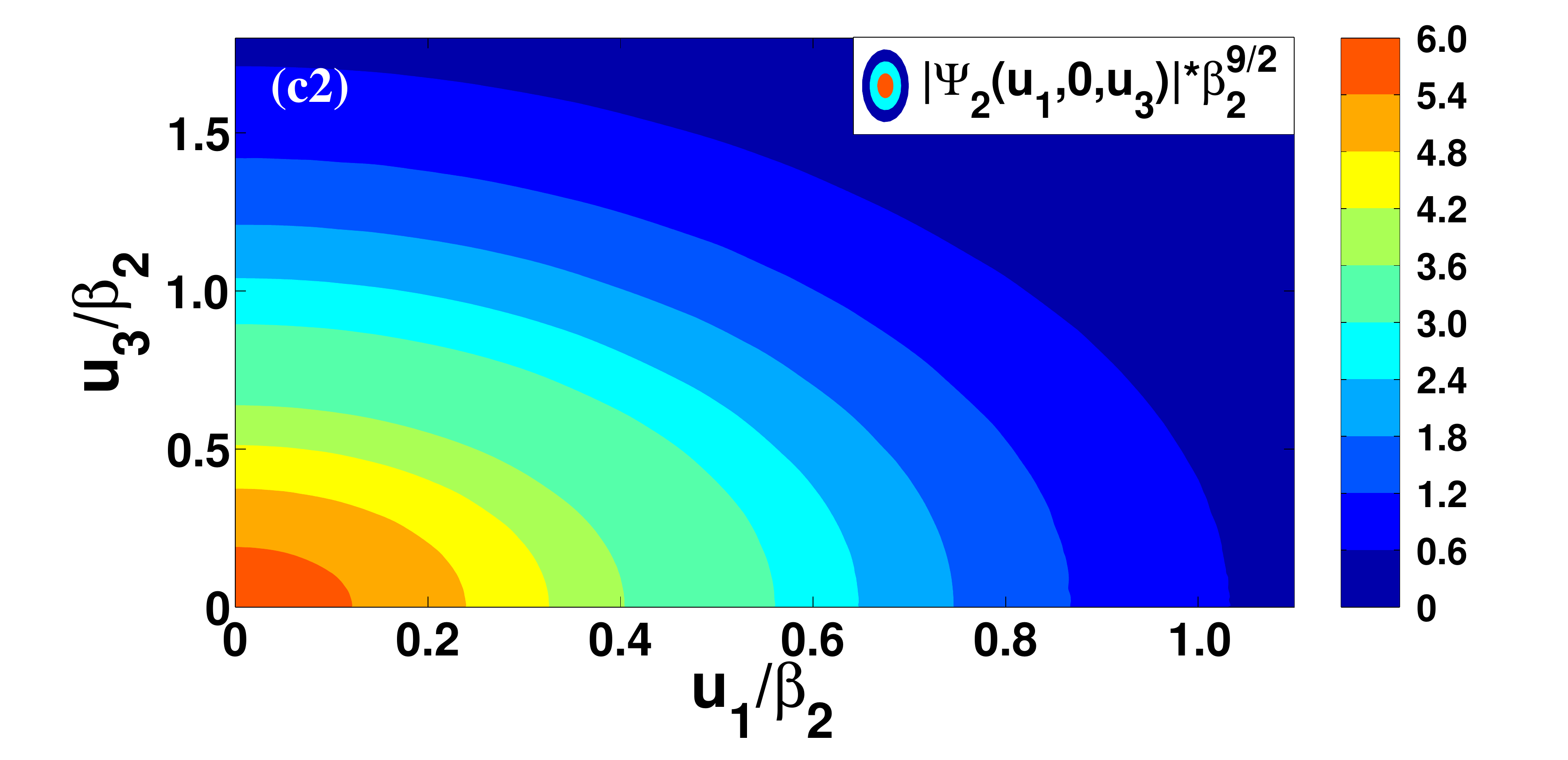}
\end{center}
\caption{(Color on-line) As in Fig.~\ref{Fig:WF_u1u2_50}, the wave-function
in given in terms of $u_1$ and $u_3$, with $u_2=0$.}
 \label{Fig:WF_u1u3_50}
\end{figure}

%FIG12
\begin{figure}[ht]
\begin{center}
\includegraphics[width=3.2in]{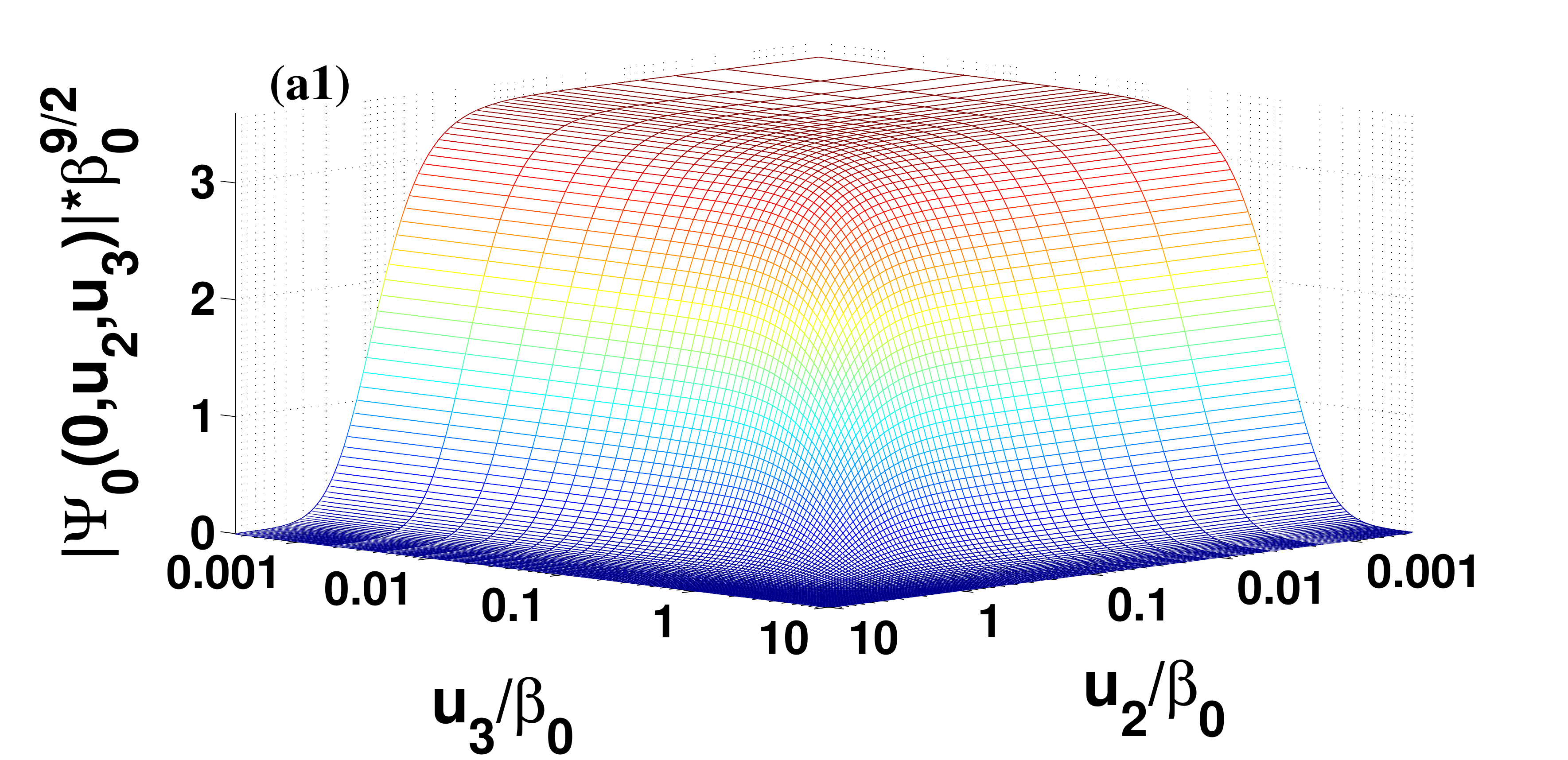}
\includegraphics[width=3.2in]{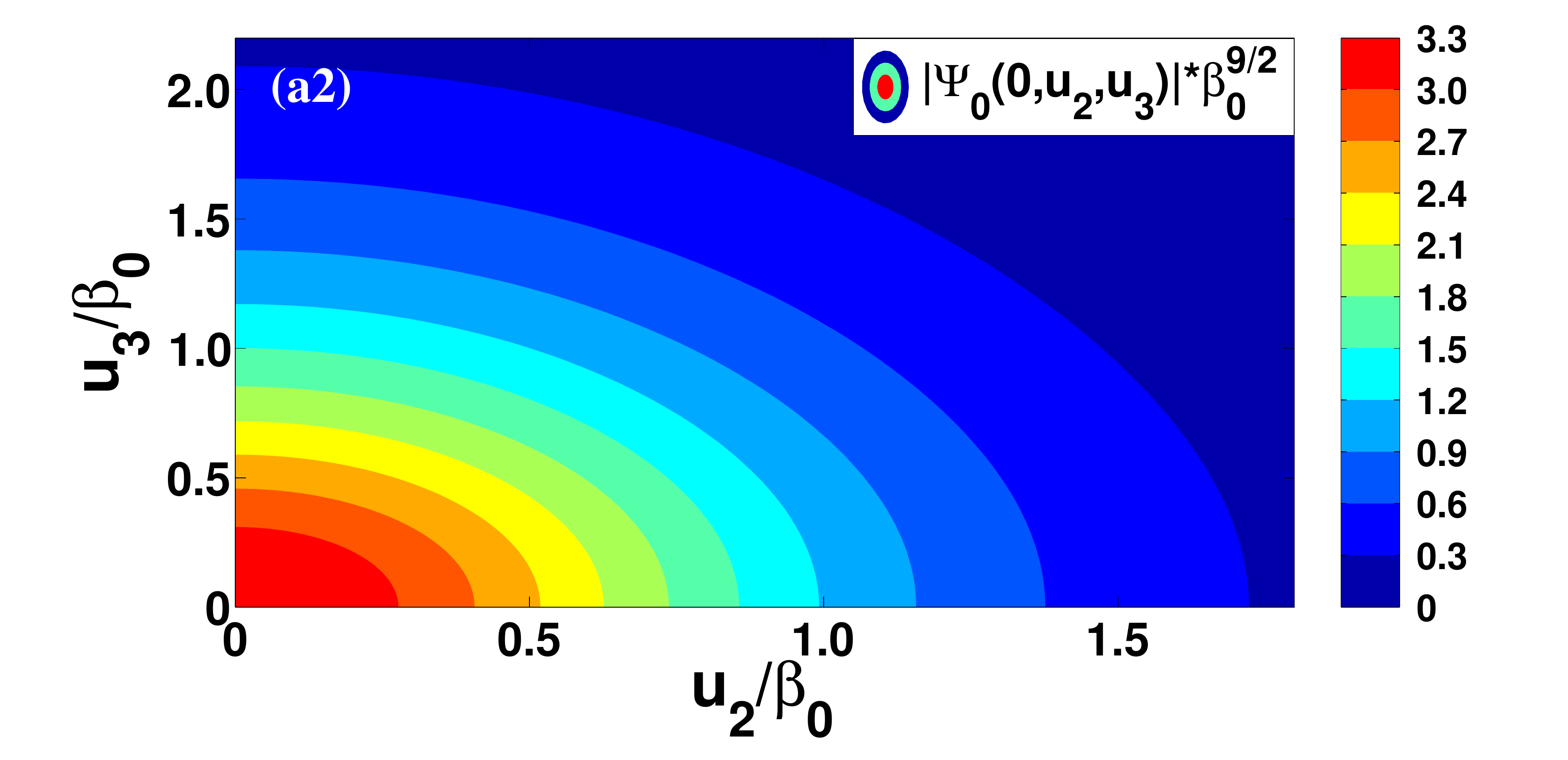} \\
\includegraphics[width=3.2in]{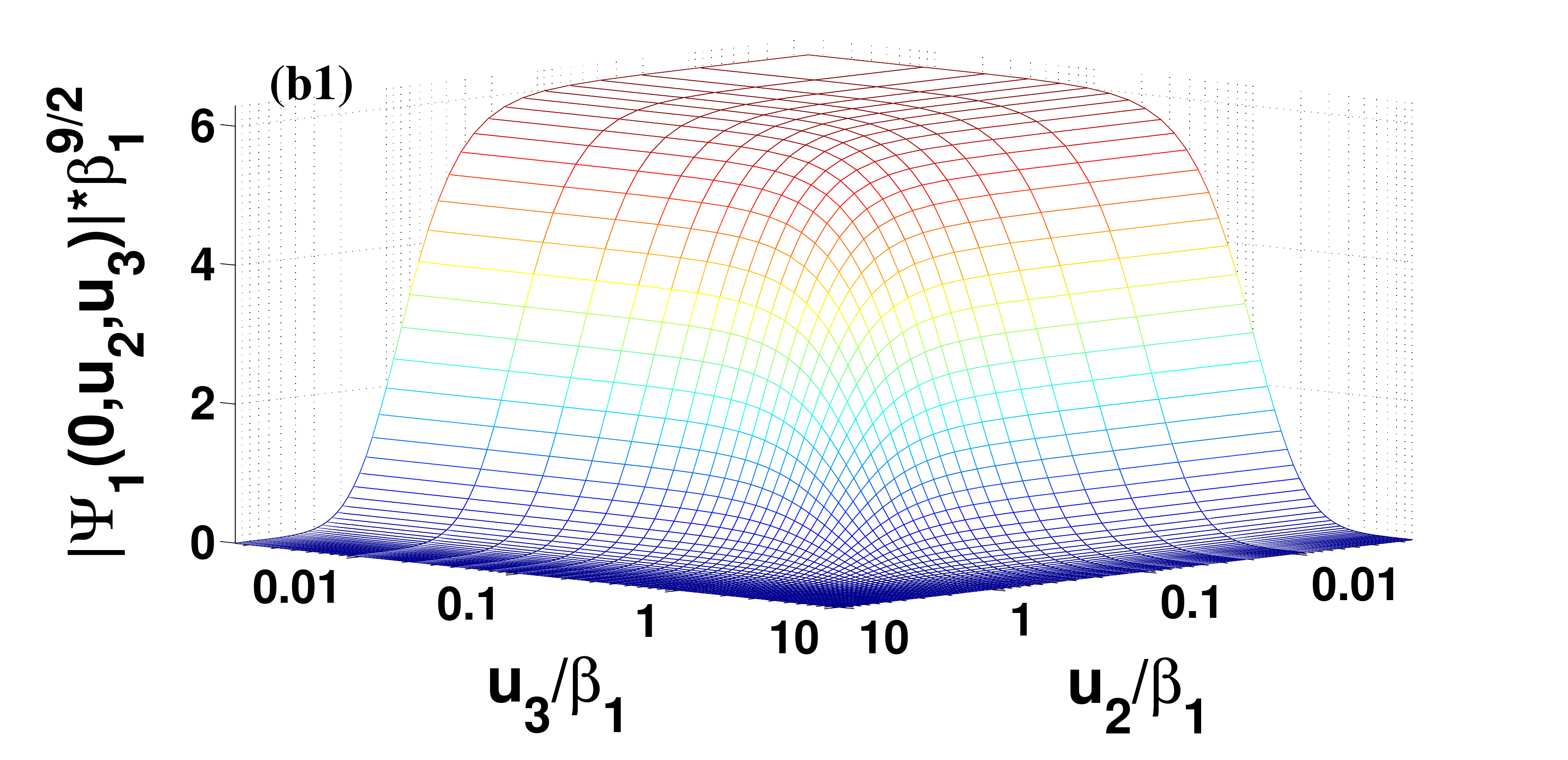}
\includegraphics[width=3.2in]{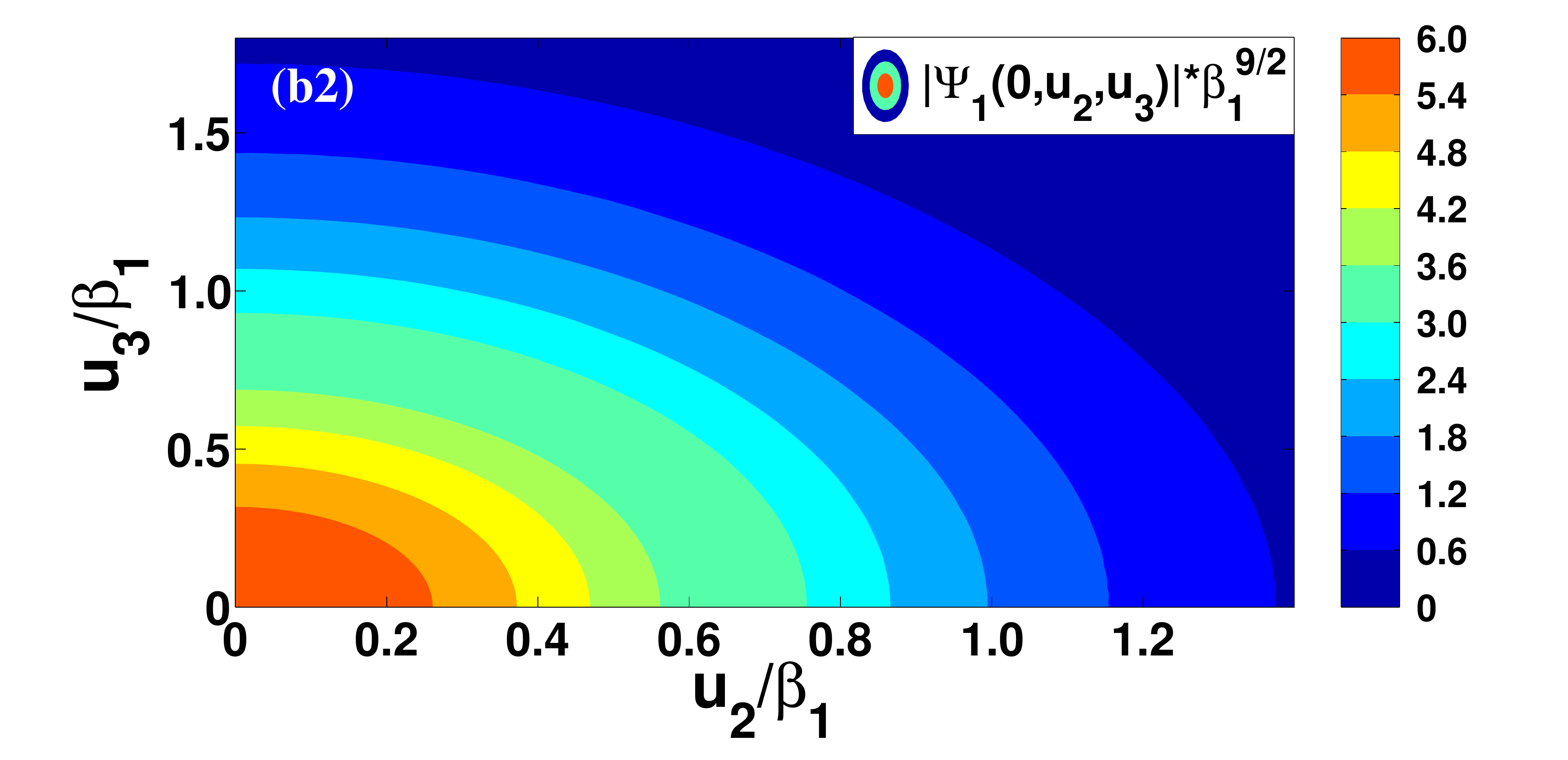} \\
\includegraphics[width=3.2in]{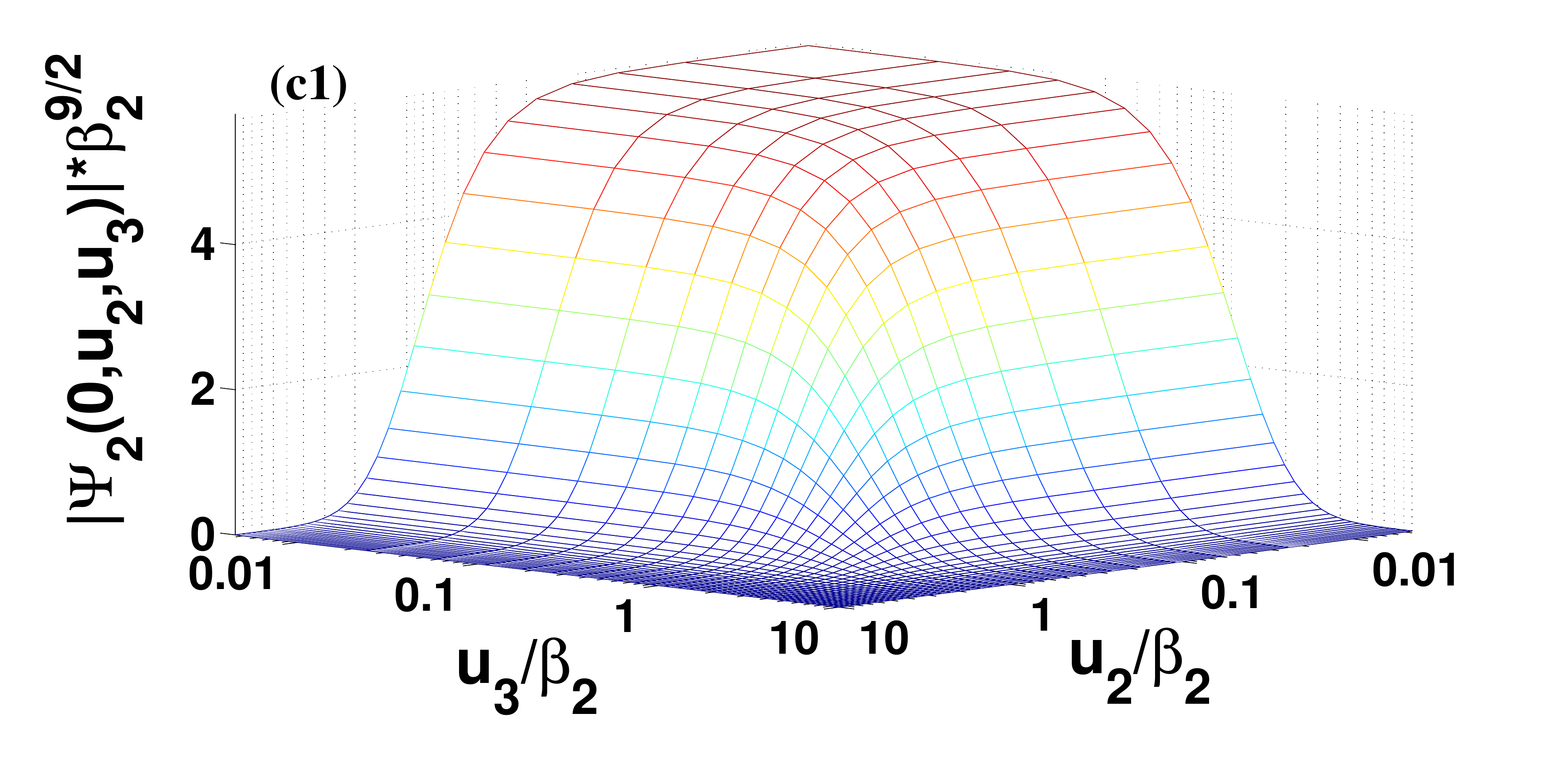}
\includegraphics[width=3.2in]{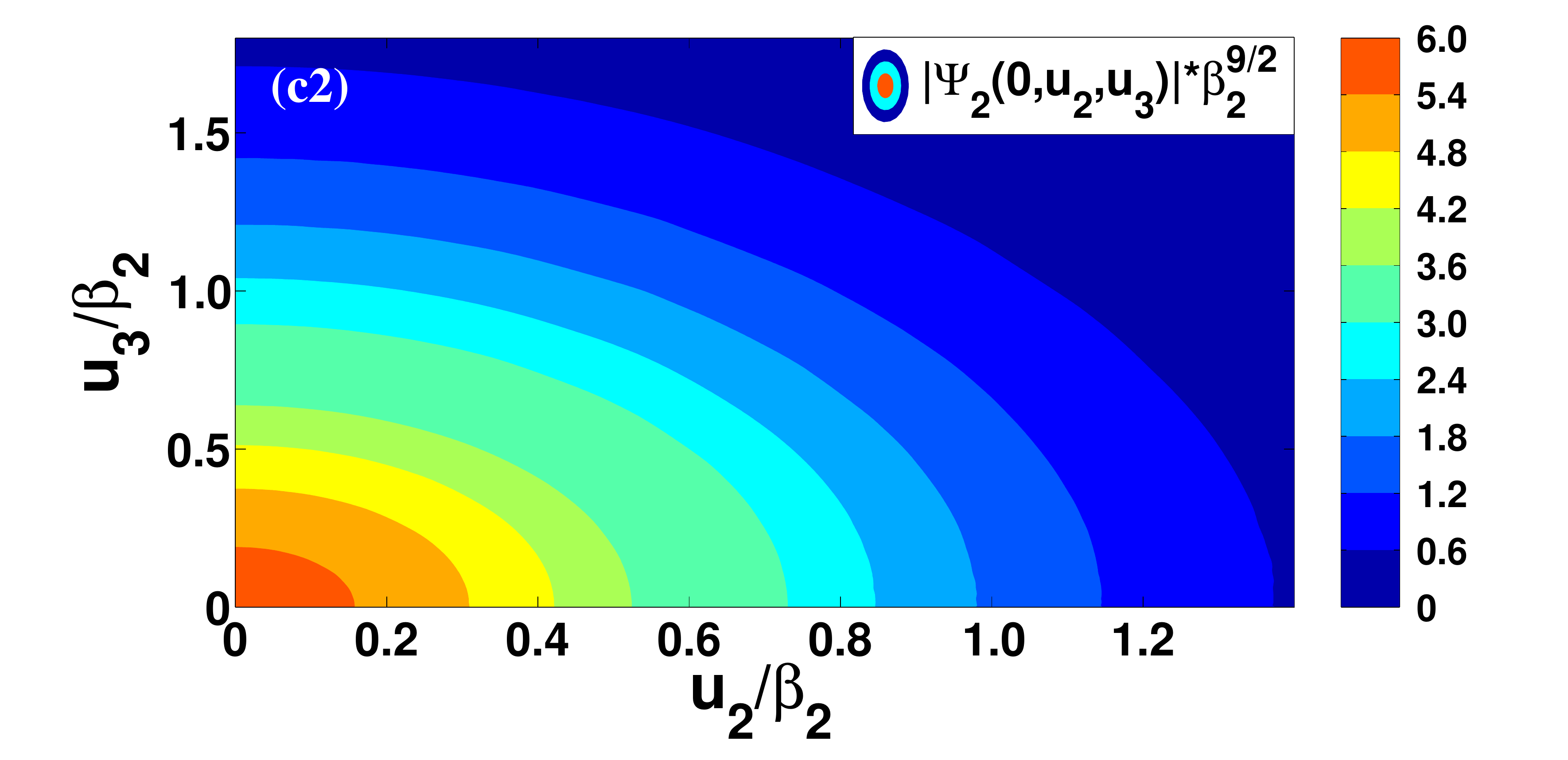}
\end{center}
\caption{(Color on-line) As in Fig.~\ref{Fig:WF_u1u2_50}, the wave-function
is given in terms of $u_2$ and $u_3$, with $u_1=0$.}
\label{Fig:WF_u2u3_50}
\vspace{-0.5cm}
\end{figure}

%FIG13
\begin{figure}[ht]
\begin{center}
\includegraphics[width=3.in]{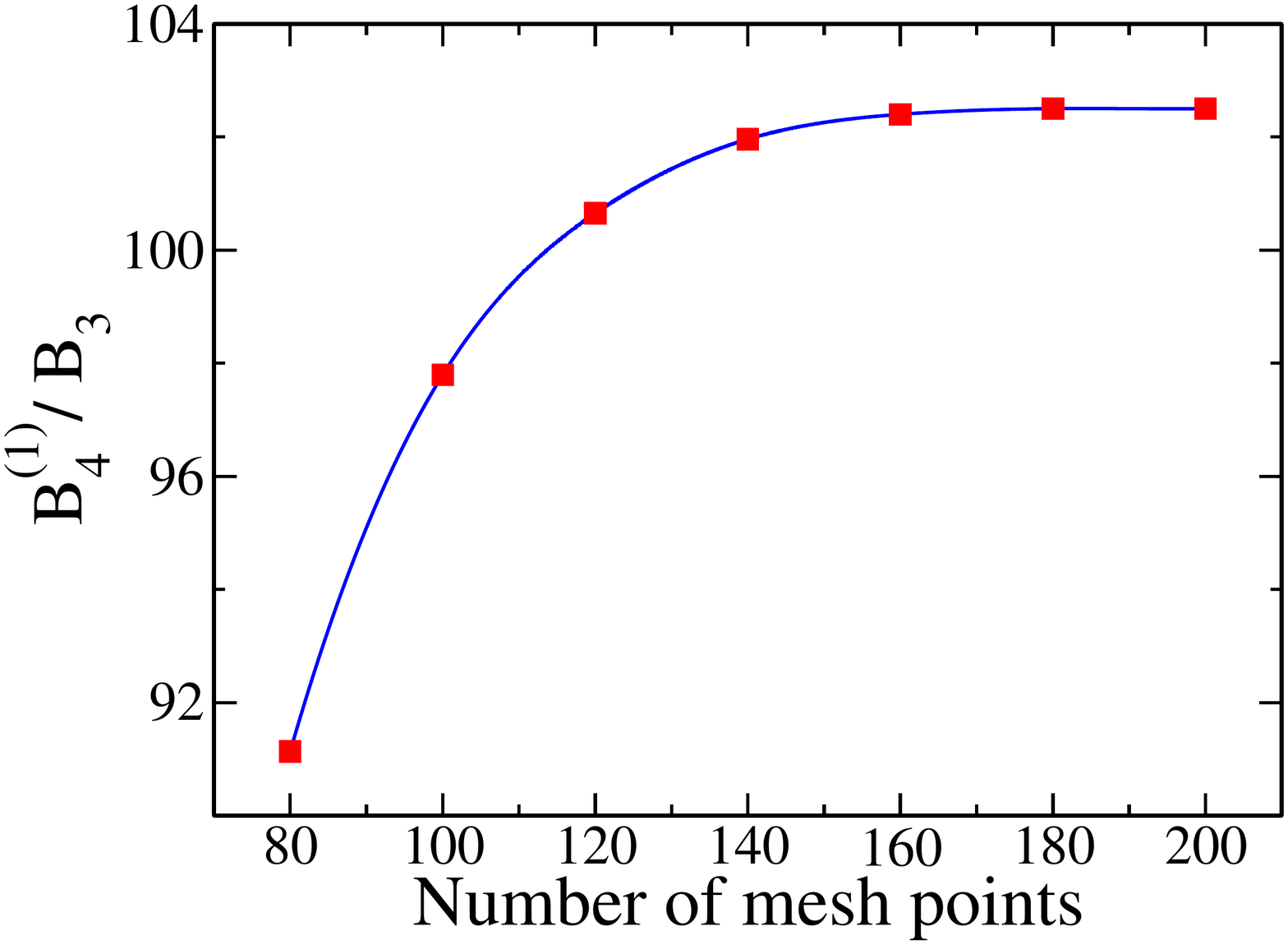}
\hspace{0.3cm}\includegraphics[width=3.in]{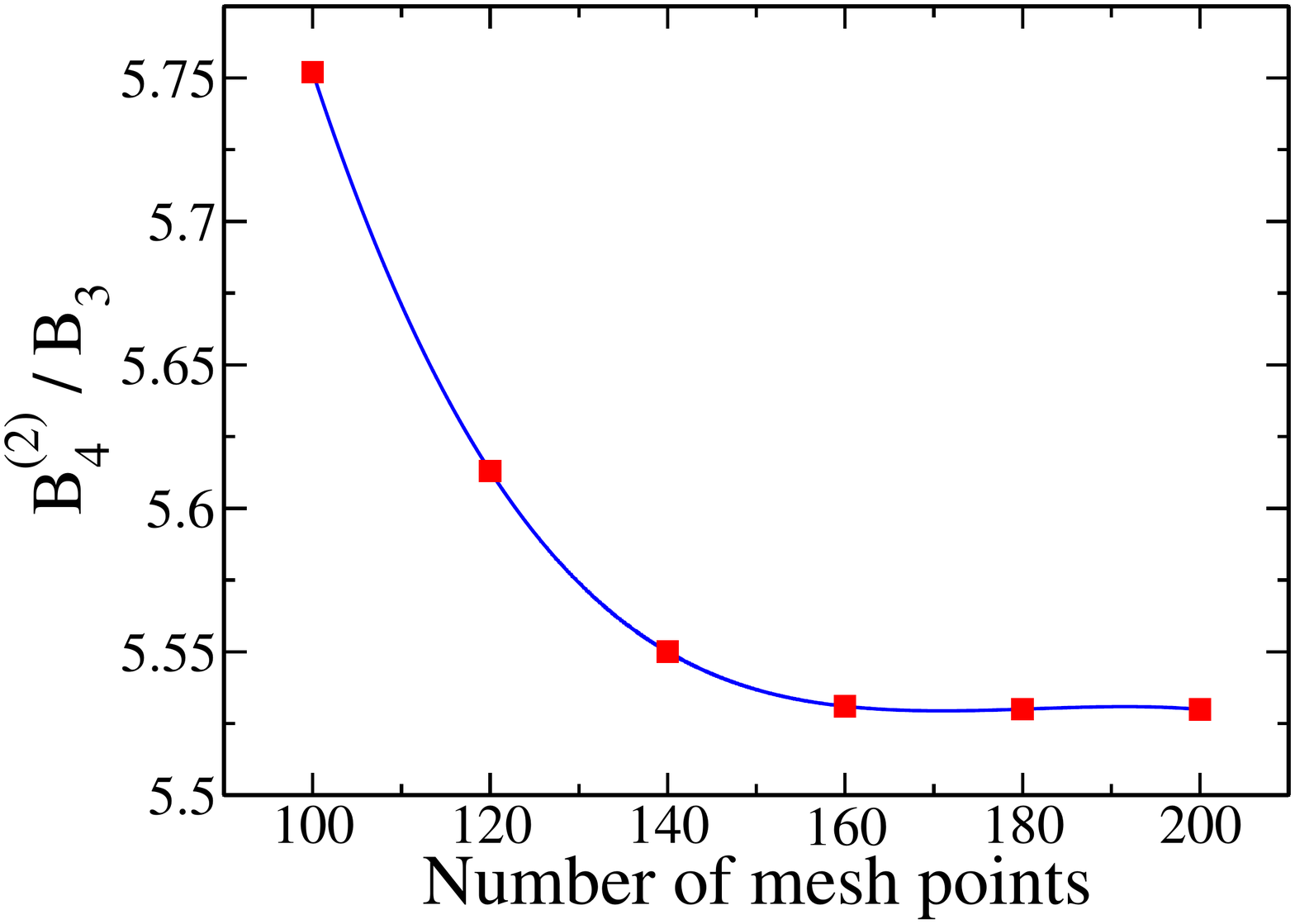}
\end{center}
\caption{(Color on-line) Convergence of numerical results for the first and second
excited state energies, in the case that $\mu_4/\mu_3 =300$.}
\label{convergence}
\end{figure}

\end{widetext}
\end{document}